\documentclass[reqno,12pt]{amsart} 
\usepackage{amssymb}
\numberwithin{equation}{section}

\def\stackreb#1#2{\ \mathrel{\mathop{#1}\limits_{#2}}}
\newcommand{\nc}{\newcommand}

\newcommand{\be}{\begin{equation}}
\newcommand{\ee}{\end{equation}}
\newcommand{\ba}{\begin{eqnarray}}
\newcommand{\ea}{\end{eqnarray}}

\nc{\rnc}{\renewcommand} \nc{\beq}{\begin{equation}}
\nc{\eeq}{\end{equation}} \nc{\beqa}{\begin{eqnarray}}
\nc{\eeqa}{\end{eqnarray}}

\begin{document}

\begin{flushright} AEI-2011-049 \end{flushright}

\title[Elliptic hypergeometry of supersymmetric dualities]
{\bf Elliptic hypergeometry of supersymmetric dualities II. \\
Orthogonal groups, knots, and vortices}

\author{V.~P.~Spiridonov}

\address{Bogoliubov Laboratory of Theoretical Physics,
JINR, Dubna, Moscow Region 141980, Russia; e-mail address:
spiridon@theor.jinr.ru}

\author{G.~S.~Vartanov}
\address{Max-Planck-Institut f\"ur Gravitationsphysik, Albert-Einstein-Institut
14476 Golm, Germany; e-mail address: vartanov@aei.mpg.de. Current address:
DESY Theory, Notkestr. 85, 22603 Hamburg, Germany}

\begin{abstract}
We consider Seiberg electric-magnetic dualities for $4d$ $\mathcal{N}=1$
SYM theories with $SO({N})$ gauge group. For all such known theories
we construct superconformal indices (SCIs) in
terms of elliptic hypergeometric integrals.
Equalities of these indices for dual theories lead both to proven earlier special function
identities and new conjectural relations for integrals.
In particular, we describe a number of new elliptic beta integrals
associated with the $s$-confining theories with the spinor matter fields.
Reductions of some dualities from $SP(2{N})$ to $SO(2{N})$ or
$SO(2{N}+1)$ gauge groups are described.
Interrelation of SCIs and the Witten anomaly is briefly discussed.
Possible applications of the elliptic hypergeometric
integrals to a two-parameter deformation of $2d$
conformal field theory and related matrix models
are indicated. Connections of the reduced SCIs with the state integrals
of the knot theory, generalized AGT duality for $(3+3)d$
theories, and a $2d$ vortex partition function are described.

\bigskip\medskip

{\normalsize

\begin{flushright} \em
Dedicated to D.I. Kazakov on the occasion of his 60th birthday
 \end{flushright}}
\end{abstract}


\maketitle

\tableofcontents

\section{Introduction}

Gauge field theories play a crucial role in the modern theory of elementary
particles. A generalization of the notion of electric-magnetic
duality from electrodynamics to non-abelian gauge theories was
suggested in the fundamental work of Goddard, Nuyts, and Olive \cite{Goddard}. In
the asymptotically free theories the spectrum of elementary excitations
in the high energy region is found from the free lagrangian. In the
infrared region the interaction becomes strong and one has to pass
to the description in terms of collective degrees of freedom (in the
usual quantum chromodynamics one should describe formation of the
hadrons out of quarks and gluons). The electric-magnetic duality
relates these two energy scales and is also referred to as the
strong-weak coupling duality transformation. To the present moment
consistent consideration of such transformations in $4d$ space-time
has been given only in the maximally extended ${\mathcal N}=4$
\cite{Osborn1979}, ${\mathcal N}=2$ \cite{Seiberg:1994aj},
and ${\mathcal N}=1$ \cite{Seiberg0,Seiberg} supersymmetric field theories.
In comparison to the dualities for ${\mathcal N}>1$ there exists a
whole zoo of different Seiberg dualities for $\mathcal{N}=1$ SYM
theories (see, e.g., surveys \cite{IN,Shifman:1995ua}).
The problem of their classification using some group-theoretical
approach is still open. For a survey of the current status of development
of supersymmetric gauge theories, see \cite{S:talk}.

Highly nontrivial generalizations of the Witten index called superconformal
indices (SCIs) were proposed recently by Kinney et al \cite{Kinney} and
R\"omelsberger \cite{Romelsberger1,Romelsberger2}. SCIs count
BPS states protected by one supercharge and its (superconformal)
conjugate which cannot be combined to form long multiplets.  They
can be considered as twisted partition functions in the Hilbert space
of BPS states which are determined by specific matrix integrals over the
classical Lie groups. SCI is a conformal manifold invariant \cite{Green:2010da}
which does not change under the marginal deformations \cite{Romelsberger2,SV4}.

In this paper we continue a systematic study of electric-magnetic
dualities for $\mathcal{N}=1$ SYM theories and $s$-confining theories
initiated in \cite{SV1}. We use for that the theory of elliptic
hypergeometric integrals (EHIs) developed by the first author in \cite{S1,S2,Sthesis}.
The crucial observation on the coincidence of SCIs
 with such integrals was done by Dolan
and Osborn in \cite{Dolan:2008qi}. In a sequel of papers \cite{S5,SV1,SV2,SV3,SV4,SV5,V}
we analyzed known supersymmetric dualities, described deep relations
between them and the properties of EHIs,
and, using these relations, discovered many new dualities.
Related questions were considered also in \cite{GPRR,GPRY}.

SCI techniques provides currently the most rigorous mathematical justification
of ${\mathcal N}=1$ supersymmetric dualities \cite{Seiberg0,Seiberg},
and it serves as a very powerful tool for getting new insights.
For instance, it has led to $\mathcal{N}=1$ dualities lying outside the conformal
window \cite{SV3}, it is useful for consideration of the AdS/CFT correspondence
for gauge groups of infinite \cite{Kinney,Nakayama,Nakayama:2006ur}
and finite \cite{Nakayama:2007jy,SV4} rank. It can be applied to
theories which are difficult to treat by usual physical tools
\cite{V}. Another interesting fact is that $4d$ SCIs can be reduced
to $3d$ partition functions \cite{Drukker:2010nc,Hama:2010av,Jafferis:2010un,Kapustin:2010xq}
yielding $3d$ dualities \cite{DSV,GY,I}.
Recently in \cite{Nakayama:2011pa} SCIs with the half-BPS superconformal
surface operator have been studied. EHIs are connected with the relativistic
Calogero-Sutherland type models where they describe either special wave functions
or the normalizations of particular wave functions \cite{S4a}.
In \cite{SV2} such a connection was conjectured to extend to all SCIs.
EHIs provide a unification of
known solvable models of statistical mechanics on $2d$ lattices
\cite{BS,S5}. In \cite{S5} it was shown that SCIs of the simple gauge group SYM
theories have the meaning of partition functions of elementary
cells of $2d$ integrable lattice models, and corresponding full partition
functions describe SCIs of particular quiver theories. In this picture, the Seiberg duality
has the meaning of a generalized Kramers-Wannier duality transformation
for partition functions. As shown in \cite{SV5}, $SL(3,\mathbb{Z})$-modular
transformation properties of EHIs are responsible for 't Hooft anomaly
matching conditions in dual theories.

SCI techniques  applies not only to $4d$ field theories,
but also to $3d$ models \cite{BK,Imamura:2011su,Kim:2009wb,Krattenthaler:2011da}.
In \cite{Krattenthaler:2011da} the equality of SCIs of
some $3d$ dual theories with $U(1)$ gauge group was proved rigorously for $N_f=1,2$
flavors, and in \cite{Kapustin:2011jm} this result was generalized to arbitrary $N_f$.
The analytical proof of the coincidence for partition functions
of some $3d$ quiver $\mathcal{N}=4$ mirror symmetric theories was
considered in \cite{Benvenuti:2011ga}.

There are several different ways of computing SCIs. The localization method
was used by Moore, Nekrasov and Shatashvili for computing the principal
contribution to the Witten index of supersymmetric theories expressed
as some contour integrals over $SU(N)$ group  \cite{MNS}.
Later this approach was generalized by Nekrasov and Shadchin \cite{NS}
for solving $\mathcal{N}=2$ supersymmetric field theories with symplectic
and orthogonal gauge groups.
For $\mathcal{N}=1$ SYM theories R\"omelsberger
\cite{Romelsberger1,Romelsberger2} derived SCIs using the operator
approach to free superconformal field theories (SCFTs) and suggested that
SCIs for Seiberg dual theories coincide.
For the asymptotically free theories in the ultraviolet region this is
formally justified. In \cite{Kinney}, Kinney et al derived SCI for
$\mathcal{N}=4$ $U(N)$-SYM theory using the representation theory for
free SCFTs \cite{Dobrev:1985vh} and targeting mostly the AdS/CFT correspondence.
SCIs for extended superconformal field theories can be derived directly from
the partition functions by imposing some restriction on the parameters \cite{BDHO}.
In \cite{Nawata:2011un}, the localization technique was used for derivation
of SCI for $\mathcal{N}=4$ SYM theory. In \cite{N,NO,Pestun}, this method was used
for computing partition functions of $\mathcal{N}=2$ SYM theories.
For related questions concerning counting the BPS operators,
see also \cite{Plethystic,Dolan:2007rq,Plethystic2}.
One can get $\mathcal{N}=2$ and $\mathcal{N}=4$ SYM theories out of
$\mathcal{N}=1$ theories by adjusting the matter fields content and
superpotentials. Analogously,
SCIs of extended theories can be obtained from $\mathcal{N}=1$
SCIs by appropriate fitting of the set of representations \cite{SV2}.

In this paper we are investigating SCIs for $4d$ $\mathcal{N}=1$ theories with
orthogonal gauge groups. The most interesting $SO(N)$-dualities arise from
the matter fields in spinor
representation. Dualities without such matter fields can be obtained
by reductions from the $SP(2N)$-gauge group cases.
Additionally, we outline possible application
of some of EHIs  (particular $4d$ SCIs) to a hypothetical
elliptic  deformation of $2d$ CFT. As an important relation between
$4d$ and $3d$ field theories, we show that reductions
of $4d$ SCIs to the hyperbolic $q$-hypergeometric level
yield the state integrals of knots
\cite{Dimofte:2011gm,Dimofte:2011jd,Dimofte:2009yn,Hikami1,Hikami2}.
Further reduction of a particular hyperbolic $q$-hypergeometric
integral emerging in this way is shown to give a $2d$ vortex partition function.

By definition SCIs count gauge invariant operators which saturate the BPS bounds
for short and semi-short multiplets. $\mathcal{N}=1$
SCFTs are based on the $SU(2,2|1)$ space-time symmetry group
which is generated by the following set of operators: $J_i,
\overline{J}_i$-- the generators of two $SU(2)$ subgroups forming
the $4d$ Lorentz group $SO(3,1)$, translations, $P_\mu$, $\mu=1,2,3,4$,
special conformal transformations, $K_\mu$,
the dilations, $H$, and also the $U(1)_R$-group generator
$R$. Apart from these bosonic generators there are supercharges
$Q_{\alpha},\overline{Q}_{\dot\alpha}$ with $\alpha, \dot{\alpha} = 1,2$
and their superconformal partners $S_{\alpha},\overline{S}_{\dot\alpha}$.
The full set of commutation relations for these operators can be found, for instance,
in \cite{SV2}. Taking a distinguished pair of intrinsically superconformal
charges \cite{Romelsberger1}, for example, $Q=\overline{Q}_{1 }$ and
$Q^{\dag}=-{\overline S}_{1}$, one has
\begin{equation}
\{Q,Q^{\dag}\}= 2{\mathcal H},\qquad Q^2=(Q^{\dag})^2=0,\qquad
\mathcal{H}=H-2\overline{J}_3-3R/2. \label{susy}\end{equation}
In this case the superconformal index is defined by the matrix integral
\begin{eqnarray}
I(p,q,f_k) = \int_{G} d \mu(g)\, \text{Tr} \Big( (-1)^{\mathcal F}
p^{\mathcal{R}/2+J_3}q^{\mathcal{R}/2-J_3}
e^{\sum_{a} g_aG^a} e^{\sum_{k}f_kF^k}e^{-\beta {\mathcal H}}\Big),
\quad \mathcal{R}= R + 2 \overline{J}_3, \label{Ind}\end{eqnarray}
where ${\mathcal F}$
is the fermion number operator and $d \mu(g)$ is the invariant
measure of the gauge group $G$. We explicitly singled out the
integration over gauge group, though most often it is assumed
to be a part of the gauge invariant trace.
To calculate the index one should not consider the whole space of states,
but only zero modes of the operator $\mathcal H$ because
contributions of states not annihilated by the supercharge $Q$
cancel each other. The chemical potentials $g_a,\, f_k$
correspond to the gauge $G$ and flavor $F$ symmetry group
generators $G^a$ and $F^k$, respectively.

$4d$ SCI coincides with the supersymmetric index on $S^3 \times S^1$ manifold.
For a latest discussion of such space-time manifestations, see
\cite{Festuccia:2011ws,S:talk}. According to the R\"omelsberger prescription
(in the form suggested in \cite{Dolan:2008qi}) one should first compute
the single particle index, given by the following general formula
\begin{eqnarray}
\text{ind}(p,q,z,y) &=& \frac{2pq - p - q}{(1-p)(1-q)}
\chi_{adj}(z)\cr &+& \sum_j
\frac{(pq)^{r_j}\chi_{R_F,j}(y)\chi_{R_G,j}(z) - (pq)^{1-r_j}\chi_{{\bar
R}_F,j}(y)\chi_{{\bar R}_G,j}(z)}{(1-p)(1-q)}.
\label{index}\end{eqnarray}
Here the first term describes the contribution of gauge fields belonging to the adjoint
representation of group $G$; the sum over $j$ corresponds to
the chiral matter superfields $\Phi_j$ transforming as the gauge
group representations $R_{G,j}$ and the flavor symmetry
group representations $R_{F,j}$ with $2r_j$ being their $R$-charges.
The functions $\chi_{adj}(z)$, $\chi_{R_F,j}(y)$ and $\chi_{R_G,j}(z)$ are the
characters of representations with $z$ and $y$ being the
maximal torus variables of $G$ and $F$ groups, respectively.
All the characters needed for this work are explicitly listed in Appendix
\ref{chSUSP}. Originally \cite{Dolan:2008qi,Romelsberger2} the index was
expressed in terms of variables $x,t$ related to our bases as $p=tx, q=tx^{-1}$.
We remark that as a result of the change of variables there appears a sign ambiguity,
the term $(pq)^{r_j}$ in \eqref{index} can be written as $(\pm\sqrt{pq})^{R_j}$,
where $R_j$ are $R$-charges of the fields, and this may influence the balancing
condition for integrals below.

To obtain the full superconformal
index, the single particle states index \eqref{index} is inserted into the
``plethystic" exponential which is then averaged over the gauge group:
\begin{equation}\label{Ind_fin}
I(p,q,y) \ = \ \int_G d \mu(z)\, \exp \bigg ( \sum_{n=1}^{\infty}
\frac 1n \text{ind}\big(p^n ,q^n, z^n , y^ n\big ) \bigg ).
\end{equation}
It appears that such matrix integrals  are expressed in terms of the new
special functions of mathematical physics known as elliptic
hypergeometric integrals which were
discovered in \cite{S1,S2,Sthesis} (see also \cite{S3} for a general survey).
Their simplest representative -- the exactly
computable elliptic beta integral \cite{S1} is the top level known
generalization of the Euler beta integral,
the Askey-Wilson and Rahman $q$-beta integrals \cite{aar}.
As found in  \cite{Dolan:2008qi}, it describes the confinement phenomenon for $4d$ $\mathcal{N}=1$
SYM theory with $SU(2)$ gauge group and $6$ quarks which
is dual to the theory of free baryons forming the absolutely
antisymmetric tensor representation of the flavor group $SU(6)$.
On the base of a very large number of explicit examples listed in \cite{SV2},
we conjectured that to every supersymmetric duality
there corresponds either an exact integration formula for elliptic beta
integrals or a nontrivial Weyl group symmetry transformation
for the higher order EHIs .

One important remark is in order. Described index computation
algorithm does not impose in advance any constraint on the fugacities,
whereas the EHI identities used for establishing
equalities of SCIs require neat fitting of parameter constraints for their
existence (see below). It would be interesting to find the arguments leading to
needed constraints for fugacities directly in formulas \eqref{index}
and \eqref{Ind_fin}.

This paper can be considered as a second part of the work \cite{SV2} since
we cover several subjects skipped in it. However, there are still some
interesting questions touched in \cite{SV2}, but
not included in this paper. In particular, we do not discuss SCIs
of quiver theories which have attracted recently some interest in \cite{C-T,S5}.

\section{Reduction of $\mathcal{N}=1$ dualities from symplectic to
orthogonal gauge groups}\label{SP-SO}

\subsection{Dualities without spinor matter}
Let us show that known  $\mathcal{N}=1$ dualities with $SO(n)$
gauge group without matter in the spinor representation can be derived as
consequences of known $SP(2N)$ gauge group dualities.
At the level of SCIs this implication is achieved  by
particular restriction of the values of a number of parameters
in the corresponding EHIs ,
as observed first by Dolan and Osborn for the simplest cases \cite{Dolan:2008qi}.
In the present section we discuss such reductions in more detail.
The spinor matter theories will be considered later on.

We start from $\mathcal{N}=1$ SYM theory with $SP(2{N})$ gauge group
and $2N_f$ quarks in the fundamental representation having the
global symmetry group $SU(2N_f) \times U(1)_R.$ The matter fields
are described in the table below, where we indicate their representation types
for the gauge and flavor groups and provide $R$-charges

\vbox{ \hskip2.5cm \vbox{\tabskip=0pt \offinterlineskip \hrule
\halign{&\vrule# &\strut \ \hfil#\  \cr
height2pt&\omit&&\omit&&\omit&&\omit&\cr &\     \ \hfil   && \
$SP(2{N})$  \  && \  $SU(2N_f)$ \ &&  \ $U(1)_R$ \ &\cr
height2pt&\omit&&\omit&&\omit&&\omit&\cr \noalign{\hrule}
height2pt&\omit&&\omit&&\omit&&\omit& \cr & \ $Q$  \ \hfil &&  $f$
\hfil     &&    $f$     \hfil &&  $1 - ({N}+1)/ N_f$ \hfil  & \cr }
\hrule} }
In this and all other tables below we skip the vector
superfield $V$ (or its dual partner $\widetilde V$, which is absent
in confining theories) described by the
adjoint representation of $G$ and singlets of the non-abelian part
of the flavor group, and having trivial hypercharges for the abelian global groups.

The dual magnetic theory constructed by Intriligator and Pouliot
\cite{Intriligator1} has the same flavor group and the gauge group
$G=SP(2\widetilde{N}),$ where $\widetilde{N} = N_f-{N}-2$, with
the matter field content described in the table below

\vbox{ \hskip2.5cm \vbox{\tabskip=0pt \offinterlineskip \hrule
\halign{&\vrule# &\strut \ \hfil#\  \cr
height2pt&\omit&&\omit&&\omit&&\omit&\cr &\     \ \hfil   && \ $SP(2
\widetilde{N})$  \  && \  $SU(2N_f)$ \ &&  \ $U(1)_R$ \  &\cr
height2pt&\omit&&\omit&&\omit&&\omit&\cr \noalign{\hrule}
height2pt&\omit&&\omit&&\omit&&\omit& \cr & \ $q$  \ \hfil &&  $f$
\hfil     &&    $\overline{f}$ \hfil            &&  $({N} +1)/ N_f$
\hfil & \cr & \ $M$  \ \hfil &&  $1$  \hfil     && $T_A$ \hfil
  &&  $2(\widetilde{N}+1)/N_f$ \hfil
& \cr } \hrule} }
\noindent
where $f$ ($\overline{f}$) denotes (anti)fundamental representation
and $T_A$ denotes the antisymmetric tensor of the second rank.

The conformal window for this duality is $3({N}+1)/2 < N_f <
3({N}+1)$; it emerges from the demand that both dual theories
are asymptotically free in the one-loop approximation.
The Seiberg electric-magnetic duality at the
infrared fixed points of these theories, which is not proven rigorously yet,
had the following justifying arguments \cite{Seiberg}:
\begin{itemize}
  \item the 't Hooft anomaly matching conditions are satisfied. They were
shown in \cite{SV5} to follow from the $SL(3,\mathbb{Z})$-group transformation
properties of EHIs;
  \item matching reduction of the number of flavors $2N_f\to 2(N_f-1)$.
Integrating out $2N_f, (2N_f-1$)-th flavor quarks by adding the mass
term in electric theory results in Higgsing the magnetic theory gauge
group with decoupling of a number of meson fields. For SCIs this is realized by
restricting a pair of parameters, $t_{2N_f}t_{2N_f-1}=pq$
\cite{SV1,SV2};
  \item matching of the moduli spaces and gauge invariant operators in
dual theories. This information is believed to be hidden in the
topological meaning of SCIs.
\end{itemize}

We need the following EHI on the $BC_n$ root system
\beq
\label{eqR} I_{n}^{(m)}(\underline{t};p,q) \ = \
\frac{(p;p)_\infty^n (q;q)_\infty^n}{2^n n!} \int_{\mathbb{T}^n}
\prod_{1 \leq i < j \leq n}
\frac{1}{\Gamma(z_i^{\pm1}z_j^{\pm1})} \prod_{j=1}^n
\frac{\prod_{i=1}^{2(m+n+2)}
\Gamma(t_iz_j^{\pm1})}{\Gamma(z_j^{\pm2})} \prod_{j=1}^n
\frac{dz_j}{2 \pi \textup{i} z_j},\eeq
where $\mathbb{T}$ is the unit circle with positive orientation, all $|t_i|<1$,
$\prod_{i=1}^{2(m+n+2)}t_i = (pq)^{m+1}$,
$$
(z;q)_\infty = \prod_{i=0}^\infty (1-zq^i), \qquad |q|<1,
$$
is the standard infinite $q$-shifted factorial \cite{aar} and
\beq \Gamma(z) \ \equiv \ \Gamma(z;p,q) \ = \
\prod_{i,j=0}^\infty \frac{1-z^{-1}p^{i+1}q^{j+1}}{1-zp^iq^j}, \quad
|p|, |q|<1,\eeq
 is the elliptic gamma function \cite{S3}. We use the convention
$$
 \Gamma(t_1,\ldots,t_k)= \Gamma(t_1)\ldots \Gamma(t_k), \quad
 \Gamma(tz^{\pm1})= \Gamma(tz) \Gamma(tz^{-1}).
$$
Then the algorithm for construction SCIs described above yields
for the electric theory $I_E^{SP(2N)} =I_{N}^{(N_f-N-2)}(t_1,\ldots,t_{2N_f};p,q)$
\cite{Dolan:2008qi,SV2}.
The dual magnetic theory has SCI of the form
$$
I_M^{SP(2\widetilde N)}=\prod_{1 \leq i < j \leq 2N_f} \Gamma(t_it_j) \
I^{(N)}_{N_f-N-2}((pq)^{1/2}/t_1,\ldots,(pq)^{1/2}/t_{2N_f};p,q).
$$
R\"omelsberger's conjecture on the equality of SCIs for dual theories
$I_E^{SP(2N)} = I_M^{SP(2\widetilde N)}$
was proven in \cite{Dolan:2008qi} on the basis of
the symmetry transformation for integrals established in \cite{Rains}. For $N=1$ the
full symmetry group of SCI is $W(E_7)$. The key transformation
generating this group was found earlier in \cite{S2}.
Its physical consequences for multiple dualities have been
studied in \cite{SV1} and the superpotentials for such
theories were investigated later in \cite{Khmelnitsky:2009vc}.
Altogether the results of \cite{Dolan:2008qi,SV1,SV2} gave a new
powerful, most rigorous from the mathematical point of view
confirmation of the Seiberg duality, complementing the tests mentioned above.

It should be stressed that this and all other equalities of SCIs
of dual theories are true or supposed to be true
only if the values of parameters in all integrals guarantee that only
sequences of poles of the integrands converging to zero are located inside
the contour of integration $\mathbb{T}$ (otherwise one should use the
nontrivial analytical continuation procedure for identities to be true
in other regions of parameters).

Consider now the Seiberg duality for $\mathcal{N}=1$ SYM
theories with orthogonal gauge group \cite{Seiberg}. The
electric theory matter fields are described in the following table
\begin{center}
\begin{tabular}{|c|c|c|c|c|}
  \hline
    & $SO({N})$ & $SU(N_f)$ & $Z_{2N_f}$ & $U(1)_R$ \\  \hline
  $Q$ & $f$  & $f$ & $k$ & $\frac{N_f-{N}+2}{N_f}$ \\  \hline
\end{tabular}
\end{center} and for the magnetic theory one has
\begin{center}
\begin{tabular}{|c|c|c|c|c|}
  \hline
    & $SO(\widetilde{N})$ & $SU(N_f)$ & $Z_{2N_f}$ & $U(1)_R$ \\  \hline
  $q$ & $f$  & $\overline{f}$ & $-k$ & $\frac{{N}-2}{N_f}$ \\
  $M$ & $1$  & $T_S$ & $2k$ & $2\frac{N_f-{N}+2}{N_f}$ \\  \hline
\end{tabular}
\end{center}
where $T_S$ denotes the absolutely symmetric tensor of second
rank and $\widetilde{N} = N_f-{N}+4$.
The  conformal window \cite{Seiberg} for this duality has the form
$3(N - 2)/2 < N_f < 3(N-2),$
which guarantees existence of the non-trivial infrared   fixed points
(one should be careful with the use of such windows since there are
examples \cite{SV3} of dualities lying outside them).

In these tables we explicitly indicated existence of the discrete $Z_{2N_f}$
symmetry  \cite{Seiberg,IN3}. In order to take it into account in
the construction of SCIs we modify the R\"omelsberger prescription
for orthogonal groups. Introduce the single particles states index
\begin{eqnarray}\label{indSO}
\text{ind}(p,q,z,y,x_k) &=& \frac{2pq - p - q}{(1-p)(1-q)}
\chi_{adj}(z)\cr &+& \sum_j
\frac{x_k (pq)^{r_j}\chi_{R_F,j}(y)\chi_{R_G,j}(z) - (pq)^{1-r_j}\chi_{{\bar
R}_F,j}(y)\chi_{{\bar R}_G,j}(z)/x_k}{(1-p)(1-q)},
\end{eqnarray}
where $x_k=e^{\pi \textup{i}k/N_f},\, k=0,\ldots,2N_f-1,$ and apply
the general formula (\ref{Ind_fin}) with the powers $x_k^n$ in the
plethystic exponential.

Orthogonal groups $SO(n)$ are qualitatively different for even $n=2N$
(root system $D_N$) and odd $n=2N+1$ (roots system $B_N$).
SCIs in the electric theory take the form
\begin{eqnarray} I_E^{SO(2{N})} &=& \frac{(p;p)_{\infty}^{{N}}
(q;q)_{\infty}^{{N}}}{2^{{N}-1} {N}!} \int_{\mathbb{T}^{{N}}}
\frac{\prod_{i=1}^{N_f} \prod_{j=1}^{{N}} \Gamma(t_i z_j^{\pm
1})}{\prod_{1 \leq i < j \leq {N}} \Gamma(z_i^{\pm 1}z^{\pm
1}_j)} \prod_{j=1}^{{N}} \frac{d z_j}{2 \pi \textup{i} z_j},
\label{iso2N}\end{eqnarray}
where the balancing condition reads $\prod_{i=1}^{N_f}t_i=\pm(pq)^{N_f/2-N+1}$, and
\beqa\label{iso2N+1} && \makebox[-5em]{}
I_E^{SO(2{N}+1)} =
\frac{(p;p)_{\infty}^{{N}} (q;q)_{\infty}^{{N}}}{2^{{N}} {N}!}
\prod_{i=1}^{N_f} \Gamma(t_i)
\int_{\mathbb{T}^{{N}}} \frac{\prod_{i=1}^{N_f} \prod_{j=1}^{{N}}
\Gamma(t_i z_j^{\pm 1})}{\prod_{j=1}^{{N}}
\Gamma(z_j^{\pm1}) \prod_{1 \leq i < j \leq {N}} \Gamma(z_i^{\pm
1}z^{\pm 1}_j)} \prod_{j=1}^{{N}} \frac{d z_j}{2 \pi \textup{i}z_j},
\eeqa
where the balancing condition is  $\prod_{i=1}^{N_f}t_i=\pm (pq)^{N_f/2-N+1/2}$.
Here $t_i:=x_k(pq)^{r_i}y_i$ and the effect of the discrete chemical
potential $k$ is reduced to
the sign value on the right-hand side of the balancing condition.

The magnetic theory SCI can be written in the form:
\beq I_M(\underline{t};p,q)^{SO(\widetilde{N})} =
\prod_{1 \leq i < j \leq N_f} \Gamma(t_it_j) \prod_{i=1}^{N_f}
\Gamma(t_i^2)I_E(\frac{\sqrt{pq}}{\underline{t}};p,q)^{SO(\widetilde{N})}.
\eeq
To show the duality relation
$ I_E(\underline{t};p,q)^{SO({N})} = I_M(\underline{t};p,q)^{SO(\widetilde{N})}$
one has to restrict parameters in the $SP(2N)$-indices
\cite{Dolan:2008qi}. First we identify \beqa
I_E(\underline{t};p,q)^{SO(2n)} =
\begin{cases}
2I_{n}^{(\frac 12 (N_f+4)-n)}(\underline{t},\underline{u};p,q), & N_f \text{ even}, \\
2I_{n}^{(\frac 12 (N_f+3)-n)}(\underline{t},\underline{v};p,q), &
N_f \text{ odd},
\end{cases}\eeqa
where  parameters $\underline{u}$ and $\underline{v}$ in $I_{n}^{(m)}$ are chosen as
$$\underline{u} \ = \ \left \{ \pm1, \pm \sqrt{p}, \pm \sqrt{q}, \pm \sqrt{pq} \right
\},\qquad \underline{v} \ = \ \left \{ \pm1, \pm \sqrt{p}, \pm
\sqrt{q}, - \sqrt{pq} \right \}.$$
Analogously,
\beqa
I_E(\underline{t};p,q)^{SO(2n+1)} =
\begin{cases}
\prod_{i=1}^{N_f} \Gamma(t_i) I_{n}^{(\frac 12 (N_f+2)-n)}(\underline{t},
\underline{u}'), & N_f \text{ even}, \\
\prod_{i=1}^{N_f} \Gamma(t_i) I_{n}^{(\frac 12
(N_f+3)-n)}(\underline{t},\underline{v}';p,q), & N_f \text{ odd},
\end{cases}\eeqa
where
$$\underline{u}' \ = \ \left \{ -1, \pm \sqrt{p}, \pm \sqrt{q}, - \sqrt{pq} \right
\}, \qquad \underline{v}' \ = \ \left \{ -1, \pm \sqrt{p}, \pm
\sqrt{q}, \pm \sqrt{pq} \right \}.$$

These relations are based on the duplication formula for the elliptic
gamma function
\beq \Gamma(z^2) = \prod_{\varepsilon=\pm1} \Gamma(\varepsilon
z,\varepsilon \sqrt{p} z, \varepsilon \sqrt{q} z, \varepsilon\sqrt{pq} z)
\eeq
and the inversion formula $\Gamma(z) \Gamma(pq/z)= 1.$
They allow one to reduce EHIs from $SP(2n)$-group to $SO(2n)$
or $SO(2n+1)$ and, simultaneously, reduces mesons from $T_A$- to
$T_S$-representation.

The same line of arguments works for checking equality of SCIs
for many other known dualities of orthogonal gauge group theories
whose matter content we list below:
\begin{itemize}
  \item the antisymmetric tensor
of the second rank (or the  adjoint representation)
and quarks in the fundamental representation, see \cite{LSS} for the duality
between interacting field theories and \cite{Csaki7,Klein} for
the $s$-confining theory;
  \item the symmetric tensor of the second rank
and quarks in the fundamental representation, see \cite{Intriligator2}
for nontrivial dual gauge group case and
\cite{Csaki7,Klein} for the $s$-confining theory;
  \item   two matter fields -- symmetric tensors
of the second rank and quarks in the fundamental
representation, see \cite{Brodie2,Klein};
  \item   one matter field -- the symmetric tensor
of the second rank, and another field, the antisymmetric tensor of
the second rank, together with the quarks in the
fundamental representation, see \cite{Brodie2,Klein}.

\end{itemize}

For brevity we are not presenting explicitly SCIs of these theories
and do not indicate how they are related to $SP(2N)$-group indices
considered in \cite{SV2} since they are easily obtained by
reductions similar to the one described above. Moreover,
one can obtain new orthogonal gauge group dualities with the flavor
group composed of several $SP(2m)$-groups and $SU(4)$ group
after a similar reduction of the duality considered in Sect.  $7$
of \cite{SV2} (as well as the related $s$-confining theory).
The general question why $SO$-dualities for theories without spinor
matter can be derived from $SP$-theories is not understood
from the physical point of view yet.

Now we would like to discuss some special cases in more detail.
Consider $G=SO(n)$ theory with $N_f=n-1$ quarks known to have
three dual pictures \cite{IN3}: electric, magnetic, and dyonic.
For $G=SO(2N+1)$ with $2N$ quarks SCI is obtained from
\eqref{iso2N+1} with $N_f$ replaced by $2N$.
The magnetic dual has $SO(3)$ gauge group with SCI
\beqa\label{Tr_mag1} &&\makebox[-2em]{}
I^{SO(3)}_{M}= \prod_{1 \leq m < s \leq
2{N}} \Gamma(t_mt_s) \prod_{i=1}^{2{N}} \Gamma(t_i^2,
\frac{\sqrt{pq}}{t_i}) \frac{(p;p)_\infty (q;q)_\infty}{2}
\int_{\mathbb{T}}\frac{\prod_{i=1}^{2{N}}
\Gamma(\frac{\sqrt{pq}}{t_i}y^{\pm1})}{\Gamma(y^{\pm1})}
\frac{dy}{2 \pi \textup{i} y}, \eeqa
the balancing condition here reads
$\prod_{m=1}^{2{N}} t_m = \sqrt{pq}.$
These expressions can also be obtained from $SP(2N)$-indices
with $N_f={N}+3$. The moduli space of vacua of the $SO(3)$-theory has two non-trivial
points leading to two dual theories. One of them is the original $SO(2{N}+1)$-electric
theory, and the second one is the $SO(2{N}+1)$-dyonic theory, which is
obtained from the electric one by adding a particular term to the superpotential
and  shifting the theta angle by $\pi$. The electric and dyonic theories
are related to each other by the ``weak-to-weak" $T$-duality transformation and,
therefore, their superconformal indices are identical, $I_D\equiv I_E$.
These duality transformations form the permutation group
$S_3$, a subgroup of the $SL(2,\mathbb{Z})$-group, interchanging the three
different theories.

The same arguments apply to $\mathcal{N}=1$ SYM theory with $SO(2{N})$
gauge group and $2{N}-1$ quarks. Restricting
seven parameters in $I_E^{SP(2N)}$ (with $N_f={N}+3$) as
$1, \pm \sqrt{p}, \pm \sqrt{q}, \pm \sqrt{pq},$
one obtains SCI of the electric theory identically coinciding with
the index of the dyonic theory. Substituting the same constraints
to  $I_M^{SP(2N)}$ one obtains SCI of the $SO(3)$-magnetic theory. In both
cases the balancing condition reads $\prod_{i=1}^{2{N}-1} t_i = 1$,
i.e. at least one of the parameters $t_i$ has modulus greater than 1,
which requires an appropriate deformation of the integration contours
for separation of relevant sequences of integrand poles.

As to the self-dual case of $SO(3)$-gauge group, its SCIs $I_E^{SO(3)}$
and $I_M^{SO(3)}$ depend on two parameters with the balancing condition
$t_1t_2 = \sqrt{pq}$.
Remarkably, after taking into account the latter constraint, the index $I_M^{SO(3)}$
becomes identically equal to $I_{E}^{SO(3)}$.
So, the electric, magnetic, and dyonic theories differ from each other only
by particular terms in the superpotential (governed by the
parameter $e=0,\pm1$ in \cite{IN3}) and have SCIs of identical shape.

According to Seiberg \cite{Seiberg}, the case $G=SO(n)$ with $N_f=n-2$
has the dual gauge group $SO(2)$, i.e. the magnetic theory coincides with  $\mathcal{N}=1$
abelian theory describing the supersymmetric photon with the gauge group $U(1)$.
This duality can be deduced from the $SP(2{N}) \leftrightarrow SP(2(N_f-{N}-2))$
duality with $N_f={N}+3$. Corresponding SCIs are obtained by imposing
appropriate constraints on the parameters, as described above.
For $G=SO(2N+1)$ SCI is given by expression
\eqref{iso2N+1} with $N_f$ replaced by $2N-1$.
The dual SCI has the form
\beqa\nonumber && I^{SO(2)}_M= \prod_{1 \leq m < s \leq
2{N}-1} \Gamma(t_mt_s) \prod_{i=1}^{2{N}-1} \Gamma(t_i^2)
\frac{(p;p)_\infty (q;q)_\infty}{2}
\int_{C} \prod_{i=1}^{2{N}-1}
\Gamma(\frac{\sqrt{pq}}{t_i}y^{\pm1}) \frac{dy}{2 \pi \textup{i}y},
 \eeqa
where is it assumed that $N\geq 2$.
Here the balancing condition reads $\prod_{m=1}^{2{N}-1} t_m = 1,$
so that at least one of the parameters should be of modulus greater than $1$.
Therefore the integration contours in $I_E$ should be deformed appropriately.
For the gauge group $SO(2N)$ we have SCI given by \eqref{iso2N} with
$N_f$ replaced by $2N-2$ and
\beqa &&
I^{SO(2)}_M= \prod_{1 \leq m < s \leq 2{N}-2}
\Gamma(t_mt_s) \prod_{i=1}^{2{N}-2}
\Gamma(t_i^2)\frac{(p;p)_\infty (q;q)_\infty}{2}
\int_{\mathbb{T}} \prod_{i=1}^{2{N}-2}
\Gamma(\frac{\sqrt{pq}}{t_i}y^{\pm1}) \frac{dy}{2 \pi \textup{i}y},
 \nonumber \eeqa
where  the balancing condition is $\prod_{m=1}^{2{N}-2} t_m = 1$ and $N>2$.
For $N=2$ both expressions
diverge and one has to apply appropriate regularization $t_1t_2\neq 1$
and residue calculus to obtain a meaningful limit $t_1t_2\to1$.
Interestingly, both magnetic SCIs are represented by the general well-poised
EHIs without the very-well-poisedness condition \cite{S3}
(which is thus not obligatory for applications in supersymmetric theories).

Consider dualities for $G=SO(n)$ and $N_f=n-3$ \cite{IN3}. Their
SCIs are obtained by a reduction of the elliptic beta integral for $SP(2N)$
group of type I as described above. For $SO(2{N}+1)$-group with $2N-2$ quarks
the index is given in \eqref{iso2N+1} with $N_f$ replaced by $2N-2$
and the balancing condition $\prod_{m=1}^{2{N}-2} t_m = (pq)^{-1/2}$
requiring a change of the integration contour.
Due to the confinement the dual index has a simple form
\beqa\label{sp1_2''} && I_M= \prod_{1
\leq m < s \leq 2{N}-2} \Gamma(t_mt_s) \prod_{i=1}^{2{N}-2}
\Gamma(t_i^2,\frac{\sqrt{pq}}{t_i}).
\end{eqnarray}
For the $SO(2{N})$-group the electric index has the form
\eqref{iso2N} with $N_f$ replaced by $2N-3$ and the balancing condition
$\prod_{m=1}^{2{N}-3} t_m = (pq)^{-1/2}.$
Its magnetic partner is
\beqa\label{sp1_2'''} && I_M =
\prod_{1 \leq m < s \leq 2{N}-3} \Gamma(t_mt_s)
\prod_{i=1}^{2{N}-3} \Gamma(t_i^2,\frac{\sqrt{pq}}{t_i}).
\end{eqnarray}
Extra terms $\prod_{i=1}^{2{N}-3}\Gamma(\frac{\sqrt{pq}}{t_i})$
appear in \eqref{sp1_2'''} from the fundamental representation,
although the dual gauge group is absent being formally defined as $SO(1)$.

Similarly one can consider the case of $G=SO(n)$ with $N_f=n-4$ \cite{IN3}.
For $SO(2N+1)$-group SCI has the form \eqref{iso2N+1} with $N_f$ replace by $2N-3$
and the balancing condition $\prod_{m=1}^{2{N}-3} t_m =(pq)^{-1}$.
In the infrared region particles confine and
\beqa\label{sp1_2''''} && I_M= \prod_{1
\leq m < s \leq 2{N}-3} \Gamma(t_mt_s) \prod_{i=1}^{2{N}-3}
\Gamma(t_i^2).
\end{eqnarray}
For $SO(2N)$-group electric SCI has the form \eqref{iso2N+1} with
$N_f$ replaced by $2N-4$ and the balancing condition
$\prod_{m=1}^{2{N}-4} t_m =(pq)^{-1}$.
Its dual has the form
\beqa\label{sp1_2} && I_M= \prod_{1 \leq m < s \leq
2{N}-4} \Gamma(t_mt_s) \prod_{i=1}^{2{N}-4} \Gamma(t_i^2).
\end{eqnarray}

\subsection{Connection to the Witten anomaly}

The even-dimensional  theories
have triangle anomalies associated with the global currents. For odd-dimensional
field theories these anomalies are absent and this fact plays a negative role
in searching corresponding dualities (because
of the absence of powerful 't Hooft anomaly matching conditions).
That is why the reduction of $4d$ SCIs to $3d$ partition functions
discovered in \cite{DSV} is important for searching $3d$ dualities,
since it inherits the information hidden in higher dimensional
anomaly matching conditions.

However, apart from the global triangle anomalies there is a non-perturbative
anomaly found by Witten \cite{W}, which is associated with the fact that the fourth
homotopy group is non-trivial for some gauge groups. For examples, it was
found that an $SU(2)$ gauge group theory with odd number of
fermions is not well defined because $\pi^4(SU(2)) = \mathbb{Z}_2$.
The same argument applies to supersymmetric field theories. Therefore
it is important to understand how this anomaly manifests itself in SCIs
and we analyze this question below.

We start from an example of the $s$-confining theory:
$4d$ $\mathcal{N}=1$ SYM theory with $SU(2)$ gauge group and 6 chiral
superfields. The confining phase
contains baryons $M_{ij}$ forming the antisymmetric tensor of the flavor
group $SU(6)$. Corresponding SCIs
were discussed in \cite{Dolan:2008qi,SV2} and they are given
by the left- and right-hand sides of the elliptic beta integral \cite{S1}.
So, the electric SCI has the form
\beq \label{W1}
I_E(s_1,\ldots,s_6) = \frac{(p;p)_\infty (q;q)_\infty}{2}
\int_{\mathbb{T}} \frac{\prod_{i=1}^6 \Gamma(s_i z^{\pm1})}
{\Gamma(z^{\pm2})} \frac{dz}{2 \pi \textup{i} z}, \eeq
with the balancing condition $\prod_{i=1}^6 s_i = pq$.
Changing the integration variable $z \rightarrow -z$ we see that
$I_E(s_1,\ldots,s_6) =I_E(-s_1,\ldots,-s_6)$. The magnetic SCI is
$I_M=\prod_{1\leq j<k\leq 6}\Gamma(s_js_k)=I_E$.

Let us set $s_6=\sqrt{pq}$. From the reflection equation for the elliptic
gamma function one has $\Gamma(\sqrt{pq} z^{\pm1}) = 1$.
Therefore the reduced SCI takes the form
\beq \label{W2}
I_{E1}(s_1,\ldots,s_5) = \frac{(p;p)_\infty (q;q)_\infty}{2}
\int_{\mathbb{T}} \frac{\prod_{i=1}^5 \Gamma(s_i z^{\pm1})}
{\Gamma(z^{\pm2})} \frac{dz}{2 \pi \textup{i} z},
 \eeq
where the balancing condition is $\prod_{i=1}^5 s_i = \sqrt{pq}$.
According to the prescription for constructing SCIs, this expression
describes $\mathcal{N}=1$ SYM theory with $SU(2)$
gauge group and 5 quarks forming a fundamental representation of
the flavor group $SU(5)$ and having the $R$-charges $2r=1/5.$
The situation looks as if one of the quarks has been integrated
out. As to the magnetic SCI, it takes the form
$$
I_{M1}(s_1,\ldots,s_5) = \prod_{1\leq i<j\leq 5}\Gamma(s_is_j)
\prod_{i=1}^5 \Gamma(\sqrt{pq}s_i)
$$
and describes a confined theory of two types of mesons --
the antisymmetric tensor representation $T_A$ of the group $SU(5)$ with the
$R$-charge 2/5 and the fundamental representation of $SU(5)$
with the $R$-charge $6/5$. As a consequence of the superconformal algebra,
formal canonical dimension of the latter field is bigger than 1,
i.e. formally the unitarity is broken, but real physical content
of formally dual theories outside conformal windows require better understanding.

So, the  electric theory has the Witten anomaly and the magnetic theory
has problems with the unitarity. Despite of the non-physical properties, these theories
are presumably dual to each other since all known duality tests are
valid for them, including the equality of SCIs. A natural question is
whether SCI feels in any way this anomaly ambiguity or not?
As argued in \cite{W}, physical observables in this anomalous theory
should vanish due to the
cancellation induced by the ``large" gauge transformations which change
the sign of the fermion determinant. This means that SCI should vanish as well,
as a gauge invariant object. However, SCI we use was computed
basically from the free field theory (in a sense, perturbatively),
and the non-perturbative effect of the large gauge transformation
do not enter it, yielding a nonzero result.

Still, we believe that SCIs catch this effect. For instance, in the above
confining theory with 5 quarks $I_{E1}(s_1,\ldots,s_5)\neq I_{E1}(-s_1,\ldots,-s_5)$,
since the balancing condition is not preserved by the reflections $s_j\to -s_j$.
There is an ambiguity in reducing
the number of quarks: one can choose $s_6=-\sqrt{pq}$ and obtain SCI of the same
shape \eqref{W2}, but with the balancing condition having the different sign
$\prod_{i=1}^5s_i=-\sqrt{pq}$. We interpret this ambiguity in reductions
together with the breaking of the reflection symmetry $s_j\to -s_j$ as
manifestations of the Witten anomaly.

For instance, if we choose in the elliptic beta integral $s_6=\sqrt{pq}$
and $s_5=-\sqrt{pq}$, we obtain the relation
\begin{eqnarray}\nonumber
&& I_{E2}=\frac{(p;p)_\infty (q;q)_\infty}{2}\int_{\mathcal{C}}\frac{\prod_{k=1}^4
\Gamma(s_kz^{\pm1}; p,q)} {\Gamma(z^{\pm2}; p,q)}\frac{dz}{2 \pi \textup{i} z}
\\ && \makebox[2em]{}
=I_{M2}=2(-p;p)_\infty (-q;q)_\infty \prod_{1\leq j<k\leq4}
\Gamma(s_js_k)\prod_{k=1}^4\Gamma(pqs_k^2; p^2,q^2),
\label{eAW}\end{eqnarray}
where $\prod_{k=1}^4s_k=-1$ and the contour $\mathcal{C}$ is chosen appropriately.
(There is a misprint in the corresponding equality given before formula (4.9)
in \cite{S3} -- the infinite products independent on $s_j$ were combined
there in an erroneous way.) If we interpret this relation
as the equality of superconformal indices for some confining theory with four quarks,
then the Witten anomaly is absent and, indeed, $I_{E2}(s_1,\ldots,s_4)=I_{E2}(-s_1,\ldots,-s_4)$.
The physical meaning of this duality is not quite clear since the standard
R\"omelsberger prescription does not apply to it. Namely, the electric theory
has four quarks, but some nontrivial topological contributions to SCI are present
leading to the non-standard balancing condition indicating on a non-marginal
deformation of the standard four quarks electric theory.

The described effect exists only for $\mathcal{N}=1$
SYM theories with $SP(2N)$ (and $SU(2)$) gauge group theories.
The $G=SO(n)$ theories do not have such a problem
since $\pi^4(SO(n)) =1$. The flavor symmetry group in this case
is $SU(N_f)$ (instead of $SU(2N_f)$) and one can integrate out
a single quark field without problems. At the level of SCIs
this is reached by restricting one of the fugacities in an appropriate way.

\subsection{$SO/SP$ gauge group theories with small number of flavors}

Here we consider relations between $\mathcal{N}=1$ SYM theories with
orthogonal and symplectic gauge groups with small number of flavors.
Take the dualities for $SP(2)$ gauge group theory with $8$ quarks.
This model was suggested in \cite{Csaki2} and studied in detail in \cite{SV1},
where it was argued that there are in total $72$ dual theories
having specific physical manifestations \cite{Khmelnitsky:2009vc}.

Electric theory SCI is described by an elliptic
analogue of the Euler-Gauss hypergeometric function introduced in \cite{S2,Sthesis}
\begin{equation}
V(t_1, \ldots , t_8;p,q) = \frac{(p;p)_\infty (q;q)_\infty}{2}\int_{{\mathbb T}}
\frac{\prod_{j=1}^8 \Gamma(t_jz^{\pm 1})}{\Gamma(z^{\pm 2})}
\frac{dz}{2\pi \textup{i}z} \label{egauss}\end{equation}
with the constraints $|t_j|<1$ for eight complex variables
$t_1, \ldots , t_8$ and the balancing condition
$\prod_{j=1}^8t_j=(pq)^2$.
This function obeys the following symmetry transformation derived in \cite{S2}
\begin{eqnarray}\label{Sp}
&& V(t_1, \ldots , t_8;p,q) = \prod_{1 \leq j < k \leq 4}
\Gamma(t_jt_k,t_{j+4}t_{k+4})\, V(s_1, \ldots ,
s_8;p,q),
\end{eqnarray}
where complex variables $s_j,\, |s_j|<1,$ are connected with
$t_j$ as follows
\begin{eqnarray}
s_j &=& \rho^{-1} t_j, \ j=1,2,3,4, \quad s_j = \rho t_j, \
j=5,6,7,8, \\ \nonumber \rho &=&
\sqrt{\frac{t_1t_2t_3t_4}{pq}}=\sqrt{\frac{pq}{t_5t_6t_7t_8}}.
\end{eqnarray}
This fundamental relation taken together with the evident $S_8$-permutational
group of symmetries in parameters $t_j$ generates the Weyl
group $W(E_7)$ \cite{Rains}.

Let us apply the following constraint on the parameters
$$
t_{3,4,5,6,7,8} \ =
\ \left\{ \pm \sqrt{p}, \pm \sqrt{q}, -1, -\sqrt{pq} \right\}.
$$
The initial electric SCI takes the form
\beq
I_E =\frac{(p;p)_\infty (q;q)_\infty}{2} \int_{\mathbb{T}} \frac{\prod_{i=1}^2 \Gamma(t_i
z^{\pm1})}{\Gamma(z^{\pm1})} \frac{dz}{2 \pi \textup{i} z},
\eeq
where $t_1t_2 = \sqrt{pq},$
while in the magnetic SCIs $S_8$-symmetry is explicitly broken
and we can get various inequivalently looking expressions.
E.g., split the initial $8$ parameters into two sets
$$
\{ \pm \sqrt{q}, -\sqrt{pq}, t_1 \} \ \ \text{and} \ \ \{ \pm
\sqrt{p}, -1, t_2 \}
$$
for which $\rho=\sqrt{t_1} (q/p)^{1/4}$. In terms of the parameters
$$
s_{1,2,3,4} = \rho^{-1} \{ \pm \sqrt{q}, -\sqrt{pq},
t_1 \} \ \text{and} \ \ \ s_{5,6,7,8} = \rho \{ \pm \sqrt{p}, -1, t_2 \}
$$
the magnetic SCI takes a quite simple form
\beq I_M = \frac{(p;p)_\infty (q;q)_\infty}{2}
\int_{\mathbb{T}} \frac{\prod_{i=1}^8 \Gamma(s_i
z^{\pm1})}{\Gamma(z^{\pm2})} \frac{dz}{2 \pi \textup{i} z}.
\eeq

After multiplication of both $I_E$ and $I_M$ by $\prod_{i=1,2}
\Gamma(t_i)$, on the electric side we obtain SCI for $\mathcal{N}=1$
SYM with $SO(3)$ gauge group with two quarks and on the magnetic
side we have SCI of a $\mathcal{N}=1$ SYM theory with $SP(2)$ gauge
group and eight quarks whose flavor fugacities are chosen in a special way.
 This relation can be generalized to arbitrary number
of colors ${N}$ and to the theories discussed in Sect. 2.1. However, the general
meaning of all such relations is not clear yet.

\section{$S$-confining theories with the spinor matter}\label{s-conf}

In this chapter we systematically
consider all known $s$-confining theories with $SO({N})$-gauge
groups and the matter in spinor representation \cite{Csaki:1996zb}.
The upper parts of the tables contain information on the charges and
field representation types of the electric models
(except of the vector superfield).
The lower parts of the tables describe the $s$-confining phase of the theory.
The models with the rank of the gauge group smaller than
4 are not considered because of different isomorphisms for
orthogonal groups: $SO(6) \simeq SU(4), SO(5) \simeq SP(4), SO(4)
\simeq SU(2) \times SU(2), SO(3) \simeq SU(2)$, and $SO(2) \simeq  U(1)$.

For the orthogonal group $SO(2{N})$ there are two types of spinor
representations: the proper spinor representation, which we denote as $s$,
and its complex conjugate which is denoted as $c$,
both representations have dimension $2^{{N}-1}$. For gauge
group $SO(2{N}+1)$ there exists only the spinor representation $s$ which has the
dimension $2^N$. Characters of the corresponding representations
can be found in the Appendix.

\subsection{Confinement for $SO(7)$ gauge group}

\subsubsection{$SU(6)$ flavor symmetry group}
The matter field content is \cite{Csaki:1996zb}
\begin{center}
\begin{tabular}{|c|c|c|c|}
  \hline
    & $SO(7)$ & $SU(6)$ & $U(1)_R$ \\  \hline
  $S$ & $s$ & $f$ & $2r=\frac{1}{6}$ \\ \hline \hline
  $S^2$ &   & $T_S$ & $\frac 13$ \\
  $S^4$ &   & $\overline{T}_A$ & $\frac 23$ \\  \hline
\end{tabular}
\end{center}

Corresponding SCIs have the form
 \beq \label{E} I_E =
\frac{(p;p)_\infty^3 (q;q)_\infty^3}{2^3 3!} \int_{{\mathbb T}^3}
\frac{\prod_{i=1}^6 \Gamma(s_i (z_1z_2z_3)^{\pm 1})
\prod_{j=1}^3\Gamma(s_i(z_j^{-2}z_1z_2z_3)^{\pm1})}
{\prod_{j=1}^3 \Gamma(z_j^{\pm2}) \prod_{1\leq j<k\leq3}
\Gamma(z_j^{\pm2}z_k^{\pm2})} \prod_{j=1}^3\frac{dz_j}{2 \pi
\textup{i} z_j}, \eeq
where $|s_i|<1$ with the balancing condition $\prod_{i=1}^6 s_i=(pq)^{1/2}$, and
\begin{eqnarray}\label{M}
I_M &=& \prod_{i=1}^6 \Gamma(s_i^2) \prod_{1 \leq i < j \leq 6}
\Gamma(s_is_j,(pq)^{\frac 12} s_i^{-1}s_j^{-1}).
\end{eqnarray}
In the limit $p=q=0$ (after proper treatment of the balancing
condition) and $s_{2,3,4,5}=0$ the equality $I_E=I_M$ is directly
verified by residue calculus.

This and all other dualities described in this paper satisfy the 't Hooft
anomaly matching conditions. According to \cite{SV5} this means that dual SCIs
have the same $SL(3,\mathbb{Z})$-modular group properties (in particular,
one can associate with these dualities some totally elliptic hypergeometric terms).

\subsubsection{$SU(5) \times U(1)$ flavor group}
The matter content is \cite{Csaki:1996zb}
\begin{center}
\begin{tabular}{|c|c|c|c|c|}
  \hline
    & $SO(7)$ & $SU(5)$ & $U(1)$ & $U(1)_R$ \\  \hline
  $S$ & $s$ & $f$ & 1 & $0$ \\
  $Q$ & $f$ & 1 & $-5$ & $1$ \\
 \hline \hline
  $Q^2$ &   & 1 & $-10$ & $2$ \\
  $S^2$ &   & $T_S$ & 2 & $0$ \\
  $S^4$ &   & $\overline{f}$ & 4 & 0 \\
  $S^2Q$ &   & $T_A$ & $-3$ & $1$ \\
  $S^4Q$ &   & $\overline{f}$ & $-1$ & 1 \\   \hline
\end{tabular}
\end{center}

Corresponding SCIs are
\beqa && I_E = \frac{(p;p)_\infty^3
(q;q)_\infty^3}{2^3 3!} \Gamma(t) \int_{{\mathbb T}^3}
\frac{\prod_{j=1}^3 \Gamma(t z_j^{\pm2})} {\prod_{j=1}^3
\Gamma(z_j^{\pm2}) \prod_{1\leq j<k\leq3}
\Gamma(z_j^{\pm2}z_k^{\pm2})} \nonumber \\ && \makebox[3.5em]{}
\times \prod_{i=1}^5 \Gamma(s_i (z_1z_2z_3)^{\pm1})
\prod_{j=1}^3\Gamma(s_i(\frac{z^2_j}{z_1z_2z_3})^{\pm1})
\prod_{j=1}^3\frac{dz_j}{2 \pi \textup{i} z_j}, \eeqa
where $|s_i|<1$ with the balancing condition $t \prod_{i=1}^5 s_i=\sqrt{pq},$ and
\beqa && I_M = \Gamma(t^2) \prod_{i=1}^5
\Gamma(\frac{\sqrt{pq}}{s_it},\frac{\sqrt{pq}}{s_i},s_i^2)
\prod_{1 \leq i < j \leq 5} \Gamma(s_is_j, t s_is_j).
\eeqa

Again, this $s$-confining duality predicts
the exact integration formula $I_E  =  I_M.$
Similar to the previous case, this identity is easily checked in the
limit $p=q= 0$ and $s_{2,3,4}=0$.

\subsubsection{$SU(4) \times SU(2) \times U(1)$ flavor group}
The matter content is \cite{Csaki:1996zb}
\begin{center}
\begin{tabular}{|c|c|c|c|c|c|}
  \hline
    & $SO(7)$ & $SU(4)$ & $SU(2)$ & $U(1)$ & $U(1)_R$ \\  \hline
  $S$ & $s$ & $f$ & 1 & 1 & $0$ \\
  $Q$ & $f$ & 1 & $f$ & $-2$ & $\frac 12$ \\
 \hline \hline
  $Q^2$ &   & 1 & $T_S$ & $-4$ & $1$ \\
  $S^2$ &   & $T_S$ & 1 & 2 & $0$ \\
  $S^2Q$ &   & $T_A$ & $f$ & 0 & $\frac 12$ \\
  $S^2Q^2$ &   & $T_A$ & $1$ & $-2$ & $1$ \\
  $S^4$ &   & 1 & 1 & 4 & $0$ \\
  $S^4Q$ &   & 1 & $f$ & $2$ & $\frac 12$ \\   \hline
\end{tabular}
\end{center}

Corresponding SCIs are
\beqa && I_E = \frac{(p;p)_\infty^3
(q;q)_\infty^3}{2^3 3!} \prod_{i=1}^2 \Gamma(t_i) \int_{{\mathbb
T}^3} \frac{\prod_{i=1}^2 \prod_{j=1}^3 \Gamma(t_i z_j^{\pm2})}
{\prod_{j=1}^3 \Gamma(z_j^{\pm2}) \prod_{1\leq j<k\leq3}
\Gamma(z_j^{\pm2}z_k^{\pm2})} \nonumber \\ && \makebox[4em]{}
\times \prod_{i=1}^4 \Gamma(s_i (z_1z_2z_3)^{\pm1})
\prod_{j=1}^3\Gamma(s_i(\frac{z^2_j}{z_1z_2z_3})^{\pm1})
\prod_{j=1}^3\frac{dz_j}{2 \pi \textup{i} z_j}, \eeqa
where $|s_i|,|t_j|<1,$ $st =\sqrt{pq}$ with $s = \prod_{i=1}^4 s_i, t =
\prod_{i=1}^2 t_i$, and
\beq \makebox[-2em]{}
I_M = \Gamma(s,t) \prod_{i=1}^2
\Gamma(st_i, t_i^2) \prod_{i=1}^4 \Gamma(s_i^2) \prod_{1
\leq i < j \leq 4} \Gamma(s_is_j, t s_is_j) \prod_{k=1}^2
\Gamma(s_is_jt_k).
 \eeq

\subsubsection{$SU(3) \times SU(3) \times U(1)$ flavor group}
The matter content is \cite{Csaki:1996zb}
\begin{center}
\begin{tabular}{|c|c|c|c|c|c|}
  \hline
    & $SO(7)$ & $SU(3)$ & $SU(3)$ & $U(1)$ & $U(1)_R$ \\  \hline
  $S$ & $s$ & $f$ & 1 & 1 & $0$ \\
  $Q$ & $f$ & 1 & $f$ & $-1$ & $\frac 13$ \\
 \hline \hline
  $Q^2$ &   & 1 & $T_S$ & $-2$ & $\frac 23$ \\
  $S^2$ &   & $T_S$ & 1 & 2 & $0$ \\
  $S^2Q$ &   & $\overline{f}=T_A$ & $f$ & 1 & $\frac 13$ \\
  $S^2Q^2$ &   & $\overline{f}$ & $\overline{f}$ & $0$ & $\frac 23$ \\
  $S^2Q^3$ &   & $T_S$ & 1 & $-1$ & $1$ \\ \hline
\end{tabular}
\end{center}

Corresponding SCIs are
\beqa && I_E = \frac{(p;p)_\infty^3
(q;q)_\infty^3}{2^3 3!} \prod_{i=1}^3 \Gamma(t_i) \int_{{\mathbb
T}^3} \frac{\prod_{i=1}^3 \prod_{j=1}^3 \Gamma(t_i z_j^{\pm2})}
{\prod_{j=1}^3 \Gamma(z_j^{\pm2}) \prod_{1\leq j<k\leq3}
\Gamma(z_j^{\pm2}z_k^{\pm2})} \nonumber \\ && \makebox[4em]{}
\times \prod_{i=1}^3 \Gamma(s_i (z_1z_2z_3)^{\pm1})
\prod_{i,j=1}^3\Gamma(s_i(\frac{z^2_j}{z_1z_2z_3})^{\pm1})
\prod_{j=1}^3\frac{dz_j}{2 \pi \textup{i} z_j}, \eeqa
 where $|s_i|,|t_j|<1,$ $st = \sqrt{pq}$  with $s =
\prod_{i=1}^3 s_i, t = \prod_{i=1}^3 t_i$, and
\beq I_M = \prod_{i=1}^3 \Gamma(s_i^2, t_i^2,
ts_i^2) \prod_{i,j=1}^3
\Gamma(sts_i^{-1}t_j^{-1},ss_i^{-1}t_j) \prod_{1 \leq i < j \leq
3} \Gamma(s_is_j, t_it_j, t s_is_j).
 \eeq

\subsection{$G=SO(8)$}

\subsubsection{$SU(4) \times SU(3) \times U(1)$ flavor group}
The matter content is \cite{Csaki:1996zb}
\begin{center}
\begin{tabular}{|c|c|c|c|c|c|}
  \hline
    & $SO(8)$ & $SU(4)$ & $SU(3)$ & $U(1)$ & $U(1)_R$ \\  \hline
  $Q$ & $f$ & $f$ & 1 & 3 & $\frac 14$ \\
  $S$ & $s$ & 1 & $f$ & $-4$ & $0$ \\
 \hline \hline
  $Q^2$ &   & $T_S$ & 1 & $6$ & $\frac 12$ \\
  $S^2$ &   & 1 & $T_S$ & $-8$ & $0$ \\
  $S^2Q^2$ &   & $T_A$ & $\overline{f}$ & $-2$ & $\frac 12$ \\
  $S^2Q^4$ &   & $1$ & $T_S$ & $4$ & $1$ \\ \hline
\end{tabular}
\end{center}

Corresponding SCIs are
\beqa && I_E = \frac{(p;p)_\infty^4
(q;q)_\infty^4}{2^3 4!} \int_{{\mathbb T}^4} \frac{\prod_{i=1}^4
\prod_{j=1}^4 \Gamma(s_i z_j^{\pm2})} {\prod_{1 \leq j < k \leq
4} \Gamma(z_j^{\pm2}z_k^{\pm2})}
 \nonumber \\ && \makebox[-3em]{}\times
\prod_{i=1}^3 \Gamma(t_i (z_1z_2z_3z_4)^{\pm1})
\prod_{i=1}^3 \prod_{1 \leq j < k \leq 4} \Gamma(t_i
\frac{z^2_jz_k^2}{z_1z_2z_3z_4} ) \prod_{j=1}^4\frac{dz_j}{2 \pi
\textup{i} z_j}, \eeqa
where $|s_i|,|t_j|<1,$ and
\beqa &&  \makebox[-2em]{}
I_M =\prod_{i=1}^3 \Gamma(t_i^2, st_i^2) \prod_{i=1}^4
\Gamma(s_i^2) \prod_{1 \leq i < j \leq 3} \Gamma(t_it_j,st_it_j)
\prod_{1 \leq i < j \leq 4}\Big( \Gamma(s_is_j)\prod_{k=1}^3
\Gamma(ts_is_jt_k^{-1}) \Big),
\eeqa
with $s = \prod_{i=1}^4 s_i, t =
\prod_{i=1}^3 t_i$, and the balancing condition $st = \sqrt{pq}.$
A simple check of the equality of these SCIs is obtained in the limit $p=q= 0$
and $s_{2,3,4}=t_2=0$.

\subsubsection{$SU(4) \times SU(2) \times U(1)_1 \times U(1)_2$ flavor group}
The matter content is \cite{Csaki:1996zb}
\begin{center}
\begin{tabular}{|c|c|c|c|c|c|c|}
  \hline
    & $SO(8)$ & $SU(4)$ & $SU(2)$ & $U(1)_1$ & $U(1)_2$ & $U(1)_R$ \\  \hline
  $Q$ & $f$ & $f$ & 1 & 1 & 0 & $\frac 14$ \\
  $S$ & $s$ & 1 & $f$ & $-2$ & 1 & $0$ \\
  $S'$& $c$ & 1 & 1 & 0 & $-2$ & 0 \\
 \hline \hline
  $Q^2$ &   & $T_S$ & 1 & 2 & $0$ & $\frac 12$ \\
  $S^2$ &   & 1 & $T_S$ & $-4$ & 2 & $0$ \\
  $S'^2$ &  & 1 & 1 & 0 & $-4$ & 0 \\
  $S^2Q^2$ &   & $T_A$ & 1 & $-2$ & 2 & $\frac 12$ \\
  $S^2Q^4$ &   & $1$ & $T_S$ & $0$ & 2 & $1$ \\
  $S'^2Q^4$&   & 1 & 1 & 4 & $-4$ & 1 \\
  $SS'Q$   &   & $f$ & $f$ & $-1$ & $-1$ & $\frac 14$ \\
  $SS'Q^3$ &   & $\overline{f}$ & $f$ & 1 & $-1$ & $\frac 34$ \\
   \hline
\end{tabular}
\end{center}

Corresponding SCIs are
\beqa && I_E = \frac{(p;p)_\infty^4
(q;q)_\infty^4}{2^3 4!} \int_{{\mathbb T}^4} \frac{\prod_{i,j=1}^4
\Gamma(s_i z_j^{\pm2})} {\prod_{1 \leq j < k \leq 4}
\Gamma(z_j^{\pm2}z_k^{\pm2})} \prod_{i=1}^2 \Gamma(t_i
(z_1z_2z_3z_4)^{\pm1}) \nonumber \\ && \makebox[1.5em]{} \times
\prod_{i=1}^2 \prod_{1 \leq j < k \leq 4} \Gamma\left(t_i
\frac{z^2_jz_k^2}{z_1z_2z_3z_4} \right) \prod_{j=1}^4 \Gamma\left(u
(\frac{z_j^2}{z_1z_2z_3z_4})^{\pm1}\right) \prod_{j=1}^4 \frac{dz_j}{2
\pi \textup{i} z_j}, \eeqa
where $|s_i|,|t_j|,|u|<1,$ $stu  =  \sqrt{pq}$ with $s =
\prod_{i=1}^4 s_i, t = \prod_{i=1}^2 t_i$, and
\beqa && \makebox[-2.5em]{}
I_M = \Gamma(u^2,su^2,t,st) \prod_{j=1}^2\Big( \Gamma(t_j^2,st_j^2) \prod_{i=1}^4
\Gamma(us_it_j,\frac{us}{s_i}t_j)\Big)
\prod_{i=1}^4 \Gamma(s_i^2) \prod_{1
\leq i < j \leq 4} \Gamma(s_is_j,ts_is_j).
 \eeqa

\subsubsection{$SU(3) \times SU(3) \times U(1)_1 \times U(1)_2$ flavor group}
The matter content is \cite{Csaki:1996zb}
\begin{center}
\begin{tabular}{|c|c|c|c|c|c|c|}
  \hline
    & $SO(8)$ & $SU(3)$ & $SU(3)$ & $U(1)_1$ & $U(1)_2$ & $U(1)_R$ \\  \hline
  $Q$ & $f$ & 1 & 1 & 0 & 6 & 1 \\
  $S$ & $s$ & $f$ & 1 & 1 & $-1$ & $0$ \\
  $S'$& $c$ & 1 & $f$ & $-1$ & $-1$ & 0 \\
 \hline \hline
  $Q^2$ &   & 1 & 1 & 0 & 12 & 2 \\
  $S^2$ &   & $T_S$ & 1 & 2 & $-2$ & $0$ \\
  $S'^2$ &  & 1 & $T_S$ & $-2$ & $-2$ & 0 \\
  $SS'Q$ &   & $f$ & $f$ & 0 & 4 & 1 \\
  $S^3S'Q$ &   & $1$ & $f$ & $2$ & 2 & $1$ \\
  $SS'^3Q$&   & $f$ & 1 & $-2$ & $2$ & 1 \\
  $S^2S'^2$   &   & $\overline{f}$ & $\overline{f}$ & $0$ & $-4$ & $0$ \\
   \hline
\end{tabular}
\end{center}

Corresponding SCIs are
\beqa && I_E = \frac{(p;p)_\infty^4
(q;q)_\infty^4}{2^3 4!} \int_{{\mathbb T}^4} \frac{\prod_{j=1}^4
\Gamma(u z_j^{\pm2})} {\prod_{1 \leq j < k \leq 4}
\Gamma(z_j^{\pm2}z_k^{\pm2})} \prod_{i=1}^3 \Gamma(s_i
(z_1z_2z_3z_4)^{\pm1}) \nonumber \\ && \makebox[0.5em]{} \times
\prod_{i=1}^3 \prod_{1 \leq j < k \leq 4} \Gamma\left(s_i
\frac{z^2_jz_k^2}{z_1z_2z_3z_4} \right) \prod_{i=1}^3 \prod_{j=1}^4
\Gamma\left(t_i \big(\frac{z_j^2}{z_1z_2z_3z_4}\big)^{\pm1}\right) \prod_{j=1}^4
\frac{dz_j}{2 \pi \textup{i} z_j}, \eeqa
where $|s_i|,|t_j|,|u|<1,$ and
\beqa &&
I_M = \Gamma(u^2) \prod_{i=1}^3 \Gamma(s_i^2,
t_i^2, sut_i,tus_i)
\prod_{i,j=1}^3 \Gamma(us_it_j,sts_i^{-1}t_j^{-1}) \prod_{1 \leq
i < j \leq 3} \Gamma(s_is_j,t_it_j), \eeqa with $s =
\prod_{i=1}^3 s_i, t = \prod_{i=1}^3 t_i$, and the balancing
condition $stu = \sqrt{pq}.$ We checked the equality
of these SCIs in the limit $p=q= 0$ and $s_{2,3}=t_2\to 0$.

\subsubsection{$SU(3) \times SU(2)_1
 \times SU(2)_2 \times U(1)_1 \times U(1)_2$ flavor  group}
The matter content is \cite{Csaki:1996zb}
\begin{center}
\begin{tabular}{|c|c|c|c|c|c|c|c|}
  \hline
    & $SO(8)$ & $SU(3)$ & $SU(2)_1$ & $SU(2)_2$ & $U(1)_1$ & $U(1)_2$ & $U(1)_R$ \\  \hline
  $Q$ & $f$ & $f$ & 1 & 1 & 0 & 4 & $0$ \\
  $S$ & $s$ & 1 & $f$ & $1$ & 1 & $-3$ & $\frac 14$ \\
  $S'$& $c$ & 1 & 1 & $f$ & $-1$ & $-3$ & $\frac 14$ \\
 \hline \hline
  $Q^2$ &   & $T_S$ & 1 & 1 & $0$ & 8 & $0$ \\
  $S^2$ &   & 1 & $T_S$ & $1$ & 2 & $-6$ & $\frac 12$ \\
  $S'^2$ &  & 1 & 1 & $T_S$ & $-2$ & $-6$ & $\frac 12$ \\
  $SS'Q$ &   & $f$ & $f$ & $f$ & 0 & $-2$ & $\frac 12$ \\
  $S^2Q^2$ &   & $\overline{f}$ & 1 & 1 & $2$ & 2 & $\frac 12$ \\
  $S'^2Q^2$&   & $\overline{f}$ & 1 & 1 & $-2$ & 2 &$\frac 12$ \\
  $SS'Q^3$   &   & $1$ & $f$ & $f$ & $0$ & 6 & $\frac 12$ \\
  $S^2S'^2$ &   & $1$ & $1$ & 1 & $0$ & $-12$ & $1$ \\
  $S^2S'^2Q^2$ && $\overline{f}$ & 1 & 1 & 0 & $-4$ & 1 \\
   \hline
\end{tabular}
\end{center}

Corresponding SCIs are
\beqa &&
I_E = \frac{(p;p)_\infty^4
(q;q)_\infty^4}{2^3 4!} \int_{{\mathbb T}^4} \frac{\prod_{i=1}^3
\prod_{j=1}^4 \Gamma(s_i z_j^{\pm2})} {\prod_{1 \leq j < k \leq
4} \Gamma(z_j^{\pm2}z_k^{\pm2})} \prod_{i=1}^2 \Gamma(t_i
(z_1z_2z_3z_4)^{\pm1}) \nonumber \\ && \makebox[0.5em]{} \times
\prod_{i=1}^2 \prod_{1 \leq j < k \leq 4} \Gamma\left(t_i
\frac{z^2_jz_k^2}{z_1z_2z_3z_4} \right) \prod_{i=1}^2 \prod_{j=1}^4
\Gamma\left(u_i \big(\frac{z_j^2}{z_1z_2z_3z_4}\big)^{\pm1}\right) \prod_{j=1}^4
\frac{dz_j}{2 \pi \textup{i} z_j}, \eeqa
where $|s_i|,|t_i|,|u_i|<1,$ $stu =  \sqrt{pq}$ with $s = \prod_{i=1}^3 s_i,\,
t = \prod_{i=1}^2 t_i,\, u = \prod_{i=1}^2 u_i$, and
\beqa &&
I_M = \Gamma(t,u,tu) \prod_{i=1}^3
\Gamma(s_i^2) \prod_{i=1}^2 \Gamma(t_i^2, u_i^2)
\prod_{i,j=1}^2 \Gamma(st_iu_j) \nonumber \\
&& \makebox[-2em]{} \times \prod_{1 \leq i < j \leq 3}
\Gamma(s_is_j) \prod_{i=1}^3 \Gamma(stu
s_i^{-1},sts_i^{-1},sus_i^{-1}) \prod_{i=1}^3 \prod_{j,k=1}^2
\Gamma(s_it_ju_k). \eeqa

\subsection{$G=SO(9)$}

\subsubsection{$SU(4)$ flavor group}\label{exspin}
The matter content is \cite{Csaki:1996zb}
\begin{center}
\begin{tabular}{|c|c|c|c|}
  \hline
    & $SO(9)$ & $SU(4)$ & $U(1)_R$ \\  \hline
  $S$ & $s$ & $f$ & $\frac 18$ \\
 \hline \hline
  $S^2$ &   & $T_S$ & $\frac 14$ \\
  $S^4$ &   & $T_{AASS}$ & $\frac 12$ \\
  $S^6$ &   & $T_S$ & $\frac 34$ \\
   \hline
\end{tabular}
\end{center}
where $T_{AASS}$ denotes the fourth rank tensor representation
symmetric in two indices and antisymmetric in other two indices,
whose character is given by the formula
$$\chi_{T_{AASS},SU(4)} (\underline{s}) \ = \
\sum_{1 \leq i < j \leq 4} s_i^2 s_j^2 + \sum_{i=1}^4 \sum_{1 \leq j
< k \leq 4; j,k \neq i} s_i^2 s_j s_k +2.$$

Corresponding SCIs are
\beqa && I_E = \frac{(p;p)_\infty^4
(q;q)_\infty^4}{2^4 4!} \int_{{\mathbb T}^4} \frac{\prod_{i=1}^4 \Gamma(s_i z^{\pm1})
\prod_{i,j=1}^4 \Gamma(s_i (\frac{z_j^2}{z})^{\pm1})
}{\prod_{i=1}^4 \Gamma(z_i^{\pm2}) \prod_{1 \leq j < k \leq 4}
\Gamma(z_j^{\pm2}z_k^{\pm2})} \nonumber \\ && \makebox[5em]{} \times
\prod_{i=1}^4 \prod_{1 \leq j < k
\leq 4} \Gamma(s_i \frac{z^2_jz_k^2}{z} ) \prod_{j=1}^4
\frac{dz_j}{2 \pi \textup{i} z_j}, \eeqa
where $z=z_1z_2z_3z_4$, $|s_i|<1,$ the balancing condition $s^2  = \sqrt{pq}$
with $s = \prod_{i=1}^4 s_i$, and
\begin{equation} \makebox[-2em]{} 
I_M = \Gamma^2(s)
\prod_{i=1}^4 \Gamma(s_i^2, ss_i^2)
\prod_{1 \leq i < j \leq 4}
\Gamma(s_is_j,ss_is_j,s_i^2s_j^2) \prod_{i=1}^4 \prod_{1 \leq j
< k \leq 4;j,k \neq i} \Gamma(s_i^2s_js_k).
 \end{equation}

\subsubsection{$SU(3) \times SU(2) \times U(1)$ flavor group}
The matter content is \cite{Csaki:1996zb}
\begin{center}
\begin{tabular}{|c|c|c|c|c|c|}
  \hline
    & $SO(9)$ & $SU(3)$ & $SU(2)$ & $U(1)$ & $U(1)_R$ \\  \hline
  $S$ & $s$ & $f$ & 1 & 1 & 0 \\
  $Q$ & $f$ & 1 & $f$ & $-3$ & $\frac 12$ \\
 \hline \hline
  $Q^2$ &   & 1 & $T_S$ & $-6$ & 1 \\
  $S^2Q$ &  & $T_S$ & $f$ & $-1$ & $\frac 12$ \\
  $S^2$ &  & $T_S$ & 1 & 2 & 0 \\
  $S^4$ &  & $\overline{T}_S$ & 1 & 4 & 0 \\
  $S^2Q^2$ &   & $\overline{f}$ & 1 & $-4$ & 1 \\
  $S^4Q^2$&   & $f$ & 1 & $-2$ & 1 \\
  $S^4Q$   &   & $f$ & $f$ & 1 & $\frac 12$ \\
   \hline
\end{tabular}
\end{center}

Corresponding SCIs are
\beqa && I_E = \frac{(p;p)_\infty^4
(q;q)_\infty^4}{2^4 4!} \prod_{i=1}^2 \Gamma(t_i) \int_{{\mathbb
T}^4} \frac{\prod_{i=1}^2 \prod_{j=1}^4 \Gamma(t_i z_j^{\pm2})
\prod_{i=1}^3 \Gamma(s_i z^{\pm1})} {\prod_{j=1}^4
\Gamma(z_j^{\pm2}) \prod_{1 \leq j < k \leq 4}
\Gamma(z_j^{\pm2}z_k^{\pm2})} \nonumber \\ && \makebox[2em]{}
\times \prod_{i=1}^3 \prod_{j=1}^4 \Gamma(s_i
(z_j^2z^{-1})^{\pm1}) \prod_{i=1}^3 \prod_{1 \leq j < k \leq 4}
\Gamma(s_i z^2_jz_k^2 z^{-1} ) \prod_{j=1}^4 \frac{dz_j}{2 \pi
\textup{i} z_j},
\eeqa
where $z=z_1z_2z_3z_4$, $|s_i|<1,$ $s^2t = \sqrt{pq}$ with
$s = \prod_{i=1}^3s_i,\, t = \prod_{i=1}^2 t_i$, and
\beqa
&& I_M = \prod_{i=1}^2 \Gamma(t_i^2) \prod_{i=1}^3
\Gamma(s_i^2,sts_i,s^2s_i^{-1},sts_i^{-1}) \prod_{i=1}^3 \prod_{j=1}^2 \Gamma(s_i^2t_j, ss_it_j) \nonumber \\
&& \makebox[2em]{} \times \Gamma(t) \prod_{1 \leq i < j \leq 3}
\Gamma(s_is_j,s^2s_i^{-1}s_j^{-1}) \prod_{1 \leq i < j \leq 3}
\prod_{k=1}^2 \Gamma(s_is_jt_k).
 \eeqa

\subsubsection{$SU(2) \times SU(4) \times U(1)$ flavor group}
The matter content is \cite{Csaki:1996zb}
\begin{center}
\begin{tabular}{|c|c|c|c|c|c|}
  \hline
    & $SO(9)$ & $SU(2)$ & $SU(4)$ & $U(1)$ & $U(1)_R$ \\  \hline
  $S$ & $s$ & $f$ & 1 & 1 & $\frac 14$ \\
  $Q$ & $f$ & 1 & $f$ & $-1$ & 0 \\
 \hline \hline
  $Q^2$ &   & 1 & $T_S$ & $-2$ & 0 \\
  $S^2Q$ &  & $T_S$ & $f$ & $1$ & $\frac 12$ \\
  $S^2$ &  & $T_S$ & 1 & 2 & $\frac 12$ \\
  $S^2Q^3$ &  & 1 & $\overline{f}$ & $-1$ & $\frac 12$ \\
  $S^2Q^2$ &   & 1 & $T_A$ & 0 & $\frac 12$ \\
  $S^4Q^3$&   & 1 & $\overline{f}$ & 1 & 1 \\
  $S^2Q^4$   &   & $T_S$ & 1 & $-2$ & $\frac 12$ \\
  $S^4$ &   & 1 & 1 & 4 & 1 \\
   \hline
\end{tabular}
\end{center}

Corresponding SCIs are
 \beqa && I_E = \frac{(p;p)_\infty^4
(q;q)_\infty^4}{2^4 4!} \prod_{i=1}^4 \Gamma(t_i) \int_{{\mathbb
T}^4} \frac{\prod_{i,j=1}^4 \Gamma(t_i z_j^{\pm2}) \prod_{i=1}^2
\Gamma(s_i z^{\pm1})} {\prod_{j=1}^4 \Gamma(z_j^{\pm2})
\prod_{1 \leq j < k \leq 4} \Gamma(z_j^{\pm2}z_k^{\pm2})}
\nonumber \\ && \makebox[2em]{} \times \prod_{i=1}^2 \prod_{j=1}^4
\Gamma(s_i (z_j^2z^{-1})^{\pm1}) \prod_{i=1}^2 \prod_{1 \leq j <
k \leq 4} \Gamma(s_i z^2_jz_k^2 z^{-1} ) \prod_{j=1}^4
\frac{dz_j}{2 \pi \textup{i} z_j}, \eeqa
where $z=z_1z_2z_3z_4$, $|s_i|<1,$ $s^2t= \sqrt{pq}$ with
$s = \prod_{i=1}^2 s_i,\, t =\prod_{i=1}^4 t_i$,  and
\beqa && I_M = \Gamma(s,st,s^2) \prod_{i=1}^2
\Gamma(s_i^2,ts_i^2) \prod_{i=1}^4
\Gamma(t_i^2,st_i^2,stt_i^{-1},s^2tt_i^{-1}) \nonumber \\
&& \makebox[4.5em]{} \times \prod_{i=1}^2 \prod_{j=1}^4
\Gamma(s_i^2t_j) \prod_{1 \leq i < j \leq 4}
\Gamma(t_it_j,st_it_j).
 \eeqa

\subsection{$G=SO(10)$}

\subsubsection{$SU(4) \times U(1)$ flavor group}
The matter content is \cite{Csaki:1996zb}
\begin{center}
\begin{tabular}{|c|c|c|c|c|}
  \hline
    & $SO(10)$ & $SU(4)$ & $U(1)$ & $U(1)_R$ \\  \hline
  $S$ & $s$ & $f$ & 1 & 0 \\
  $Q$ & $f$  & 1  & $-8$ & 1 \\
 \hline \hline
  $Q^2$ &   & 1 & $-16$ & 2 \\
  $S^2Q$ &   & $T_{S}$ & $-6$ & 1 \\
  $S^4$ &   & $T_{AASS}$ & 4 & 0 \\
  $S^6Q$&   & $T_S$ & $-2$ & 1 \\
   \hline
\end{tabular}
\end{center}

Corresponding SCIs are
\beqa && I_E = \frac{(p;p)_\infty^5
(q;q)_\infty^5}{2^4 5!} \int_{{\mathbb T}^5} \frac{\prod_{i=1}^4
\Gamma(s_i z) \prod_{i=1}^4 \prod_{j=1}^5 \Gamma(s_i
z_j^2z^{-1}) }{\prod_{1 \leq j < k \leq 5}
\Gamma(z_j^{\pm2}z_k^{\pm2})} \nonumber \\ && \makebox[2em]{}
\times \prod_{i=1}^4 \prod_{1 \leq j < k \leq 5} \Gamma(s_i
zz_j^{-2}z_k^{-2}) \prod_{j=1}^5 \Gamma(t z_j^{\pm2})
\prod_{j=1}^5 \frac{dz_j}{2 \pi \textup{i} z_j}, \eeqa where
$z=z_1z_2z_3z_4z_5$, $|s_i|,|t|<1,$ and
\beqa && \makebox[-2.5em]{}
I_M =\Gamma(t^2) \Gamma^2(s)  \prod_{i=1}^4
\Gamma(ts_i^2, sts_i^2) \prod_{1 \leq i < j \leq 4}
\Gamma(ts_is_j,sts_is_j,s_i^2s_j^2)
\prod_{i=1}^4 \prod_{1 \leq j < k \leq 4;j,k \neq i} \Gamma(s_i^2s_js_k),
\eeqa with $s = \prod_{i=1}^4 s_i$ and the balancing condition
$s^2t  = \sqrt{pq}.$
A simple check of the equality of these integrals is obtained
in the limit $p=q=0$ and $s_{2,3}=0$.

\subsubsection{$SU(3) \times SU(3) \times U(1)$ flavor group}
The matter content is \cite{Csaki:1996zb}
\begin{center}
\begin{tabular}{|c|c|c|c|c|c|}
  \hline
    & $SO(10)$ & $SU(3)$ & $SU(3)$ & $U(1)$ & $U(1)_R$ \\  \hline
  $S$ & $s$ & $f$ & 1 & 1 & 0 \\
  $Q$ & $f$  & 1 & $f$ & $-2$ & $\frac 13$ \\
 \hline \hline
  $Q^2$ &   & 1 & $T_S$ & $-4$ & $\frac 23$ \\
  $S^2Q$ &   & $T_{S}$ & $f$ & 0 & $\frac 13$ \\
  $S^2Q^3$ &   & $\overline{f}$ & 1 & $-4$ & 1 \\
  $S^4$&   & $\overline{T}_S$ & 1 & 4 & 0 \\
  $S^4Q^2$ & & $f$ & $\overline{f}$ & 0 & $\frac 23$ \\
   \hline
\end{tabular}
\end{center}

Corresponding SCIs are
 \beqa && I_E = \frac{(p;p)_\infty^5
(q;q)_\infty^5}{2^4 5!} \int_{{\mathbb T}^5} \frac{\prod_{i=1}^3
\Gamma(s_i z) \prod_{i=1}^3 \prod_{j=1}^5 \Gamma(s_i
z_j^2z^{-1}) }{\prod_{1 \leq j < k \leq 5}
\Gamma(z_j^{\pm2}z_k^{\pm2})} \nonumber \\ && \makebox[1em]{}
\times \prod_{i=1}^3 \prod_{1 \leq j < k \leq 5} \Gamma(s_i
zz_j^{-2}z_k^{-2}) \prod_{i=1}^3 \prod_{j=1}^5 \Gamma(t_i
z_j^{\pm2}) \prod_{j=1}^5 \frac{dz_j}{2 \pi \textup{i} z_j},
\eeqa
where $z=z_1z_2z_3z_4z_5$, $|s_i|,|t_j|<1,$ $s^2t  =  \sqrt{pq}$
with $s = \prod_{i=1}^3 s_i,\, t = \prod_{i=1}^3 t_i$, and
\beqa &&  \makebox[-2.5em]{}
I_M =\prod_{i=1}^3 \Gamma(t_i^2, s^2s_i^{-2},sts_i^{-1})
\prod_{i,j=1}^3
\Gamma(s_i^2t_j,sts_it_j^{-1})
 \prod_{1 \leq i < j \leq 3}\big(
\Gamma(t_it_j,s^2s_i^{-1}s_j^{-1}) \prod_{k=1}^3 \Gamma(s_is_jt_k)\big).
\eeqa

\subsubsection{$SU(2) \times SU(5) \times U(1)$ flavor group}
The matter content is \cite{Csaki:1996zb}
\begin{center}
\begin{tabular}{|c|c|c|c|c|c|}
  \hline
    & $SO(10)$ & $SU(2)$ & $SU(5)$ & $U(1)$ & $U(1)_R$ \\  \hline
  $S$ & $s$ & $f$ & 1 & 5 & $\frac 14$ \\
  $Q$ & $f$  & 1 & $f$ & $-4$ & 0 \\
 \hline \hline
  $Q^2$ &   & 1 & $T_S$ & $-8$ & 0 \\
  $S^2Q$ &   & $T_{S}$ & $f$ & 6 & $\frac 12$ \\
  $S^2Q^3$ &   & 1 & $\overline{T}_A$ & $-2$ & $\frac 12$ \\
  $S^2Q^5$&   & $T_S$ & 1 & $-10$ & $\frac 12$ \\
  $S^4$ & & $1$ & 1 & 20 & $1$ \\
  $S^4Q^4$ && 1 & $\overline{f}$ & 4 & 1 \\
   \hline
\end{tabular}
\end{center}

Corresponding SCIs are
\beqa && I_E = \frac{(p;p)_\infty^5
(q;q)_\infty^5}{2^4 5!} \int_{{\mathbb T}^5} \frac{\prod_{i=1}^2
\Gamma(s_i z) \prod_{i=1}^2 \prod_{j=1}^5 \Gamma(s_i
z_j^2z^{-1}) }{\prod_{1 \leq j < k \leq 5}
\Gamma(z_j^{\pm2}z_k^{\pm2})} \nonumber \\ && \makebox[1.5em]{}
\times \prod_{i=1}^2 \prod_{1 \leq j < k \leq 5} \Gamma(s_i
zz_j^{-2}z_k^{-2}) \prod_{i,j=1}^5 \Gamma(t_i z_j^{\pm2})
\prod_{j=1}^5 \frac{dz_j}{2 \pi \textup{i} z_j}, \eeqa
 where
$z=z_1z_2z_3z_4z_5$, $|s_i|,|t_j|<1,$ $s^2t =  \sqrt{pq}$
with $s = \prod_{i=1}^2s_i,\, t = \prod_{i=1}^5 t_i$, and
 \beqa && \makebox[-2.5em]{}
I_M =\Gamma(st,s^2) \prod_{j=1}^2 \Gamma(ts_j^2) \prod_{i=1}^5
\Big(\Gamma(t_i^2,st_i,s^2tt_i^{-1})
 \prod_{j=1}^2\Gamma(s_j^2t_i)\Big) \prod_{1 \leq i < j \leq 5}
\Gamma(t_it_j,stt_i^{-1}t_j^{-1}).
\eeqa

\subsubsection{$SU(3) \times U(1)_1 \times U(1)_2$ flavor group}
The matter content is \cite{Csaki:1996zb}
\begin{center}
\begin{tabular}{|c|c|c|c|c|c|}
  \hline
    & $SO(10)$ & $SU(3)$ & $U(1)_1$ & $U(1)_2$ & $U(1)_R$ \\  \hline
  $S$ & $s$ & $f$ & 1 & 0 & 0 \\
  $\overline{S}$ & $c$ & 1 & $-3$ & 1 & 0 \\
  $Q$ & $f$  & 1 & 0 & $-2$ & 1 \\
 \hline \hline
  $Q^2$ &   & 1 & 0 & $-4$ & 2 \\
  $S^2Q$ &   & $T_{S}$ & 2 & $-2$ & 1 \\
  $S\overline{S}$ &   & $f$ & $-2$ & 1 & 0 \\
  $S^3\overline{S}Q$&   & $T_{AS}$ & 0 & $-1$ & 1 \\
  $S^2\overline{S}{}^2$ & & $T_S$ & $-4$ & 2 & 0 \\
  $S^4$ && $\overline{T}_S$ & 4 & 0 & 0 \\
  $S^5\overline{S}$ && $T_S$ & 2 & 1 & 0 \\
  $S^4\overline{S}{}^2Q$ && $f$ & $-2$ & 0 & 1 \\
  $\overline{S}{}^2Q$ && 1 & $-6$ & 0 & 1 \\
  $S^3\overline{S}{}^3Q^2$ && 1 & $-6$ & $-1$ & 2 \\
   \hline
\end{tabular}
\end{center}
where $T_{AS}$ stands for the rank three tensor
representation which is symmetric in the first two indices
and antisymmetric in the last two indices.

Corresponding SCIs are
 \beqa && I_E = \frac{(p;p)_\infty^5
(q;q)_\infty^5}{2^4 5!} \int_{{\mathbb T}^5} \frac{\Gamma (tz^{-1})\prod_{i=1}^3
\Big(\Gamma(s_i z) \prod_{j=1}^5\Gamma(s_iz_j^2z^{-1})\Big)\prod_{j=1}^5\Gamma(tzz_j^{-2}) }
{\prod_{1 \leq j < k \leq 5}
\Gamma(z_j^{\pm2}z_k^{\pm2})}
 \nonumber \\ && \makebox[1.5em]{}\times
\prod_{1 \leq j < k \leq 5}\Big( \Gamma(t z_j^2z_k^2z^{-1})
 \prod_{i=1}^3 \Gamma(s_izz_j^{-2}z_k^{-2})
\Big) \prod_{j=1}^5 \Gamma(uz_j^{\pm2}) \frac{dz_j}{2 \pi \textup{i} z_j},
\eeqa
where $z=z_1z_2z_3z_4z_5$, $|s_i|,|t|,|u|<1,$ $s^2t^2u  = \sqrt{pq}$ with $s =
\prod_{i=1}^3 s_i$, and
\beqa &&
I_M= \Gamma(u^2,t^2u,st^3u^2) \Gamma^2 (stu) \prod_{i=1}^3
\Gamma(ts_i,us_i^2,t^2s_i^2,s^2s_i^{-2},sts_i^2,st^2us_i) \nonumber \\
&& \makebox[2em]{} \times \prod_{1 \leq i < j \leq 3}
\Gamma(us_is_j, t^2s_is_j,sts_is_j,s^2s_i^{-1}s_j^{-1})
\prod_{i,j=1;i \neq j}^3 \Gamma(tus^2_is_j).
\eeqa

\subsubsection{$SU(2) \times SU(3) \times U(1)_1 \times U(1)_2$ flavor group}
The matter content is \cite{Csaki:1996zb}
\begin{center}
\begin{tabular}{|c|c|c|c|c|c|c|}
  \hline
    & $SO(10)$ & $SU(2)$ & $SU(3)$ & $U(1)_1$ & $U(1)_2$ & $U(1)_R$ \\  \hline
  $S$ & $s$ & $f$ & 1 & 1 & 1 & 0 \\
  $\overline{S}$ & $c$ & 1 & 1 & $-2$ & 1 & $\frac 12$ \\
  $Q$ & $f$  & 1 & $f$ & 0 & $-2$ & 0 \\
 \hline \hline
  $Q^2$ &   & 1 & $T_S$ & 0 & $-4$ & 0 \\
  $S^2Q$ &   & $T_{S}$ & $f$ & 2 & 0 & 0 \\
  $\overline{S}{}^2Q$ &   & 1 & $f$ & $-4$ & 0 & 1 \\
  $S\overline{S}$ && $f$ & 1 & $-1$ & 2 & $\frac 12$ \\
  $S^2\overline{S}{}^2$ & & $T_S$ & 1 & $-2$ & 4 & 1 \\
  $S^2Q^3$ && 1 & 1 & 2 & $-4$ & 0 \\
  $S^3\overline{S}Q$ && $f$ & $f$ & 1 & 2 & $\frac 12$ \\
  $S^4$ && 1 & 1 & 4 & 4 & 0 \\
  $S\overline{S}Q^2$ && $f$ & $\overline{f}$ & $-1$ & $-2$ & $\frac 12$ \\
  $S^2\overline{S}{}^2Q^2$ && 1 & $\overline{f}$ & $-2$ & 0 & 1 \\
  $S^3\overline{S}Q^3$ && $f$ & 1 & 1 & $-2$ & $\frac 12$ \\
   \hline
\end{tabular}
\end{center}

Corresponding SCIs are
 \beqa && I_E = \frac{(p;p)_\infty^5
(q;q)_\infty^5}{2^4 5!} \int_{{\mathbb T}^5} \frac{\Gamma (tz^{-1})\prod_{i=1}^2
\Big(\Gamma(s_i z) \prod_{j=1}^5\Gamma(s_iz_j^2z^{-1})\Big)\prod_{j=1}^5\Gamma(tzz_j^{-2}) }
{\prod_{1 \leq j < k \leq 5}
\Gamma(z_j^{\pm2}z_k^{\pm2})}
 \nonumber \\ && \makebox[1.5em]{}\times
\prod_{1 \leq j < k \leq 5}\Big( \Gamma(t z_j^2z_k^2z^{-1})
 \prod_{i=1}^2 \Gamma(s_izz_j^{-2}z_k^{-2})
\Big) \prod_{j=1}^5 \Gamma(uz_j^{\pm2}) \frac{dz_j}{2 \pi \textup{i} z_j},
\eeqa
where $z=z_1z_2z_3z_4z_5$, $|s_i|,|t|,|u_j|<1,$  $s^2t^2u  =  \sqrt{pq}$
 with $s = \prod_{i=1}^3 s_i$, and
 \beqa &&
I_M = \Gamma(s^2,su,st^2) \prod_{i=1}^2
\Gamma(ts_i,t^2s_i^2,stus_i) \prod_{i=1}^3
\Gamma(su_i,u_i^2,t^2u_i,st^2uu_i^{-1}) \nonumber \\
&& \makebox[4em]{} \times \prod_{i=1}^2 \prod_{j=1}^3
\Gamma(s_i^2u_j,sts_iu_j,tus_iu_j^{-1}) \prod_{1 \leq i < j \leq
3} \Gamma(u_iu_j).
 \eeqa

\subsubsection{$SU(2)_1 \times SU(2)_2 \times U(1)_1 \times U(1)_2$ flavor group}
The matter content is \cite{Csaki:1996zb}
\begin{center}
\begin{tabular}{|c|c|c|c|c|c|c|}
  \hline
    & $SO(10)$ & $SU(2)_1$ & $SU(2)_2$ & $U(1)_1$ & $U(1)_2$ & $U(1)_R$ \\  \hline
  $S$ & $s$ & $f$ & 1 & 1 & 1 & 0 \\
  $\overline{S}$ & $c$ & 1 & $f$ & $-1$ & 1 & 0 \\
  $Q$ & $f$  & 1 & 1 & 0 & $-8$ & 1 \\
 \hline \hline
  $Q^2$ &   & 1 & 1 & 0 & $-16$ & 2 \\
  $S^2Q$ &   & $T_{S}$ & 1 & 2 & $-6$ & 1 \\
  $\overline{S}{}^2Q$ &   & 1 & $T_S$ & $-2$ & $-6$ & 1 \\
  $S\overline{S}$ && $f$ & $f$ & 0 & 2 & 0 \\
  $S^4$ && 1 & 1 & 4 & 4 & 0 \\
  $\overline{S}{}^4$ && 1 & 1 & $-4$ & 4 & 0 \\
  $S^2\overline{S}{}^2$ && $T_S$ & $T_S$ & 0 & 4 & 0 \\
  $S^3\overline{S}Q$&   & $f$ & $f$ & 2 & $-4$ & 1 \\
  $S\overline{S}{}^3Q$ && $f$ & $f$ & $-2$ & $-4$ & 1 \\
  $S^2\overline{S}{}^2Q^2$ & & 1 & 1 & 0 & $-12$ & 2 \\
  $S^4\overline{S}{}^2Q$ && $T_S$ & 1 & 2 & $-2$ & 1 \\
  $S^2\overline{S}{}^4Q$ && 1 & $T_S$ & $-2$ & $-2$ & 1 \\
  $S^3\overline{S}{}^3$ && $f$ & $f$ & 0 & 6 & 0 \\
  $S^6\overline{S}{}^2$ && 1 & 1 & 4 & 8 & 0 \\
  $S^2\overline{S}{}^6$ && 1 & 1 & $-4$ & 8 & 0 \\
   \hline
\end{tabular}
\end{center}

Corresponding SCIs are
\beqa && I_E = \frac{(p;p)_\infty^5
(q;q)_\infty^5}{2^4 5!} \int_{{\mathbb T}^5} \frac{\prod_{i=1}^2
\Gamma(s_i z,t_iz^{-1}) \prod_{i=1}^2 \prod_{j=1}^5 \Gamma(s_i
z_j^2z^{-1},t_izz_j^{-2}) }{\prod_{1 \leq j < k \leq 5}
\Gamma(z_j^{\pm2}z_k^{\pm2})} \nonumber \\ && \makebox[2em]{}
\times \prod_{i=1}^2 \prod_{1 \leq j < k \leq 5} \Gamma(s_i
zz_j^{-2}z_k^{-2},t_iz_j^2z_k^2z^{-1}) \prod_{j=1}^5 \Gamma(u
z_j^{\pm2})\frac{dz_j}{2 \pi \textup{i} z_j},
\eeqa
 where $z=z_1z_2z_3z_4z_5$, $|s_i|,|t_i|,|u|<1,$ $s^2t^2u = \sqrt{pq}$
with $s = \prod_{i=1}^3 s_i$, and
\beqa &&
I_M = \Gamma(s^2,t^2,u^2,st,su,tu,s^3t,st^3,stu^2,s^2tu,st^2u)
\nonumber \\
&& \makebox[-7em]{} \times \prod_{i=1}^2
\Gamma(us_i^2,ut_i^2,st_i^2,ts_i^2,stus_i^2,stut_i^2)
\prod_{i,j=1}^2
\Gamma(s_it_j,s_i^2t_j^2,sus_it_j,tus_it_j,sts_it_j).
 \eeqa

\subsubsection{$SU(5) \times U(1)_1 \times U(1)_2$ flavor group}
The matter content is \cite{Csaki:1996zb}
\begin{center}
\begin{tabular}{|c|c|c|c|c|c|}
  \hline
    & $SO(10)$ & $SU(5)$ & $U(1)_1$ & $U(1)_2$ & $U(1)_R$ \\  \hline
  $S$ & $s$ & 1 & 1 & 5 & $\frac 14$ \\
  $\overline{S}$ & $c$ & 1 & $-1$ & 5 & $\frac 14$ \\
  $Q$ & $f$  & $f$ & 0 & $-4$ & 0 \\
 \hline \hline
  $Q^2$ &   & $T_S$ & 0 & $-8$ & 0 \\
  $S^2Q$ &  & $f$ & 2 & 6 & $\frac 12$ \\
  $\overline{S}{}^2Q$ &   & $f$ & $-2$ & 6 & $\frac 12$ \\
  $S\overline{S}$&   & 1 & 0 & 10 & $\frac 12$ \\
  $S^2Q^5$ & & 1 & 2 & $-10$ & $\frac 12$ \\
  $\overline{S}{}^2Q^5$ && 1 & $-2$ & $-10$ & $\frac 12$ \\
  $S\overline{S}Q^2$ && $T_A$ & 0 & 2 & $\frac 12$ \\
  $S\overline{S}Q^4$ && $\overline{f}$ & 0 & $-6$ & $\frac 12$ \\
  $S^2\overline{S}{}^2$ && 1 & 0 & 20 & 1 \\
  $S^2\overline{S}{}^2Q^4$ && $\overline{f}$ & 0 & 4 & 1 \\
   \hline
\end{tabular}
\end{center}

Corresponding SCIs are
\beqa && \makebox[-2em]{}
I_E = \frac{(p;p)_\infty^5
(q;q)_\infty^5}{2^4 5!} \int_{{\mathbb T}^5} \frac{\Gamma(s z,t z^{-1})
\prod_{j=1}^5 \Gamma(s z_j^2z^{-1}) }{\prod_{1 \leq j < k \leq
5} \Gamma(z_j^{\pm2}z_k^{\pm2})}
 \nonumber \\ && \makebox[0em]{}
\times \prod_{1 \leq j < k \leq 5} \Gamma(s zz_j^{-2}z_k^{-2},tz_j^2z_k^2z^{-1})
\prod_{i,j=1}^5 \Gamma(u_i z_j^{\pm2})
\prod_{j=1}^5  \Gamma(t zz_j^{-2})\frac{dz_j}{2 \pi \textup{i} z_j}, \eeqa where
$z=z_1z_2z_3z_4z_5$, $|s|,|t|,|u_i|<1,$  $s^2t^2u = \sqrt{pq}$
with $u = \prod_{i=1}^5 u_i$, and
\begin{equation} \makebox[-2em]{}
I_M =\Gamma(st,su,tu,s^2t^2) \prod_{i=1}^5
\Gamma(u_i^2,s^2u_i,t^2u_i,stuu_i^{-1},s^2t^2uu_i^{-1})
\prod_{1 \leq i < j \leq 5}\Gamma(u_iu_j,stu_iu_j).
\end{equation}

\subsection{$G=SO(11)$}

\subsubsection{$SU(6) \times U(1)$ flavor group}
The matter content is \cite{Csaki:1996zb}
\begin{center}
\begin{tabular}{|c|c|c|c|c|}
  \hline
    & $SO(11)$ & $SU(6)$ & $U(1)$ & $U(1)_R$ \\  \hline
  $S$ & $s$ & 1 & 3 & $\frac 14$ \\
  $Q$ & $f$  & $f$  & $-2$ & 0 \\
 \hline \hline
  $Q^2$ &   & $T_S$ & $-4$ & 0 \\
  $S^2Q^2$ &   & $T_A$ & 2 & $\frac 12$ \\
  $S^2Q^5$ &   & $\overline{f}$ & $-4$ & $\frac 12$ \\
  $S^4$&  & 1 & 12 & 1 \\
  $S^4Q^5$ && $\overline{f}$ & 2 & 1 \\
  $S^2Q$ && $f$ & 4 & $\frac 12$ \\
  $S^2Q^6$ && 1 & $-6$ & $ \frac 12$ \\
   \hline
\end{tabular}
\end{center}

Corresponding SCIs are
\beqa && I_E = \frac{(p;p)_\infty^5
(q;q)_\infty^5}{2^5 5!} \prod_{i=1}^6 \Gamma(t_i) \int_{{\mathbb
T}^5} \frac{\Gamma(s z^{\pm1}) \prod_{j=1}^5 \Gamma(s
(z_j^2z^{-1})^{\pm1})}{\prod_{j=1}^5 \Gamma(z_j^{\pm2})
\prod_{1 \leq j < k \leq 5} \Gamma(z_j^{\pm2}z_k^{\pm2})}
\nonumber \\ && \makebox[3em]{} \times \prod_{1 \leq j < k \leq 5}
\Gamma(s (zz_j^{-2}z_k^{-2})^{\pm1}) \prod_{i=1}^6 \prod_{j=1}^5
\Gamma(t_i z_j^{\pm2}) \prod_{j=1}^5 \frac{dz_j}{2 \pi
\textup{i} z_j},
\eeqa
where $z=z_1z_2z_3z_4z_5$, $|s|,|t_i|<1,$ $s^4t =  \sqrt{pq}$ with $t = \prod_{i=1}^6 t_i$, and
\beq I_M = \Gamma(s^4,s^2t) \prod_{i=1}^6 \Gamma(s^2t_i, t_i^2,
s^2tt_i^{-1}, s^4tt_i^{-1}) \prod_{1 \leq i < j \leq 6}
\Gamma(t_it_j,s^2t_it_j).
\eeq

\subsubsection{$SU(2)_1 \times SU(2)_2 \times U(1)$ flavor group}
This $s$-confining theory was found in \cite{Maru:1998hp}. The matter content is
\begin{center}
\begin{tabular}{|c|c|c|c|c|c|}
  \hline
    & $SO(11)$ & $SU(2)_1$ & $SU(2)_2$ & $U(1)$ & $U(1)_R$ \\  \hline
  $S$ & $s$ & $f$ & 1 & 1 & 0 \\
  $Q$ & $f$  & 1 & $f$ & $-4$ & $\frac 12$ \\
 \hline \hline
  $Q^2$ &   & 1 & $T_S$ & $-8$ & 1 \\
  $S^2Q^2$ &   & $T_S$ & 1 & $-6$ & 1 \\
  $S^2Q$ &   & $T_S$ & $f$ & $-2$ & $\frac 12$ \\
  $S^2$&  & 1 & 1 & 2 & 0 \\
  $S^4$ && $T_{4S}$ & 1 & 4 & 0 \\
  $S^{4'}$&& 1 & 1 & 4 & 0 \\
  $S^4Q^2$ && 1 & $T_S$ & $-4$ & 1 \\
  $S^4Q^{2'}$ && $T_S$ & 1 & $-4$ & 1 \\
  $S^4Q$ && $T_S$ & $f$ & 0 & $\frac 12$ \\
  $S^6Q^2$ && $T_S$ & 1 & $-2$ & 1 \\
  $S^6Q$ && $T_S$ & $f$ & 2 & $\frac 12$ \\
  $S^8$ && 1 & 1 & 8 & 0 \\
  $S^8Q$&& 1 & $f$ & 4 & $\frac 12$ \\
  $S^4Q$&& 1 & $f$ & 0 & $\frac 12$ \\
  $S^6$ && 1 & 1 & 6 & 0 \\
   \hline
\end{tabular}
\end{center}
Here $T_{4S}$ denotes the totally symmetric tensor of the fourth rank.

Corresponding SCIs are
\beqa && I_E = \frac{(p;p)_\infty^5
(q;q)_\infty^5}{2^5 5!} \prod_{i=1}^2 \Gamma(t_i) \int_{{\mathbb
T}^5} \frac{\prod_{i=1}^2 \Gamma(s_i z^{\pm1}) \prod_{i=1}^2
\prod_{j=1}^5 \Gamma(s_i (z_j^2z^{-1})^{\pm1})}{\prod_{j=1}^5
\Gamma(z_j^{\pm2}) \prod_{1 \leq j < k \leq 5}
\Gamma(z_j^{\pm2}z_k^{\pm2})} \nonumber \\ && \makebox[3.5em]{}
\times \prod_{i=1}^2 \prod_{1 \leq j < k \leq 5} \Gamma(s_i
(zz_j^{-2}z_k^{-2})^{\pm1}) \prod_{i=1}^2 \prod_{j=1}^5
\Gamma(t_i z_j^{\pm2}) \prod_{j=1}^5 \frac{dz_j}{2 \pi
\textup{i} z_j},
\eeqa
where $z=z_1z_2z_3z_4z_5$, $|s|,|t_i|<1,$ $s^4t = \sqrt{pq}$
with $s = \prod_{i=1}^2 s_i, \ t =\prod_{i=1}^2 t_i$, and
\beqa && I_M = \Gamma(s, t, st, s^3, s^3t, s^4) \Gamma^2(s,
s^2t) \prod_{i=1}^2 \Gamma(t_i^2, st_i, ts_i^2, s^2t_i,
ss_i^2) \nonumber
\\ && \makebox[-0.5em]{} \times
\prod_{i=1}^2 \Gamma(s^2t_i, s^2t_i^2, sts_i^2, s^2ts_i^2, s^3t_i,
s_i^4, s^4t_i) \prod_{i,j=1}^2 \Gamma(s_i^2t_j, ss_i^2t_j,
s^2s_i^2t_j).
\eeqa

\subsection{$G=SO(12)$}

\subsubsection{$SU(7) \times U(1)$ flavor group} The matter
content is \cite{Csaki:1996zb}
\begin{center}
\begin{tabular}{|c|c|c|c|c|}
  \hline
    & $SO(12)$ & $SU(7)$ & $U(1)$ & $U(1)_R$ \\  \hline
  $S$ & $s$ & 1 & 7 & $\frac 14$ \\
  $Q$ & $f$  & $f$  & $-4$ & 0 \\
 \hline \hline
  $Q^2$ &   & $T_S$ & $-8$ & 0 \\
  $S^2Q^2$ &   & $T_A$ & 6 & $\frac 12$ \\
  $S^2Q^6$ &   & $\overline{f}$ & $-10$ & $\frac 12$ \\
  $S^4$&  & 1 & 28 & 1 \\
  $S^4Q^6$ && $\overline{f}$ & 4 & 1 \\
   \hline
\end{tabular}
\end{center}

Corresponding SCIs are
\beqa && I_E = \frac{(p;p)_\infty^6
(q;q)_\infty^6}{2^5 6!} \int_{{\mathbb T}^6} \frac{\Gamma(s
z^{\pm1})}{\prod_{1 \leq j < k \leq 6}
\Gamma(z_j^{\pm2}z_k^{\pm2})} \nonumber \\ && \makebox[-2.5em]{}
\times \prod_{1 \leq j < k \leq 6} \Gamma(s
(zz_j^{-2}z_k^{-2})^{\pm1}) \prod_{i=1}^7 \prod_{j=1}^6
\Gamma(t_i z_j^{\pm2}) \prod_{j=1}^6 \frac{dz_j}{2 \pi
\textup{i} z_j},
\eeqa
where $z=z_1z_2z_3z_4z_5z_6$, $|s|,|t_i|<1,$ $s^4t = \sqrt{pq}$
with $t = \prod_{i=1}^7 t_i$, and
\beq
I_M = \Gamma(s^4) \prod_{i=1}^7 \Gamma(t_i^2,
s^2tt_i^{-1}, s^4tt_i^{-1}) \prod_{1 \leq i < j \leq 6}
\Gamma(t_it_j, s^2t_it_j).
 \eeq

\subsubsection{$SU(2) \times SU(3) \times U(1)$ flavor group}
The matter content is \cite{Csaki:1996zb}
\begin{center}
\begin{tabular}{|c|c|c|c|c|c|}
  \hline
    & $SO(12)$ & $SU(2)$ & $SU(3)$ & $U(1)$ & $U(1)_R$ \\  \hline
  $S$ & $s$ & $f$ & 1 & 3 & $\frac 18$ \\
  $Q$ & $f$  & 1 & $f$  & $-8$ & 0 \\
 \hline \hline
  $Q^2$ &   & 1 & $T_S$ & $-16$ & 0 \\
  $S^2$ &   & 1 & 1 & 6 & $\frac 14$ \\
  $S^2Q^2$ &   & $T_S$ & $\overline{f}$ & $-10$ & $\frac 14$ \\
  $S^4$&  & $T_{4S}$ & 1 & 12 & $\frac 12$ \\
  $S^4Q^2$ && 1 & $T_S$ & $-4$ & $\frac 12$ \\
  $S^4Q^{2'}$ && $T_S$ & $\overline{f}$ & $-4$ & $\frac 12$ \\
  $S^6$ && 1 & 1 & 18 & $\frac 34$ \\
  $S^6Q^2$ && $T_S$ & $\overline{f}$ & 2 & $\frac 34$ \\
  $S^8Q^2$ && 1 & $T_S$ & 8 & 1 \\
   \hline
\end{tabular}
\end{center}

Corresponding SCIs are
\beqa && I_E = \frac{(p;p)_\infty^6
(q;q)_\infty^6}{2^5 6!} \int_{{\mathbb T}^6} \frac{\prod_{i=1}^2
\Gamma(s_i z^{\pm1})}{\prod_{1 \leq j < k \leq 6}
\Gamma(z_j^{\pm2}z_k^{\pm2})} \nonumber \\ && \makebox[-3.5em]{}
\times \prod_{i=1}^2 \prod_{1 \leq j < k \leq 6} \Gamma(s_i
(zz_j^{-2}z_k^{-2})^{\pm1}) \prod_{i=1}^3 \prod_{j=1}^6
\Gamma(t_i z_j^{\pm2}) \prod_{j=1}^6 \frac{dz_j}{2 \pi
\textup{i} z_j},
\eeqa
where $z=z_1z_2z_3z_4z_5z_6$,
$|s_i|,|t_i|<1,$ $s^4t = \sqrt{pq}$ with $s = \prod_{i=1}^2 s_i,\ t = \prod_{i=1}^3
t_i$, and
 \beqa && I_M = \Gamma(s, s^2, s^3)
\prod_{i=1}^2 \Gamma(ss_i^2, s_i^4) \prod_{i=1}^3 \Gamma(t_i^2,
s^2t_i^2,
stt_i^{-1}, s^2tt_i^{-1}, s^4t_i^2, s^3tt_i^{-1}) \nonumber \\
&& \makebox[2em]{} \times \prod_{i=1}^2 \prod_{j=1}^3
\Gamma(sts_i^2t_j^{-1}, s^2ts_i^2t_j^{-1}, ts_i^2t_j^{-1})
\prod_{1 \leq i < j \leq 3} \Gamma(t_it_j, s^2t_it_j,s^4t_it_j).
 \eeqa

\subsubsection{$SU(3) \times U(1)_1 \times U(1)_2$ flavor group}
The matter content is \cite{Csaki:1996zb}
\begin{center}
\begin{tabular}{|c|c|c|c|c|c|}
  \hline
    & $SO(12)$ & $SU(3)$ & $U(1)_1$ & $U(1)_2$ & $U(1)_R$ \\  \hline
  $S$ & $s$ & 1 & 1 & 3 & $\frac 18$ \\
  $S'$ & $c$ & 1 & $-1$ & 3 & $\frac 18$ \\
  $Q$ & $f$  & $f$ & 0 & $-8$ & 0 \\
 \hline \hline
  $Q^2$ &   & $T_S$ & 0 & $-16$ & 0 \\
  $SS'Q^3$ &   & 1 & 0 & $-18$ & $\frac 14$ \\
  $S^2Q^2$ &   & $\overline{f}$ & 2 & $-10$ & $\frac 14$ \\
  $S^{'2}Q^2$& & $\overline{f}$ & $-2$ & $-10$ & $\frac 14$ \\
  $SS'Q$ && $f$ & 0 & $-2$ & $\frac 14$ \\
  $S^4$ && 1 & 4 & 12 & $\frac 12$ \\
  $S^{'4}$&& 1 & $-4$ & 12 & $\frac 12$ \\
  $S^2S^{'2}$&& 1 & 0 & 12 & $\frac 12$ \\
  $S^3S'Q^3$ && 1 & 2 & $-12$ & $\frac 12$ \\
  $SS^{'3}Q^3$ && 1 & $-2$ & $-12$ & $\frac 12$ \\
  $S^2S^{'2}Q^2$ && $T_S$ & 0 & $-4$ & $\frac 12$ \\
  $S^2S^{'2}Q^{2'}$ && $\overline{f}$ & 0 & $-4$ & $\frac 12$ \\
  $S^3S'Q$ && $f$ & 2 & 4 & $\frac 12$ \\
  $SS^{'3}Q^2$ && $f$ & $-2$ & 4 & $\frac 12$ \\
  $S^3S^{'3}Q^3$ && 1 & 0 & $-6$ & $\frac 34$ \\
  $S^3S^{'3}Q$ && $f$ & 0 & 10 & $\frac 34$ \\
  $S^4S^{'2}Q^2$&& $\overline{f}$ & 2 & 2 & $\frac 34$ \\
  $S^2S^{'4}Q^2$&& $\overline{f}$ & $-2$ & 2 & $\frac 34$ \\
  $S^4S^{'4}$ && 1 & 0 & 24 & 1 \\
  $S^4S^{'4}Q^2$ && $\overline{f}$ & 0 & 8 & 1 \\
   \hline
\end{tabular}
\end{center}

Corresponding SCIs are
 \beqa && I_E = \frac{(p;p)_\infty^6
(q;q)_\infty^6}{2^5 6!} \int_{{\mathbb T}^6} \frac{\Gamma(s
z^{\pm1})}{\prod_{1 \leq j < k \leq 6}
\Gamma(z_j^{\pm2}z_k^{\pm2})} \prod_{1 \leq j < k \leq 6}
\Gamma(s (zz_j^{-2}z_k^{-2})^{\pm1}) \nonumber \\ &&
\makebox[-4em]{} \times \prod_{i=1}^6 \Gamma(t
(z_i^2z^{-1})^{\pm1}) \prod_{1 \leq i < j < k \leq 6} \Gamma(t
(z_i^2z_j^2z_k^2z^{-1})^{\pm1}) \prod_{i=1}^3 \prod_{j=1}^6
\Gamma(u_i z_j^{\pm2}) \prod_{j=1}^6 \frac{dz_j}{2 \pi
\textup{i} z_j},
\eeqa
where $z=z_1z_2z_3z_4z_5z_6$, $|s|,|t|,|u_i|<1,$ $(st)^4u  =  \sqrt{pq}$
 with $u = \prod_{i=1}^3 u_i$, and
\beqa &&
I_M = \Gamma(stu,s^4,t^4,s^2t^2,st^3u,s^3tu,s^3t^3u,s^4t^4) \prod_{1 \leq i < j \leq 3} \Gamma(u_iu_j,s^2t^2u_iu_j) \nonumber \\
&& \makebox[3em]{} \times \prod_{i=1}^3
\Gamma(s^3tu_i,st^3u_i,s^3t^3u_i,s^4t^2uu_i^{-1},s^2t^4uu_i^{-1},s^4t^4uu_i^{-1}) \\
\nonumber && \makebox[4.5em]{} \times \prod_{i=1}^3
\Gamma(u_i^2,s^2uu_i^{-1},t^2uu_i^{-1},stu_i,s^2t^2u_i^2,s^2t^2uu_i^{-1}).
\eeqa

\subsection{$G=SO(13)$}

\subsubsection{$SU(4) \times U(1)$ flavor group}
The matter content is \cite{Csaki:1996zb}
\begin{center}
\begin{tabular}{|c|c|c|c|c|}
  \hline
    & $SO(13)$ & $SU(4)$ & $U(1)$ & $U(1)_R$ \\  \hline
  $S$ & $s$ & 1 & 1 & $\frac 18$ \\
  $Q$ & $f$  & $f$  & $-2$ & 0 \\
 \hline \hline
  $Q^2$ &   & $T_S$ & $-4$ & 0 \\
  $S^2Q^3$ &   & $\overline{f}$ & $-4$ & $\frac 14$ \\
  $S^2Q^2$ &   & $T_A$ & $-2$ & $\frac 14$ \\
  $S^4Q^4$&  & 1 & $-4$ & $\frac 12$ \\
  $S^4Q^3$ && $\overline{f}$ & $-2$ & $\frac 12$ \\
  $S^4Q^2$ && $T_S$ & 0 & $\frac 12$ \\
  $S^4Q$ && $f$ & 2 & $\frac 12$ \\
  $S^4$ && 1 & 4 & $\frac 12$ \\
  $S^6Q^3$ && $\overline{f}$ & 0 & $\frac 34$ \\
  $S^6Q^2$ && $T_A$ & 2 & $\frac 34$ \\
  $S^8Q^3$ && $\overline{f}$ & 2 & 1 \\
  $S^8$ && 1 & 8 & 1 \\
   \hline
\end{tabular}
\end{center}

Corresponding SCIs are
\beqa && I_E = \frac{(p;p)_\infty^6
(q;q)_\infty^6}{2^6 6!} \prod_{i=1}^4 \Gamma(t_i) \int_{{\mathbb
T}^6} \frac{\Gamma(s z^{\pm1}) \prod_{j=1}^6 \Gamma(s
(z_j^2z^{-1})^{\pm1})}{\prod_{j=1}^6 \Gamma(z_j^{\pm2})
\prod_{1 \leq j < k \leq 6} \Gamma(z_j^{\pm2}z_k^{\pm2})}
\\ \nonumber && \makebox[-2.5em]{} \times \prod_{1 \leq j < k \leq 6}
\Gamma(s (zz_j^{-2}z_k^{-2})^{\pm1}) \prod_{1 \leq i < j < k
\leq 6} \Gamma(s z_i^2z_j^2z_k^2z^{-1}) \prod_{i=1}^4
\prod_{j=1}^6 \Gamma(t_i z_j^{\pm2}) \prod_{j=1}^6 \frac{dz_j}{2
\pi \textup{i} z_j},
\eeqa
 where $z=z_1z_2z_3z_4z_5z_6$, $|s|,|t_i|<1,$ $s^8t =  \sqrt{pq}$ with $t = \prod_{i=1}^4
t_i$, and
\beqa && I_M = \Gamma(s^4, s^4t, s^8)
\prod_{i=1}^4 \Gamma(t_i^2, s^2tt_i^{-1}, s^4tt_i^{-1}, s^4t_i^2,
s^4t_i, s^6tt_i^{-1}, s^8tt_i^{-1}) \nonumber \\ &&
\makebox[5em]{} \times \prod_{1 \leq i < j \leq 4} \Gamma(t_it_j,
s^2t_it_j, s^4t_it_j, s^6t_it_j).
\eeqa

\subsection{$G=SO(14)$}

\subsubsection{$SU(5) \times U(1)$ flavor group}
The matter content is \cite{Csaki:1996zb}
\begin{center}
\begin{tabular}{|c|c|c|c|c|}
  \hline
    & $SO(14)$ & $SU(5)$ & $U(1)$ & $U(1)_R$ \\  \hline
  $S$ & $s$ & 1 & 5 & $\frac 18$ \\
  $Q$ & $f$  & $f$  & $-8$ & 0 \\
 \hline \hline
  $Q^2$ &   & $T_S$ & $-16$ & 0 \\
  $S^2Q^3$ &   & $\overline{T}_A$ & $-14$ & $\frac 14$ \\
  $S^4Q^2$ &   & $T_S$ & 4 & $\frac 12$ \\
  $S^4Q^4$&  & $\overline{f}$ & $-12$ & $\frac 12$ \\
  $S^6Q^3$ && $\overline{T}_A$ & 6 & $\frac 34$ \\
  $S^8$ && 1 & 40 & 1 \\
  $S^8Q^4$ && $\overline{f}$ & 8 & 1 \\
   \hline
\end{tabular}
\end{center}

Corresponding SCIs are
\beqa && I_E = \frac{(p;p)_\infty^7
(q;q)_\infty^7}{2^6 7!} \int_{{\mathbb T}^7} \frac{\Gamma(s z)
\prod_{j=1}^7 \Gamma(s z_j^2z^{-1})}{\prod_{1 \leq j < k \leq 7}
\Gamma(z_j^{\pm2}z_k^{\pm2})}
\\ \nonumber && \makebox[-6em]{} \times \prod_{1 \leq j < k \leq 7}
\Gamma(s zz_j^{-2}z_k^{-2}) \prod_{1 \leq i < j < k \leq 7}
\Gamma(s z_i^2z_j^2z_k^2z^{-1}) \prod_{i=1}^5 \prod_{j=1}^7
\Gamma(t_i z_j^{\pm2}) \prod_{j=1}^7 \frac{dz_j}{2 \pi
\textup{i} z_j}, \eeqa where $z=z_1z_2z_3z_4z_5z_6z_7$,
$|s|,|t_i|<1,$ $s^8t =  \sqrt{pq}$ with $t = \prod_{i=1}^5 t_i$, and
\beq I_M = \Gamma(s^8) \prod_{i=1}^5
\Gamma(t_i^2, s^4tt_i^{-1}, s^8tt_i^{-1}, s^4t_i^2) \prod_{1 \leq i < j \leq 5} \Gamma(t_it_j,
s^2tt_i^{-1}t_j^{-1}, s^4t_it_j, s^6tt_i^{-1}t_j^{-1}).
 \eeq

To summarize, the formulas of this section lead to conjectures for exact
evaluations of certain  EHIs  on $B_N$ and $D_N$ root systems
constructed from the characters of various representations necessarily
including the spinor representation, which require now rigorous mathematical proofs.

\section{Self-dual theories with the spinor matter}\label{self-dual}

We start by presenting a basic example of the self-dual
$\mathcal{N}=1$ SYM theory based on the orthogonal gauge group with
some number of fields in spinor representation. It was considered first in
\cite{Csaki2}, further examples have been described in
\cite{Distler:1996ub,Karch}. First we consider the theory with
$SO(8)$ gauge group and the flavor group
$SU(4)_l \times SU(4)_r \times U(1)_B.$
The matter content of this theory is
\begin{center}
\begin{tabular}{|c|c|c|c|c|c|}
  \hline
    & $SO(8)$ & $SU(4)_l$ & $SU(4)_r$ & $U(1)_B$ & $U(1)_R$ \\  \hline
  $S$ & $s$ & $f$ & 1 & $1$ & $\frac{1}{4}$ \\
  $Q$ & $f$ & 1 & $f$ & $-1$ & $\frac{1}{4}$ \\ \hline
\end{tabular}
\end{center}

In \cite{Csaki2} there were found $5$ theories  dual to the original
electric theory. We reconsidered these theories using SCI technique
and found that there are, actually, only $3$ dual theories. Other
theories have the fields which can be integrated out and, in particular,
their contribution to 't Hooft anomaly matching conditions is trivial (none).
The matter fields of dual theories are listed below in the table,
where the double lines separate dual theories.
\begin{center}
\begin{tabular}{|c|c|c|c|c|c|}
  \hline
    & $SO(8)$ & $SU(4)_l$ & $SU(4)_r$ & $U(1)_B$ & $U(1)_R$ \\  \hline
  $s$ & $s$ & $\overline{f}$ & 1 & $1$ & $\frac{1}{4}$ \\
  $v$ & $f$ & $1$ & $f$ & $-1$ & $\frac{1}{4}$ \\
  $M$ & 1 & $T_S$ & 1 & 2 & $\frac 12$ \\
  $N$ & 1 & $T_S$ & 1 & $-2$ & $\frac 32$ \\
   \hline \hline
  $s$ & $s$ & $f$ & 1 & $1$ & $\frac{1}{4}$ \\
  $v$ & $f$ & $1$ & $\overline{f}$ & $-1$ & $\frac{1}{4}$ \\
  $\widetilde{M}$ & 1 & 1 & $T_S$ & $-2$ & $\frac 12$ \\
  $\widetilde{N}$ & 1 & 1 & $T_S$ & 2 & $\frac 32$ \\
   \hline \hline
  $s$ & $s$ & $\overline{f}$ & 1 & $1$ & $\frac{1}{4}$ \\
  $v$ & $f$ & $1$ & $\overline{f}$ & $-1$ & $\frac{1}{4}$ \\
  $M$ & 1 & $T_S$ & 1 & 2 & $\frac 12$ \\
  $N$ & 1 & $T_S$ & 1 & $-2$ & $\frac 32$ \\
  $\widetilde{M}$ & 1 & 1 & $T_S$ & $-2$ & $\frac 12$ \\
  $\widetilde{N}$ & 1 & 1 & $T_S$ & 2 & $\frac 32$ \\ \hline
\end{tabular}
\end{center}

Corresponding SCIs are given by the following expressions \beqa
&& I_E = \frac{(p;p)^4_\infty (q;q)^4_\infty}{2^3 4!}
\int_{\mathbb{T}^4} \frac{\prod_{i,j=1}^4
\Gamma(s_iz_j^{\pm2})}{\prod_{1 \leq i < j \leq 4}
\Gamma(z_i^{\pm2}z_j^{\pm2})} \nonumber \\ && \makebox[-2em]{}
\times \prod_{i=1}^4 \Gamma(t_i Z^{\pm1}) \prod_{i=1}^4 \prod_{1
\leq j < k \leq 4} \Gamma(t_i z_j^2z_k^2Z^{-1}) \prod_{j=1}^4
\frac{d z_j}{2 \pi \textup{i} z_j},\eeqa where $Z=z_1z_2z_3z_4$ and
the balancing condition reads $\prod_{i=1}^4 s_it_i  = pq.$
Magnetic SCIs are
 \beqa && I_M^{(1)} = \prod_{i=1}^4 \Gamma(t_i^2,
St_i^2) \frac{(p;p)^4_\infty (q;q)^4_\infty}{2^3 4!}
\int_{\mathbb{T}^4} \frac{\prod_{i,j=1}^4
\Gamma(s_iz_j^{\pm2})}{\prod_{1 \leq i < j \leq 4}
\Gamma(z_i^{\pm2}z_j^{\pm2})} \nonumber \\ && \makebox[1em]{}
\times \prod_{i=1}^4 \Gamma(\frac{\sqrt{T}}{t_i} Z^{\pm1})
\prod_{i=1}^4 \prod_{1 \leq j < k \leq 4}
\Gamma(\frac{\sqrt{T}}{t_i} z_j^2z_k^2Z^{-1}) \prod_{j=1}^4
\frac{d z_j}{2 \pi \textup{i} z_j},\eeqa for the first magnetic
theory; \beqa && I_M^{(2)} = \prod_{i=1}^4 \Gamma(s_i^2, Ts_i^2)
\frac{(p;p)^4_\infty (q;q)^4_\infty}{2^3 4!} \int_{\mathbb{T}^4}
\frac{\prod_{i,j=1}^4 \Gamma(\frac{\sqrt{S}}{s_i}
z_j^{\pm2})}{\prod_{1 \leq i < j \leq 4}
\Gamma(z_i^{\pm2}z_j^{\pm2})} \nonumber \\ && \makebox[1em]{}
\times \prod_{i=1}^4 \Gamma(t_i Z^{\pm1}) \prod_{i=1}^4 \prod_{1
\leq j < k \leq 4} \Gamma(t_i z_j^2z_k^2Z^{-1}) \prod_{j=1}^4
\frac{d z_j}{2 \pi \textup{i} z_j},\eeqa for the second magnetic
theory; \beqa && I_M^{(3)} = \prod_{i=1}^4 \Gamma(s_i^2, t_i^2,
Ts_i^2, St_i^2) \frac{(p;p)^4_\infty (q;q)^4_\infty}{2^3 4!}
\int_{\mathbb{T}^4} \frac{\prod_{i,j=1}^4
\Gamma(\frac{\sqrt{S}}{s_i} z_j^{\pm2})}{\prod_{1 \leq i < j
\leq 4} \Gamma(z_i^{\pm2}z_j^{\pm2})} \nonumber \\ &&
\makebox[2.4em]{} \times \prod_{i=1}^4 \Gamma(\frac{\sqrt{T}}{t_i}
Z^{\pm1}) \prod_{i=1}^4 \prod_{1 \leq j < k \leq 4}
\Gamma(\frac{\sqrt{T}}{t_i} z_j^2z_k^2Z^{-1}) \prod_{j=1}^4
\frac{d z_j}{2 \pi \textup{i} z_j},\eeqa for the third magnetic
theory.

The situation with other self-dual theories is not so clear, e.g.
the self-duality of \cite{Csaki2,Karch} based on $SO(12)$ gauge group
with one field in the spinor representation and $8$ quarks in the
fundamental representation seems to be incorrect.
First, the representations and charges of the dual
quarks and spinor representation fields are not changed. Second,
the fields $M_4$ and $M_8$ (taken from the second section of
\cite{Karch}) can be integrated out and their contributions to
anomalies cancel out leading thus back to the original theory.

\section{Seiberg dualities for $SO(N)$ gauge group with the spinor matter} \label{dual}

\subsection{$G=SO(5)$ and $F=SU(N_f) \times SO(4) \times U(1)$} \label{DK}

A duality with $SU(N_f) \times SU(4) \times U(1)$ flavor group
was studied in \cite{Distler:1996ub} with the claim that it
can be derived from a more general duality, which we shall consider later
in Sect. \ref{DualSO10}. Using SCI technique we show that this statement
is incorrect. In our language, the duality of Sect. \ref{DualSO10}
reduces to the duality discussed below which is based on $SO(4)$-flavor subgroup
instead of $SU(4)$.

Let us describe the corrected duality from \cite{Distler:1996ub}.
The electric theory is represented by the following table
\begin{center}
\begin{tabular}{|c|c|c|c|c|c|}
  \hline
    & $SO(5)$ & $SU(N_f)$ & $SO(4)$ & $U(1)$ & $U(1)_R$ \\  \hline
  $Q$ & $f$ & $f$ & 1 & $-1$ & $1-\frac{3}{N_f+2}$ \\
  $S$ & $s$  & 1  & $f$ & $\frac{N_f}{2}$ & $1-\frac{3}{N_f+2}$ \\ \hline
\end{tabular}
\end{center}
while the magnetic theory is \begin{center}
\begin{tabular}{|c|c|c|c|c|c|}
  \hline
    & $SU(N_f)$ & $SU(N_f)$ & $SO(5) \simeq SP(4)$ & $U(1)$ & $U(1)_R$ \\  \hline
  $q$ & $\overline{f}$ & $\overline{f}$ & 1 & 1 & $\frac{3}{N_f+2} - \frac{1}{N_f}$ \\
  $q'$ & $f$  & 1  & 1 & $-N_f$ & $-1+\frac{6}{N_f+2}+\frac{1}{N_f}$ \\
  $w$ & $T_S$  & 1  & 1 & 0 & $\frac{2}{N_f}$ \\
  $t$ & $\overline{f}$  & 1  & $f$ & 0 & $1-\frac{1}{N_f}$ \\
  $Y$ & 1  & $f$  & 1 & $N_f-1$ & $3-\frac{9}{N_f+2}$ \\
  $M$ & 1  & $T_S$  & 1 & $-2$ & $2-\frac{6}{N_f+2}$ \\
  $N$ & 1  & $1$  & $f$ & $N_f$ & $2-\frac{6}{N_f+2}$ \\ \hline
\end{tabular}
\end{center}

The indices are
\beqa \label{SO5_e} && I_E = \frac{(p;p)_\infty^2
(q;q)_\infty^2}{2^2 2!} \prod_{i=1}^{N_f} \Gamma(s_i)  \int_{{\mathbb T}^2}
 \frac{\Gamma(t u_1^{\pm1}u_2^{\pm1}
(z_1z_2)^{\pm 1}, t u_1^{\pm1}u_2^{\pm1} \left( z_1 z_2^{-1} \right)^{\pm 1})}{\Gamma(z_1^{\pm2}z_2^{\pm2},z_1^{\pm2},z_2^{\pm2})}
\nonumber
\\ && \makebox[7em]{} \times
\prod_{i=1}^{N_f} \prod_{j=1}^2 \Gamma(s_iz_j^{\pm
2}) \prod_{j=1}^2 \frac{dz_j}{2 \pi \textup{i} z_j},\eeqa
where the balancing condition is
$s^2t^4  = (pq)^{N_f-1}$ with $s=\prod_{i=1}^{N_f} s_i$, and
\beqa \label{SO5_m} && I_M = \frac{(p;p)_\infty^{N_f-1} (q;q)_\infty^{N_f-1}}{N_f!}
\Gamma((pq)^{\frac{N_f-1}{2}}s^{-1}) \prod_{j=1,2}
\Gamma((pq)^{\frac{N_f-1}{2}}s^{-1}u_j^{\pm1})
\\ \nonumber && \makebox[3em]{} \times \prod_{1 \leq i < j \leq N_f}
\Gamma(s_is_j) \prod_{i=1}^{N_f} \Gamma(s_i^2) \prod_{i=1}^{N_f}
\Gamma((pq)^{\frac{N_f-1}{2}} s^{-1}s_i) \\ \nonumber &&
\makebox[4em]{} \times \int_{{\mathbb T}^{N_f-1}} \prod_{1 \leq i < j
\leq N_f} \frac{\Gamma((pq)^{\frac{1}{N_f}}y_iy_j)}
{\Gamma(y_iy_j^{-1},y_i^{-1}y_j)} \prod_{i=1}^{N_f}
\Gamma((pq)^{\frac{1}{N_f}} y_i^2)
\\ \nonumber && \makebox[0em]{} \times \prod_{i,j=1}^{N_f}
\Gamma((pq)^{\frac{N_f-1}{2N_f}} s_i^{-1}y_j^{-1}) \prod_{j=1}^{N_f}
\Gamma((pq)^{\frac 12 (2+\frac{1}{N_f}
-N_f)}sy_j,(pq)^{\frac{N_f-1}{2N_f}}y_j^{-1})
\\ \nonumber && \makebox[6em]{} \times \prod_{i=1,2}
\prod_{j=1}^{N_f} \Gamma((pq)^{\frac{N_f-1}{2N_f}} u_i^{\pm1} y_j^{-1})
\prod_{j=1}^{N_f-1} \frac{dy_j}{2 \pi \textup{i} y_j},\eeqa where
$\prod_{j=1}^{N_f} y_j=1$.

A simple explanation of the inconsistency of the duality of \cite{Distler:1996ub}
consists in the mismatch of the number of independent fugacities (parameters)
in the dual indices, for the $SU(4)$-flavor subgroup there will be
an extra parameter in the electric theory in comparison with
the magnetic one. In principle, as described in \cite{SV1}, the integrands
entering indices may have different number of parameters,
but there should be some additional multipliers to the integrals which
cancel the contribution of these extra parameters.

\subsection{$SO(7)$ gauge group with $N_f$ fundamentals}

The $\mathcal{N}=1$ SYM electric theory described in this section was historically
the first model including a matter field in the spinor representation with known
dual theory. It was discovered by Pouliot \cite{Pouliot:1995zc}
and it is based on $SO(7)$ gauge group with the following matter content
\begin{center}
\begin{tabular}{|c|c|c|c|}
  \hline
   & $SO(7)$ & $SU(N_f)$ & $U(1)_R$ \\
\hline
  $Q$ & $s$ & $f$ & $1 - \frac{5}{N_f}$ \\
 \hline
\end{tabular}
\end{center}
where $s$ means the spinor representation.
Pouliot found the following dual magnetic theory
\begin{center}
\begin{tabular}{|c|c|c|c|}
  \hline
   & $SU(N_f-4)$ & $SU(N_f)$ & $U(1)_R$ \\ \hline
  $q$ & $\overline{f}$ & $\overline{f}$ & $\frac{5}{N_f} - \frac{1}{N_f-4}$ \\
  $w$ & $T_S$ & 1 & $\frac{2}{N_f-4}$ \\
  $M$ & 1 & $T_S$ & $2 - \frac{10}{N_f}$ \\
 \hline
\end{tabular}
\end{center}
where the number of flavors is constrained by the conformal window
$6 \leq N_f \leq 15.$

According to this duality one should have equality of the following SCIs
\beq I_E = \frac{(p;p)^3_\infty (q;q)^3_\infty}{2^3
3!} \int_{\mathbb{T}^3}
\frac{\prod_{i=1}^{N_f} \Gamma(t_i (z_1z_2z_3)^{\pm 1})
\prod_{i=1}^{N_f} \prod_{j=1}^3 \Gamma(t_i \left(
\frac{z_j^2}{z_1z_2z_3} \right)^{\pm1})}{\prod_{1 \leq i < j
\leq 3} \Gamma(z_i^{\pm2}z_j^{\pm2}) \prod_{j=1}^3
\Gamma(z_j^{\pm2})} \prod_{j=1}^3 \frac{d z_j}{2 \pi \textup{i}
z_j},\eeq
with $|t|, |t_j|<1$, and the balancing condition $\prod_{m=1}^{N_f}t_m=(pq)^{(N_f-5)/2}$,
and (with $\prod_{j=1}^{N_f-4} y_j= 1$)
\beqa && I_M = \prod_{1
\leq i < j \leq N_f} \Gamma(t_it_j) \prod_{j=1}^{N_f}
\Gamma(t_j^2) \frac{(p;p)^{N_f-5}_\infty (q;q)^{N_f-5}_\infty}{(N_f-4)!}
\int_{\mathbb{T}^{N_f-5}} \prod_{j=1}^{N_f-5}
\frac{d y_j}{2 \pi \textup{i} y_j} \nonumber \\ && \makebox[-3em]{} \times \prod_{1 \leq i < j
\leq N_f-4}
\frac{\Gamma((pq)^{\frac{1}{N_f-4}}y_iy_j)}{\Gamma(y_iy_j^{-1},
y_i^{-1}y_j)}
\prod_{j=1}^{N_f-4} \Gamma((pq)^{\frac{1}{N_f-4}}y_j^2)
\prod_{i=1}^{N_f} \Gamma(S^{\frac{1}{N_f-4}}t_i^{-1}y_j^{-1}).
\label{preG2}\eeqa

To stress the non-trivial character of $SO(N)$-dualities with spinor matter and promote them,
we describe the duality for $\mathcal{N}=1$ SYM theory with the
$G_2$ gauge group proposed in \cite{Pouliot:1995zc}.
Pouliot's idea to derive this model consists in the
following: $G_2$ is a subgroup of $SO(7)$ and the corresponding duality
can be obtained from the $SO(7)$-group case with $N_f$ fields in the spinor
representation after giving masses to some mesons or integrating out one
of the quarks. In our language one should calculate accurately the
limit $t_{N_f} \rightarrow 1$ in the electric and magnetic SCIs.
In the magnetic SCI one has the diverging multiplier $\Gamma(t_{N_f}^2)$
in front of the integral, which is the only piece of SCI problematic
for this limit. Therefore we can plug $t_{N_f}=1$ in
other terms and see that the rank of the magnetic gauge group is not changed,
i.e. no Higgs mechanism applies from the physical point of view.

On the electric side one has a divergency coming from the poles approaching
the integration contour and it is necessary to use the residue calculus
\cite{die-spi:elliptic}.
Let us slightly deform the contour ${\mathbb T}$ for the $3$rd integration variable
and pick up the residues of the poles at $z_3 = t_{N_f}(z_1z_2)^{\pm1}$ and their
reciprocals.
Now divide both electric and magnetic SCIs by $\Gamma(t_{N_f}^2)$ and take
the limit $t_{N_f}\to1$. Then the electric index can be
rewritten as integral (13.3) of \cite{SV2} with $N_f$ replaced
by $N_f-1$ which describes SCI of the $G_2$
gauge group theory with $N_f-1$ fundamental quarks:
\begin{equation}\label{G2S} I_E=\frac{(p;p)_\infty^2
(q;q)_\infty^2}{2^23} \prod_{m=1}^{N_f-1} \Gamma(t_m)
\int_{{\mathbb T}^2} \frac{\prod_{k=1}^3 \prod_{m=1}^{N_f-1}
\Gamma(t_mz_k^{\pm1})} {\prod_{1\leq j<k\leq3}
\Gamma(z_j^{\pm1}z_k^{\pm1})} \prod_{k=1}^2\frac{dz_k}{2\pi
\textup{i} z_k},
\end{equation}
where $z_1z_2z_3=1$ and the balancing condition reads
$\prod_{m=1}^{N_f-1} t_m=(pq)^{(N_f-5)/2}.$
The magnetic index \eqref{preG2}
reduces to integral (13.4) of  \cite{SV2}:
\begin{eqnarray}\label{G2SM} && I_M =\frac{(p;p)_{\infty}^{N_f-5}
(q;q)_{\infty}^{N_f-5}}{(N_f-4)!} \prod_{1 \leq j < k
\leq N_f-1} \Gamma(t_jt_k) \prod_{j=1}^{N_f-1} \Gamma(t_j^2) \\
\nonumber && \makebox[0em]{} \times \int_{{\mathbb T}^{N_f-5}}
 \prod_{1 \leq j < k \leq N_f-4}
\frac{\Gamma((pq)^{r_s}z_jz_k)}{\Gamma(z_j^{-1}z_k,z_jz_k^{-1})}
\prod_{j=1}^{N_f-4} \Gamma((pq)^{r_s}z_j^2)
\\ \nonumber && \makebox[-2.2em]{} \times
\prod_{j=1}^{N_f-4} \Gamma((pq)^{(1-r_s)/2}z_j^{-1})
\prod_{k=1}^{N_f-1}
\Gamma((pq)^{(1-r_s)/2}t_k^{-1}z_j^{-1})\prod_{j=1}^{N_f-5}
\frac{dz_j}{2 \pi \textup{i} z_j},
\end{eqnarray}
where $\prod_{j=1}^{N_f-4}z_j=1.$
Another possibility of deriving this $G_2$-duality out of the standard
Seiberg duality for $SU(3)$-gauge group has been described in \cite{SV2}.

\subsection{$G=SO(7)$ and $F=SU(N_f) \times U(1)$}

The electric
theory is represented in the following table \cite{Cho:1997kr}
\begin{center}
\begin{tabular}{|c|c|c|c|c|}
  \hline
    & $SO(7)$ & $SU(N_f)$ & $U(1)$ & $U(1)_R$ \\  \hline
  $Q$ & $f$ & $f$ & $-1$ & $\frac{N_f-4}{N_f}$ \\
  $S$ & $s$  & 1 & $N_f$ & $0$ \\ \hline
\end{tabular}
\end{center}
while the magnetic theory is
\begin{center}
\begin{tabular}{|c|c|c|c|c|}
  \hline
    & $SU(N_f-3)$ & $SU(N_f)$ & $U(1)$ & $U(1)_R$ \\  \hline
  $q$ & $f$ & $\overline{f}$ & $\frac{2N_f-3}{N_f-3}$ & $\frac{3(N_f-4)}{N_f(N_f-3)}$ \\
  $q'$ & $f$  & 1 & $\frac{N_f}{N_f-3}$ & $\frac{N_f-4}{N_f-3}$ \\
  $w$ & $\overline{T}_S$ & 1 & $-\frac{2N_f}{N_f-3}$ & $\frac{2}{N_f-3}$
  \\
  $M$ & 1 & $T_S$ & $-2$ & $\frac{2(N_f-4)}{N_f}$ \\
  $L$ & 1 & 1 & $2N_f$ & 0 \\
  \hline
\end{tabular}
\end{center}
where $5 \leq N_f \leq 13.$ For the electric theory we have
\beqa && I_E =
\frac{(p;p)_\infty^3 (q;q)_\infty^3}{2^3 3!} \prod_{i=1}^{N_f}
\Gamma(s_i) \int_{\mathbb{T}^3} \frac{\Gamma(tz^{\pm1})
\prod_{j=1}^3 \Gamma(t (z_j^2z^{-1})^{\pm1})}{\prod_{j=1}^3
\Gamma(z_j^{\pm2}) \prod_{1 \leq i < j \leq 3}
\Gamma(z_i^{\pm2}z_j^{\pm2})} \nonumber \\ && \makebox[8em]{}
\times \prod_{i=1}^{N_f} \prod_{j=1}^3 \Gamma(s_iz_j^{\pm2})
\prod_{j=1}^{3} \frac{dz_j}{2 \pi \textup{i} z_j},
\eeqa
where $z=z_1z_2z_3$  and the balancing
condition reads $st  =(pq)^{\frac 12 (N_f-4)}$ with $s=\prod_{i=1}^{N_f} s_i$.
In the magnetic theory we have (with $\prod_{j=1}^{N_f-3}y_j=1$)
\beqa && \makebox[1em]{} I_M =
\Gamma(t^2) \prod_{i=1}^{N_f} \Gamma(s_i^2) \prod_{1 \leq i < j
\leq N_f} \Gamma(s_is_j) \frac{(p;p)_\infty^{N_f-4}
(q;q)_\infty^{N_f-4}}{(N_f-3)!} \nonumber \\ && \makebox[-3em]{} \times
\int_{\mathbb{T}^{N_f-4}} \prod_{1 \leq i < j \leq N_f-3}
\frac{\Gamma(s^{\frac{2}{(N_f-4)(N_f-3)}}t^{-\frac{2(N_f-5)}{(N_f-3)(N_f-4)}}y_i^{-1}y_j^{-1})}{\Gamma(y_iy_j^{-1},y_i^{-1}y_j)}
\prod_{i=1}^{N_f} \prod_{j=1}^{N_f-3}
\Gamma((st^2)^{\frac{1}{N_f-3}}s_i^{-1}y_j)  \nonumber \\ &&
\makebox[2em]{} \times \prod_{j=1}^{N_f-3}
\Gamma((st^2)^{\frac{1}{N_f-3}}y_j,s^{\frac{2}{(N_f-4)(N_f-3)}}t^{-\frac{2(N_f-5)}{(N_f-3)(N_f-4)}}y_i^{-2})
\prod_{j=1}^{N_f-4} \frac{dy_j}{2 \pi \textup{i} y_j}.
\eeqa

\subsection{$G=SO(7)$ and $F=SU(N_f) \times SU(2) \times U(1)$}

The electric theory is  \cite{Cho:1997kr}
\begin{center}
\begin{tabular}{|c|c|c|c|c|c|}
  \hline
    & $SO(7)$ & $SU(N_f)$ & $SU(2)$ & $U(1)$ & $U(1)_R$ \\  \hline
  $Q$ & $f$ & $f$ & 1 & $-2$ & $1-5/N_f$ \\
  $S$ & $s$  & 1 & $f$ & $N_f$ & 1 \\ \hline
\end{tabular}
\end{center}
while the magnetic theory is
\begin{center}
\begin{tabular}{|c|c|c|c|c|c|}
  \hline
    & $SU(N_f-2)$ & $SU(N_f)$ & $SO(3)$ & $U(1)$ & $U(1)_R$ \\  \hline
  $q$ & $f$ & $\overline{f}$ & 1 & 2 & $\frac{2(N_f-5)}{N_f(N_f-2)}$ \\
  $q'$ & $f$  & 1 & $f$ & 0 & $\frac{N_f-3}{N_f-2}$ \\
  $\widetilde{q}$ & $\overline{f}$  & 1 & 1 & $-2N_f$ & $-\frac{N_f-3}{N_f-2}$ \\
  $w$ & $\overline{T}_S$ & 1 & 1 & 0 & $\frac{2}{N_f-2}$
  \\
  $M$ & 1 & $T_S$ & 1 & $-4$ & $2-10/N_f$ \\
  $L$ & 1 & 1 & $f$ & $2N_f$ & 2 \\
  $N$ & 1 & $f$ & 1 & $2(N_f-1)$ & $3-5/N_f$ \\
  \hline
\end{tabular}
\end{center}
where $4 \leq N_f \leq 12.$ The electric theory SCI is
\beqa && I_E =
\frac{(p;p)_\infty^3 (q;q)_\infty^3}{2^3 3!} \prod_{i=1}^{N_f}
\Gamma(s_i) \int_{\mathbb{T}^3} \frac{\Gamma(yx^{\pm1}
z^{\pm1}) \prod_{j=1}^3 \Gamma(yx^{\pm1}
(z_j^2z^{-1})^{\pm1})}{\prod_{j=1}^3 \Gamma(z_j^{\pm2})
\prod_{1 \leq i < j \leq 3} \Gamma(z_i^{\pm2}z_j^{\pm2})}
\nonumber \\ && \makebox[8em]{} \times \prod_{i=1}^{N_f} \prod_{j=1}^3
\Gamma(s_i z_j^{\pm2}) \prod_{j=1}^{3} \frac{dz_j}{2 \pi
\textup{i} z_j},
 \eeqa
where $z=z_1z_2z_3$ and the balancing condition reads
$sy^2  = (pq)^{\frac 12 (N_f-3)}$ with $s=\prod_{i=1}^{N_f} s_i$.
In the magnetic theory we have (with $\prod_{j=1}^{N_f-2}y_j=1$)
\beqa && I_M = \Gamma(y^2,y^2x^{\pm1}) \prod_{i=1}^{N_f}
\Gamma(s_i^2,y^2s_i) \prod_{1 \leq i < j \leq N_f}
\Gamma(s_is_j) \frac{(p;p)_\infty^{N_f-3}
(q;q)_\infty^{N_f-3}}{(N_f-2)!} \nonumber \\ && \makebox[-1.5em]{} \times
\int_{\mathbb{T}^{N_f-3}} \prod_{1 \leq i < j \leq N_f-2}
\frac{\Gamma((pq)^{\frac{1}{N_f-2}}
y_i^{-1}y_j^{-1})}{\Gamma(y_iy_j^{-1},y_i^{-1}y_j)}
\prod_{i=1}^{N_f} \prod_{j=1}^{N_f-2} \Gamma((pq)^{\frac{N_f-5}{2(N_f-2)}}
s_i^{-1}y_j)
\\ && \makebox[1em]{} \times \prod_{j=1}^{N_f-2}
\Gamma((pq)^{\frac{1}{N_f-2}} y_j^{-2}, (pq)^{\frac{N_f-1}{2(N_f-2)}}
y^{-1} y_j^{-1}, (pq)^{\frac{N_f-3}{2(N_f-2)}} y_j,
(pq)^{\frac{N_f-3}{2(N_f-2)}} x^{\pm1} y_j) \prod_{j=1}^{N_f-3}
\frac{dy_j}{2 \pi \textup{i} y_j}.
\nonumber\eeqa

\subsection{$G=SO(8)$ and $F=SU(N_f) \times U(1)$}
The electric theory is \cite{Pouliot:1995sk}
\begin{center}
\begin{tabular}{|c|c|c|c|c|}
  \hline
    & $SU(N_f-4)$ & $SU(N_f)$ & $U(1)$ & $U(1)_R$ \\  \hline
  $Q$ & $\overline{f}$ & $f$ & $2N_f-4$ & $\frac{6(N_f-5)}{(N_f+1)(N_f-4)}$ \\
  $S$ & $T_S$  & 1 & $-2N_f$ & $\frac{12}{(N_f+1)(N_f-4)}$ \\ \hline
\end{tabular}
\end{center}
while the magnetic theory is
\begin{center}
\begin{tabular}{|c|c|c|c|c|}
  \hline
    & $SO(8)$ & $SU(N_f)$ & $U(1)$ & $U(1)_R$ \\  \hline
  $q$ & $f$ & $\overline{f}$ & $4-N_f$ & $\frac{N_f-5}{N_f+1}$ \\
  $p$ & $s$  & 1 & $N_f(N_f-4)$ & $\frac{N_f-5}{N_f+1}$ \\
  $M$ & 1 & $T_S$ & $2N_f-8$ & $\frac{12}{N_f+1}$ \\
  $U$ & 1 & 1 & $-2N_f(N_f-4)$ & $\frac{12}{N_f+1}$ \\
  \hline
\end{tabular}
\end{center}
where $6 \leq N_f \leq 16.$ The electric theory SCI is
\beqa && I_E =
\frac{(p;p)_\infty^{N_f-5} (q;q)_\infty^{N_f-5}}{(N_f-4)!}
\int_{\mathbb{T}^{N_f-5}} \prod_{1 \leq i < j \leq N_f-4}
\frac{\Gamma(uz_iz_j)}{\Gamma(z_iz_j^{-1},z_i^{-1}z_j)}
\nonumber \\ && \makebox[2em]{} \times \prod_{j=1}^{N_f-4}
\Gamma(uz_j^2) \prod_{i=1}^{N_f} \prod_{j=1}^{N_f-4}
\Gamma(s_iz_j^{-1}) \prod_{j=1}^{N_f-5} \frac{dz_j}{2 \pi
\textup{i} z_j},
 \eeqa
where $\prod_{j=1}^{N_f-4} z_j = 1$ and
the balancing condition reads $su^{N_f-2} = (pq)^3$
with $s=\prod_{i=1}^{N_f}s_i$. In the magnetic theory we have
(with $z=z_1z_2z_3z_4$)
\beqa &&\makebox[-2em]{}
 I_M = \Gamma(u^{n-4}) \prod_{1\leq i < j \leq N_f} \Gamma(us_is_j) \prod_{i=1}^{N_f}
\Gamma(us_i^2)
\frac{(p;p)_\infty^4 (q;q)_\infty^4}{2^3 4!} \int_{\mathbb{T}^4}
\frac{\Gamma(s^{\frac 16}u^{-\frac 13 (N_f-5)}z^{\pm1})}{\prod_{1
\leq i < j \leq 4} \Gamma(z_i^{\pm2}z_j^{\pm2})}
 \nonumber \\ &&\makebox[-1em]{} \times
\prod_{1 \leq i < j \leq 4} \Gamma(s^{\frac
16}u^{-\frac 13(N_f-5)}z_i^2z_j^2z^{-1}) \prod_{i=1}^{N_f}
\prod_{j=1}^4 \Gamma(s^{\frac 16}u^{\frac
16(N_f-5)}s_i^{-1}z_j^{\pm2})\prod_{j=1}^4 \frac{dz_j}{2 \pi \textup{i} z_j}.
\eeqa .

\subsection{$G=SO(8)$ and $F=SU(N_f) \times U(1)_1 \times U(1)_2$}

The electric theory is \cite{Cho:1997kr}
\begin{center}
\begin{tabular}{|c|c|c|c|c|c|}
  \hline
    & $SO(8)$ & $SU(N_f)$ & $U(1)_1$ & $U(1)_2$ & $U(1)_R$ \\  \hline
  $Q$ & $f$ & $f$ & $-2$ & 0 & $1-6/N_f$ \\
  $S$ & $s$  & 1 & $N_f$ & 1 & 1 \\
  $S'$ & $c$  & 1 & $N_f$ & $-1$ & 1 \\
  \hline
\end{tabular}
\end{center}
while the magnetic theory is
\begin{center}
\begin{tabular}{|c|c|c|c|c|c|}
  \hline
    & $SU(N_f-3)$ & $SU(N_f)$ & $U(1)_1$ & $U(1)_2$ & $U(1)_R$ \\  \hline
  $q$ & $f$ & $\overline{f}$ & 2 & 0 & $\frac{5N_f-18}{N_f(N_f-3)}$ \\
  $q'$ & $f$  & 1 & 0 & 2 & $\frac{N_f-4}{N_f-3}$ \\
  $q''$ & $f$  & 1 & 0 & $-2$ & $\frac{N_f-4}{N_f-3}$ \\
  $\widetilde{q}$ & $\overline{f}$  & 1 & $-2N_f$ & 0 & $-\frac{N_f-4}{N_f-3}$ \\
  $w$ & $\overline{T}_S$ & 1 & 0 & 0 & $\frac{2}{N_f-3}$
  \\
  $M$ & 1 & $T_S$ & $-4$ & 0 & $2-12/N_f$ \\
  $L_1$ & 1 & 1 & $2N_f$ & 2 & 2 \\
  $L_2$ & 1 & 1 & $2N_f$ & $-2$ & 2 \\
  $N$ & 1 & $f$ & $2(N_f-1)$ & 0 & $3-6/N_f$ \\
  \hline
\end{tabular}
\end{center}
where $5 \leq N_f \leq 15.$ The electric theory SCI is \beqa && I_E =
\frac{(p;p)_\infty^4 (q;q)_\infty^4}{2^3 4!} \int_{\mathbb{T}^4}
\frac{\Gamma(t z^{\pm1}) \prod_{1 \leq i < j \leq 4} \Gamma(t
z_i^2z_j^2z^{-1})}{\prod_{1 \leq i < j \leq 4}
\Gamma(z_i^{\pm2}z_j^{\pm2})} \nonumber \\ && \makebox[1.5em]{}
\times \prod_{j=1}^4 \Gamma(u (z_j^2z^{-1})^{\pm1})
\prod_{i=1}^{N_f} \prod_{j=1}^4 \Gamma(s_i z_j^{\pm2})
\prod_{j=1}^{4} \frac{dz_j}{2 \pi \textup{i} z_j}, \eeqa
where $z=z_1z_2z_3z_4$ and the balancing condition reads
$stu =  (pq)^{\frac 12 (N_f-4)}$ with $s=\prod_{i=1}^{N_f} s_i$.
In the magnetic theory we have (with $\prod_{j=1}^{N_f-3}y_j=1$)
\beqa && I_M =
\Gamma(u^2,t^2) \prod_{i=1}^{N_f} \Gamma(s_i^2,tus_i) \prod_{1
\leq i < j \leq N_f} \Gamma(s_is_j) \frac{(p;p)_\infty^{N_f-4}
(q;q)_\infty^{N_f-4}}{(N_f-3)!} \nonumber \\ && \makebox[-1em]{} \times
\int_{\mathbb{T}^{N_f-4}} \prod_{1 \leq i < j \leq N_f-3}
\frac{\Gamma((pq)^{\frac{1}{N_f-3}}
y_i^{-1}y_j^{-1})}{\Gamma(y_iy_j^{-1},y_i^{-1}y_j)}
\prod_{i=1}^{N_f} \prod_{j=1}^{N_f-3} \Gamma((pq)^{\frac{N_f-4}{2(N_f-3)}}
s_i^{-1}y_j) \nonumber
\\ && \makebox[-2em]{} \times \prod_{j=1}^{N_f-3}
\Gamma((pq)^{\frac{1}{N_f-3}} y_j^{-2}, (pq)^{\frac{N_f-2}{2(N_f-3)}}
(tu)^{-1} y_j^{-1}, (pq)^{\frac{N_f-4}{2(N_f-3)}} (tu^{-1})^{\pm1}
y_j) \prod_{j=1}^{N_f-4} \frac{dy_j}{2 \pi \textup{i} y_j}.
\eeqa

\subsection{$G=SO(9)$ and $F=SU(N_f) \times U(1)$}

The electric  theory is \cite{Cho:1997kr}
\begin{center}
\begin{tabular}{|c|c|c|c|c|}
  \hline
    & $SO(9)$ & $SU(N_f)$ & $U(1)$ & $U(1)_R$ \\  \hline
  $Q$ & $f$ & $f$ & $-2$ & $1-5/N_f$ \\
  $S$ & $s$  & 1 & $N_f$ & 0 \\
  \hline
\end{tabular}
\end{center}
while the magnetic theory is
\begin{center}
\begin{tabular}{|c|c|c|c|c|}
  \hline
    & $SU(N_f-4)$ & $SU(N_f)$ & $U(1)$ & $U(1)_R$ \\  \hline
  $q$ & $f$ & $\overline{f}$ & 2 & $\frac{4(N_f-5))}{N_f(N_f-4)}$ \\
  $q'$ & $f$  & 1 & 0 & $\frac{N_f-5}{N_f-4}$ \\
  $\widetilde{q}$ & $\overline{f}$  & 1 & $-2N_f$ & $\frac{N_f-3}{N_f-4}$ \\
  $w$ & $\overline{T}_S$ & 1 & 0 & $\frac{2}{N_f-4}$
  \\
  $M$ & 1 & $T_S$ & $-4$ & $2-10/N_f$ \\
  $L$ & 1 & 1 & $2N_f$ & 0 \\
  $N$ & 1 & $f$ & $2(N_f-1)$ & $1-5/N_f$ \\
  \hline
\end{tabular}
\end{center}
where $6 \leq N_f \leq 18.$
The electric theory SCI is
\beqa && I_E =
\prod_{i=1}^{N_f} \Gamma(s_i) \frac{(p;p)_\infty^4
(q;q)_\infty^4}{2^4 4!} \int_{\mathbb{T}^4} \frac{\Gamma(t
z^{\pm1}) \prod_{1 \leq i < j \leq 4} \Gamma(t
z_i^2z_j^2z^{-1})}{\prod_{1 \leq i < j \leq 4}
\Gamma(z_i^{\pm2}z_j^{\pm2})} \nonumber \\ && \makebox[3.5em]{}
\times \frac{\prod_{j=1}^4 \Gamma(t
(z_j^2z^{-1})^{\pm1})}{\prod_{j=1}^4 \Gamma(z_j^{\pm2})}
\prod_{i=1}^{N_f} \prod_{j=1}^4 \Gamma(s_i z_j^{\pm2})
\prod_{j=1}^{4} \frac{dz_j}{2 \pi \textup{i} z_j},
 \eeqa
where $z=z_1z_2z_3z_4$ and the balancing condition reads
$st^2 =(pq)^{\frac 12 (N_f-5)}$ with $s = \prod_{i=1}^{N_f} s_i$.
In the magnetic theory we have (with $\prod_{j=1}^{N_f-4}y_j=1$)
\beqa && I_M =
\Gamma(t^2) \prod_{i=1}^{N_f} \Gamma(s_i^2,t^2s_i) \prod_{1 \leq
i < j \leq N_f} \Gamma(s_is_j) \frac{(p;p)_\infty^{N_f-5}
(q;q)_\infty^{N_f-5}}{(N_f-4)!} \nonumber \\ && \makebox[-0.5em]{} \times
\int_{\mathbb{T}^{N_f-5}} \prod_{1 \leq i < j \leq N_f-4}
\frac{\Gamma((pq)^{\frac{1}{N_f-4}}
y_i^{-1}y_j^{-1})}{\Gamma(y_iy_j^{-1},y_i^{-1}y_j)}
\prod_{i=1}^{N_f} \prod_{j=1}^{N_f-4} \Gamma((pq)^{\frac{N_f-5}{2(N_f-4)}}
s_i^{-1}y_j) \nonumber
\\ && \makebox[0.5em]{} \times \prod_{j=1}^{N_f-4}
\Gamma((pq)^{\frac{1}{N_f-4}} y_j^{-2}, (pq)^{\frac{N_f-3}{2(N_f-4)}}
t^{-2} y_j^{-1}, (pq)^{\frac{N_f-5}{2(N_f-4)}} y_j)
\prod_{j=1}^{N_f-5} \frac{dy_j}{2 \pi \textup{i} y_j}.
\eeqa

\subsection{$G=SO(10)$ and $F=SU(N_f) \times U(1)$} \label{DualSO10}

The electric theory is  \cite{Pouliot:1996zh}
\begin{center}
\begin{tabular}{|c|c|c|c|c|}
  \hline
    & $SO(10)$ & $SU(N_f)$ & $U(1)$ & $U(1)_R$ \\  \hline
  $Q$ & $f$ & $f$ & $-1$ & $1-\frac{8}{N_f+2}$ \\
  $P$ & $s$  & 1  & $\frac{N_f}{2}$ & $1-\frac{8}{N_f+2}$ \\ \hline
\end{tabular}
\end{center} while the magnetic theory is \begin{center}
\begin{tabular}{|c|c|c|c|c|}
  \hline
    & $SU(N_f-5)$ & $SU(N_f)$ & $U(1)$ & $U(1)_R$ \\  \hline
  $w$ & $T_S$ & 1 & 0 & $\frac{2}{N_f-5}$ \\
  $q$ & $\overline{f}$ & $\overline{f}$  & 1 & $\frac{8}{N_f+2} - \frac{1}{N_f-5}$ \\
  $q'$ & $f$ & 1 & $-N_f$ & $-1+\frac{16}{N_f+2} + \frac{1}{N_f-5}$ \\
  $M$ & 1 & $T_S$ & $-2$ & $2-\frac{16}{N_f+2}$ \\
  $Y$ & 1 & $f$ & $N_f-1$ & $3-\frac{24}{N_f+2}$ \\
  \hline
\end{tabular}
\end{center}
where $7 \leq N_f \leq 21.$ The SCIs are
\beqa \label{w1} && I_E = \frac{(p;p)_\infty^5
(q;q)_\infty^5}{2^4 5!} \int_{{\mathbb T}^5} \frac{\Gamma(t Z)
\prod_{j=1}^5 \Gamma(t z^2_j Z^{-1}) \prod_{1 \leq i < j \leq 5}
\Gamma(t Zz^{-2}_iz_j^{-2})}{\prod_{1 \leq i < j \leq 5}
\Gamma(z_i^{\pm2}z_j^{\pm2})} \nonumber \\ && \makebox[9em]{}
\times \prod_{i=1}^{N_f} \prod_{j=1}^5 \Gamma(s_iz_j^{\pm2})
\prod_{j=1}^5 \frac{dz_j}{2 \pi \textup{i} z_j},\eeqa
where $s t^2 = (pq)^{\frac{N_f}{2} -3}$, $s=\prod_{i=1}^{N_f} s_i$,
$Z =z_1z_2z_3z_4z_5$, and (with $\prod_{j=1}^{N_f-5} y_j=1$)
\beqa \label{w2} && I_M = \prod_{1 \leq i < j \leq
N_f} \Gamma(s_is_j) \prod_{i=1}^{N_f} \Gamma(s_i^2,t^2s_i)
\frac{(p;p)_\infty^{N_f-6} (q;q)_\infty^{N_f-6}}{(N_f-5)!} \\ \nonumber &&
\makebox[-4em]{} \times \int_{{\mathbb T}^{N_f-6}} \prod_{1 \leq i < j
\leq N_f-5}
\frac{\Gamma((pq)^{\frac{1}{N_f-5}}y_iy_j)}{\Gamma(y_i^{-1}y_j,y_iy_j^{-1})}
\prod_{i=1}^{N_f-5}
\Gamma((pq)^{\frac{1}{N_f-5}}y_i^2,(pq)^{\frac{N_f-4}{2(N_f-5)}}t^{-2}y_i)
\\ \nonumber && \makebox[3em]{} \times \prod_{i=1}^{N_f} \prod_{j=1}^{N_f-5}
\Gamma((pq)^{\frac{N_f-6}{2(N_f-5)}}
s_i^{-1}y_j^{-1}) \prod_{j=1}^{N_f-6} \frac{dy_j}{2 \pi \textup{i}y_j}.
\eeqa

An interesting fact is that fixing $s_1=1$ and $t=\sqrt{pq}$
in both integrals, we come to SCIs of the original Seiberg duality
between $SO(9)$ and $SO(N_f-5)$ gauge theories with $N_f$ quarks
in the fundamental representation \cite{Seiberg}.
A connection between these dualities  was understood first from
the physical point of view in \cite{Pouliot:1996zh}, and
our observation is that SCIs are connected as well after imposing
appropriate constraints.
The residue calculus similar to that of \cite{die-spi:elliptic}
should be applied to the electric theory. In the limit $s_1 \rightarrow 1$
the integration contour is pinched by the poles coming from the term
$\prod_{j=1}^5 \Gamma(s_1 z_j^{\pm2})$.
Picking up residues of the poles at $z_j=s_1^{\pm 1/2}$ we come to SCI of
$\mathcal{N}=1$ SYM theory with $SO(9)$ gauge group and $N_f$ quarks in the fundamental
representation. In the magnetic SCI we have the multiplier
$\Gamma(t^2s_1)$ vanishing in the discussed limit and
further steps are a little tricky. For $N_f>5$ and $N_f$ odd it is convenient
first to rescale  $y_i \rightarrow (pq)^{-1/2(N_f-4)} y_i,\,
i=1,\dots,N_f-5$. Then the first residue comes from the pole
at $y_j=\sqrt{pq}$, and other residues come from the poles
$y_{2i+1}=y_{2i},\, i=1,\ldots,(N_f-5)/2$. Accurately computing all these sequential
residues one can verify that the resulting integral describes
SCI of the magnetic theory with $SO(N_f-5)$ gauge
group having $N_f$ quarks in the fundamental representation
and the gauge singlet baryon field in the $T_S$-representation
of the flavor group $SU(N_f)$.

There is another nice reduction of dual theories observed in
\cite{Pouliot:1996zh}. If we take $N_f=8$ then we
can obtain $S$-duality for $\mathcal{N}=2$ SYM theory with $SU(2)$
gauge group and $4$ hypermultiplets studied in detail in
\cite{Seiberg:1994aj}. From the mathematical point of view we need
to apply the following constraints in (\ref{w1}) and (\ref{w2})
$$
s_1s_5 = 1, \ \ \ s_2s_6 = 1, \ \ \ s_3s_7=1, \ \ \ s_4 =1
$$
and then compute the residues of poles $z_1=s_1, z_2=s_2, z_3=s_3, z_4=s_4$ (and
all their permutations) which leads to the equality of reduced SCIs
\beqa &&
I_E' = \frac{(p;p)_\infty (q;q)_\infty}{2} \int_{\mathbb{T}} \frac{\Gamma(s_8z^{\pm2},
\sqrt{pq} s_8^{-\frac 12} (s_1s_2s_3)^{\pm \frac 12} z^{\pm1})}{\Gamma(z^{\pm2})}
\\ \nonumber && \makebox[2.5em]{} \times
\prod_{i=1}^3 \Gamma(\sqrt{pq} s_8^{-\frac 12} (s_i
(s_1s_2s_3)^{-\frac 12}))^{\pm1}
z^{\pm1})\frac{dz}{2 \pi \textup{i}z}\eeqa
 and
\beqa && I_M' =
\frac{(p;p)_\infty (q;q)_\infty}{2} \int_{\mathbb{T}}
\frac{\Gamma(s_8z^{\pm2}) \Gamma^2(\sqrt{pq} s_8^{-\frac 12}
z^{\pm1})}{\Gamma(z^{\pm2})}
\prod_{i=1}^3 \Gamma(\sqrt{pq} s_8^{-\frac 12}
s_i^{\pm1} z^{\pm1}) \frac{dz}{2 \pi
\textup{i} z}.\eeqa
The equality $I_E'=I_M'$ is a particular case of the identity
obtained in \cite{bultF4} with
$$
b=s_8,\quad t_4 = \sqrt{pq} s_8^{-\frac12},
\quad t_i = \sqrt{pq} s_8^{-\frac 12} s_i,\; i=1,2,3.
$$

One can reduce also the duality considered in this section to
the dualities studied in \cite{Distler:1996ub}. If we give vacuum expectation
values to $k$ fundamental quarks in the electric theory, it breaks
the gauge group to $SO(10-k)$ while in the magnetic side the gauge group
remains the same \cite{Distler:1996ub}, see Sect. \ref{DK} for a particular example
when $k=5$. But
these dualities should be considered with a care since,
as we have shown in Sect. \ref{DK}, instead of $SU(4)$ flavor symmetry group one
has $SO(4)$ symmetry group.
From SCIs point of view we should restrict some
of the parameters to form the divergency $\propto\Gamma(1)$ in (\ref{w2}).
Appearance of such a term in the magnetic index requires the residue calculus
on the electric side.
For example, the model considered in Sect. \ref{DK} is obtained from (\ref{w1}) and (\ref{w2})
by taking in these expressions the following limits
\beq s_{N_f-4}s_{N_f-3} = s_{N_f-2}s_{N_f-1} = s_{N_f} = 1,\eeq
with the subsequent replacement $N_f \rightarrow N_f+5$ and identification $s_{N_f-2} = u_1, s_{N_f-4} = u_2$.

A more general duality was proposed in \cite{Berkooz:1997bb,Kawano:1996bd}
having on the electric side the same $SO(10)$ gauge group with $N_f$ vectors
and $N_k$ spinors. The magnetic dual side was conjectured to be a
quiver gauge theory with $SU$ and $SP$ gauge groups. Again,
as in Sect. \ref{DK}, we have not found evidence for this duality from SCIs technique
point of view, except of the obvious cases when the dual gauge group
contains only one simple component or when the theory $s$-confines,
in which cases one obtains known dualities. Anyway, we are not considering quiver gauge theories
in this work, so we leave open the detailed analysis of repairing the
general duality of \cite{Berkooz:1997bb,Kawano:1996bd}.
As in Sect. \ref{DK}, perhaps this question may be settled by a reduction
from an even more general unknown duality for $SO(N),\,  N>10,$ gauge group.

\section{Matrix models and an elliptic deformation of $2d$ CFT}\label{MM}

Main inspiration for this section comes from
paper \cite{Schiappa:2009cc}, where a $q$-deformed $2d$
CFT and the corresponding matrix model description
in terms of the Jackson integrals was proposed. From
EHIs' point of view there is a natural way
to propose a generalization of CFT to the elliptic and different $q$-deformed levels.
$q$-Extensions of the Virasoro algebra have been considered already
some time ago \cite{Awata1,FV,LP} (see also \cite{Awata2,Awata3,F}
for a recent discussion).
Here we propose expressions for the three- and four-point
correlation functions presumably associated with new
hypothetical $q$-deformations and an elliptic deformation of $2d$ CFT.
For that we employ various known generalizations of the Selberg integral,
 the basic integral appearing
in calculations of the three-point correlation function in $2d$ CFT.

\subsection{Elliptic Selberg integral}
The following elliptic generalization of the Selberg
integral attached to the root system $BC_N$ was discovered in
\cite{die-spi:elliptic,die-spi:selberg}:
 \beqa \label{ellSel} &&
\frac{(p;p)^N_{\infty}(q;q)^N_{\infty}}{2^N N!} \int_{\mathbb{T}^N}
\prod_{1 \leq i < j \leq N}
\frac{\Gamma(tz_i^{\pm1}z_j^{\pm1})}{\Gamma(z_i^{\pm 1}z_j^{\pm
1})} \prod_{j=1}^N \frac{\prod_{i=1}^6
\Gamma(t_iz_j^{\pm1})}{\Gamma(z_j^{\pm2})} \prod_{j=1}^N
\frac{dz_j}{2\pi \textup{i} z_j} \nonumber \\ && \makebox[4em]{} =
\prod_{j=1}^N \left( \frac{\Gamma(t^j)}{\Gamma(t)} \prod_{1
\leq i < k \leq 6} \Gamma(t^{j-1}t_it_k)\right), \eeqa
where $|t|, |t_j|<1$ and the balancing condition reads
$t^{2(N-1)} \prod_{i=1}^6 t_i= pq.$
This integral describes $\mathcal{N}=1$ $s$-confining SYM theory
with $SP(2N)$ gauge group, one chiral superfield in the $T_A$-representation of
$SP(2N)$, and $6$ quarks \cite{SV2}. This physical application provides a matrix
model interpretation of formula (\ref{ellSel}).
Note also that this integral describes
the normalization of a particular eigenstate of a relativistic
Calogero-Sutherland type model \cite{S4a}.

We \textit{postulate} that the chiral part of the three-point correlation function
of a hypothetical elliptic deformation of $2d$ CFT based on an
elliptic extension of the Virasoro algebra is given by integral (\ref{ellSel})
admitting exact evaluation. This proposition fits the fact that in all
known variations of $2d$ CFT the three-point function is computable
exactly. Note that in \cite{F} a simple
elliptic deformation of the free bosonic field algebra was proposed,
but its relevance to our construction is not clear, in particular,
the number and meaning of the parameters $t_j$ are not evident in this case.

\subsection{$q$-Selberg integral}
Different reductions of EHIs  were
systematically investigated in \cite{rai:limits} (see also \cite{bult1}).
First we reduce integral (\ref{ellSel}) to the trigonometric
level and then to the standard Selberg integral.
The limit $p \rightarrow 0$ is not straightforward due to the
balancing condition which we get rid of
by substituting in (\ref{ellSel}) $t_6 =pq/(t^{2(N-1)}T)$,
where $T=\prod_{i=1}^5t_i$, and obtain
\beqa  &&
\frac{(p;p)^N_{\infty}(q;q)^N_{\infty}}{2^N N!}
\int_{\mathbb{T}^N} \prod_{1 \leq i < j \leq N}
\frac{\Gamma(tz_i^{\pm1}z_j^{\pm1})}{\Gamma(z_i^{\pm 1}z_j^{\pm
1})} \prod_{j=1}^N \frac{\prod_{i=1}^5
\Gamma(t_iz_j^{\pm1})}{\Gamma(t^{2(N-1)}Tz_j^{\pm1})
\Gamma(z_j^{\pm2})} \prod_{j=1}^N \frac{dz_j}{2\pi \textup{i} z_j}
\nonumber \\ && \makebox[2em]{} = \prod_{j=1}^N \left(
\frac{\Gamma(t^j)}{\Gamma(t)} \prod_{1 \leq i < k \leq 5}
\Gamma(t^{j-1}t_it_k) \prod_{i=1}^5
\frac{1}{\Gamma(t^{N+j-2}T/t_i)} \right). \eeqa
After setting $p=0$ with fixed $t_i$ and subsequently $t_5=0$
one obtains the trigonometric $q$-Selberg integral of Gustafson \cite{Gu}
\beqa\label{qSel}
&& \frac{1}{2^N N!} \int_{\mathbb{T}^N} \prod_{1 \leq i < j \leq N}
\frac{(z_i^{\pm1}z_j^{\pm1};q)_{\infty}}{(tz_i^{\pm1}z_j^{\pm1};q)_{\infty}}
\prod_{j=1}^N \frac{(z_j^{\pm2};q)_{\infty}}{\prod_{i=1}^4
(t_iz_j^{\pm1};q)_{\infty}} \prod_{j=1}^N \frac{dz_j}{2\pi \textup{i} z_j} \nonumber \\
&& \makebox[1em]{} = \prod_{j=1}^N \left(
\frac{(t;q)_{\infty}}{(t^j;q)_{\infty}(q;q)_{\infty}}
(t^{N+j-2}t_1t_2t_3t_4;q)_{\infty} \prod_{1 \leq i < k \leq 4}
\frac{1}{(t^{j-1}t_it_k;q)_{\infty}} \right).\eeqa

Again, as above, we \textit{postulate} that the three-point correlation function
of a hypothetical $2d$ CFT based on a (yet unknown) $q$-deformed Virasoro algebra
is given by function (\ref{qSel}). Note that it is described by the standard
contour integral and not the Jackson $q$-integral, as suggested in \cite{Schiappa:2009cc}.

\subsection{Reduction to the Selberg integral}
To obtain the Selberg integral one should carefully
take the limit $q \rightarrow 1^-$. To simplify the left-hand side
of formula (\ref{qSel}) we use the relation
\beq
\stackreb{\lim}{q \rightarrow 1^{-}}
\frac{(q^az;q)_{\infty}}{(z;q)_{\infty}} \ = \ (1-z)^{-a},\eeq
and the duplication formula
$
(z^2;q)_{\infty} =  (\pm z, \pm q^{\frac 12} z;q)_{\infty}.
$
To simplify the right-hand side expression we replace infinite products by
the Jackson $q$-gamma function
\beq
\Gamma_q(x) =
\frac{(q;q)_\infty}{(q^x;q)_\infty} (1-q)^{1-x}, \qquad
\Gamma_q(x)\stackreb{=}{q\to 1}\Gamma_{rat}(x),
\eeq
where $\Gamma_{rat}(x)$ is the Euler gamma function.
Now we denote the parameters entering (\ref{qSel}) as
\beq\label{restr}
t=q^{\gamma}, \quad t_1 = q^{\alpha-\frac 12}, \quad t_2 =
-q^{\beta-\frac 12}, \quad t_3=q^{\frac 12}, \quad t_4 = -q^{\frac 12}.
\eeq
On the left-hand side of (\ref{qSel}) we change also the integration
variables $z_j = e^{i \theta_j}$ and denote $x_i =(1 + \cos \theta_i)/2$.
Finally, for fixed $\alpha, \beta,\gamma$, we can take safely the limit
$q \to 1^-$, which brings us to the standard Selberg integral \cite{aar}
\beqa \label{Sel} &&
\int_0^1 \ldots
\int_0^1 \prod_{j=1}^N x_j^{\alpha-1} (1-x_j)^{\beta-1} \prod_{1
\leq i < j \leq N} |x_i-x_j|^{2\gamma} dx_1 \ldots dx_N \nonumber \\
&& \makebox[2em]{} = \prod_{j=1}^N \frac{\Gamma_{rat}(\alpha+(j-1)\gamma)
\Gamma_{rat}(\beta+(j-1)\gamma)
\Gamma_{rat}(1+j\gamma)}{\Gamma_{rat}(\alpha+\beta+(n+j-2)\gamma)
\Gamma_{rat}(1+\gamma)},\eeqa
where the integral converges for
$\Re \alpha, \Re \beta>0, \Re \gamma
> - \min (1/N,\Re \alpha/(N-1), \Re\beta/(N-1)).$
Expression (\ref{Sel}) defines the $\beta$-deformed  matrix integral
and gives the three-point function of the
standard undeformed $2d$ CFT, see, e.g.,  Sect. 4.1 of \cite{Schiappa:2009cc}.

\subsection{A higher order elliptic Selberg integral}
A two parameter extension of the elliptic Selberg integral
(\ref{ellSel}) is given by the integral
\beqa &&  \makebox[-3em]{}
V(t_1,\ldots,t_8;t;p,q)= \frac{(p;p)^N_{\infty}(q;q)^N_{\infty}}{2^N N!}
\int_{\mathbb{T}^N} \prod_{1 \leq i < j \leq N}
\frac{\Gamma(tz_i^{\pm1}z_j^{\pm1})}{\Gamma(z_i^{\pm 1}z_j^{\pm 1})}
\prod_{j=1}^N \frac{\prod_{i=1}^8
\Gamma(t_iz_j^{\pm1})}{\Gamma(z_j^{\pm2})} \prod_{j=1}^N
\frac{dz_j}{2\pi \textup{i} z_j}, \label{V}\eeqa
where the balancing condition reads
$t^{2(N-1)} \prod_{i=1}^8 t_i = (pq)^2.$ The symmetry transformation properties
of this integral were found in \cite{S2} for $N=1$ and in \cite{Rains}
for general $N$. We are not presenting them here explicitly for
brevity (for $N=1$ they are described by formula
\eqref{Sp}). We conjecture that integral \eqref{V} coincides with the four-point
correlation function for an elliptic deformation of $2d$ CFT
for which the elliptic Selberg integral defines the three-point function.
Then the $s$-$t$-channels duality for this four-point function
is described by known symmetries of \eqref{V}.

Again, taking appropriately the (trigonometric)
limit $p\to0$ we can come to the
two parameter extension of the $q$-Selberg integral with further
degeneration to the rational level \cite{rai:limits}. For arbitrary $N$ and
a special choice of one of the parameters, there emerges the $_2F_1$-hypergeometric
function describing the chiral part of the four point correlation function
(see formula (4.9) in \cite{Schiappa:2009cc}). General $_2F_1$-function is obtained
also for $N=1$, we skip explicit description of these well known results.
In \cite{LP}, the four point correlation function of a $q$-deformed CFT was
connected to a $q$-analog of the $_2F_1$-hypergeometric function. We conjecture
that an appropriate elliptic analog of the latter correlation function will
be expressed in terms of the $V$-function of \cite{S3} given by $N=1$ case of \eqref{V}.
Apart from the mentioned limit $p\to 0$, there exists a different
degenerating limit for the elliptic Selberg integral to the hyperbolic
$q$-hypergeometric level \cite{ds:unit}, which was discussed recently
in detail in \cite{DSV} where one of the resulting integrals was
interpreted as the partition function of a particular $3d$ $\mathcal{N}=2$
supersymmetric field theory model (it is also expected to play a
proper role in $2d$ CFT deformations).

\section{Connection to the knot theory}\label{Knot}

In this section we discuss the connection of partition functions for some
$3d$ supersymmetric field theories and non-supersymmetric CS theories with
the complexified gauge groups to topological invariants of the knot theory
\cite{Dimofte:2011gm,Dimofte:2011jd,Dimofte:2009yn,Hikami1,Hikami2}.
In \cite{DSV}, the theory of hyperbolic $q$-hypergeometric integrals
has been exploited for checking and searching for $3d$ supersymmetric
dualities. Earlier it was proposed in \cite{Hikami1} that the state integrals
for knots are also defined in terms of such integrals.
In an independent approach to state integrals \cite{Dimofte:2011gm}, Dimofte
proposed a new expression for the figure-eight knot state integral
and conjectured that it coincides with the one of \cite{Hikami1}.
Using the approach of \cite{DSV} we prove here this conjecture, as well as
some other similar identities needed in \cite{DGG}.

The hyperbolic $q$-hypergeometric integrals can be rigorously obtained
as reductions of the EHIs  \cite{rai:limits}
(for an earlier formal approach see, e.g., \cite{ds:unit},
and for a detailed explicit analysis of reducing many
integrals see \cite{BultPhD}).
The reduction procedure inherits certain pieces of the {\em unique}
symmetry properties of the original integrals and yields {\em many}
nontrivial identities at the hyperbolic level.
The resulting hyperbolic integrals and identities emerge in various
physical problems. Here we stress that they describe partition functions for
$3d$ supersymmetric theories living on the squashed
three-sphere and the state integrals for the knots.
As the most recent example of their relevance, we mention a generalization
of the AGT duality \cite{Alday:2009aq} to the duality inspired by the
$(3+3)$-dimensional theories \cite{Dimofte:2011jd,Drukker:2010jp,Hosomichi:2010vh},
with the non-supersymmetric CS theory living on a  $3d$ manifold $\mathcal{M}$
on the one side and $3d$ $\mathcal{N}=2$ supersymmetric
theory living on the squashed sphere on the other side.

\subsection{The figure-eight knot}
We start from the notation for hyperbolic gamma function used in \cite{DSV,S5}.
This function appeared in \cite{fad:mod} under the name
``noncompact quantum dilogarithm". For $q=e^{2\pi \textup{i} \omega_1/\omega_2}$
and $\tilde q =e^{-2\pi \textup{i} \omega_2/\omega_1}$ with $|q|<1$ we define
$$
\gamma(u;\omega_1,\omega_2)
= \frac{(e^{2\pi \textup{i} u/\omega_1}\tilde q; \tilde q)_\infty}
{(e^{2\pi \textup{i}  u/\omega_2}; q)_\infty},
\quad
\gamma^{(2)}(u;\omega_1,\omega_2) \
= \ e^{-\pi \textup{i} B_{2,2}(u)/2} \gamma(u;\omega_1,\omega_2),
$$
where $B_{2,2}(u;\omega_1,\omega_2)$ is the
second order Bernoulli polynomial,
$$
B_{2,2}(u;\omega_1,\omega_2) =
\frac{u^2}{\omega_1\omega_2} - \frac{u}{\omega_1} -
\frac{u}{\omega_2} + \frac{\omega_1}{6\omega_2} +
\frac{\omega_2}{6\omega_1} + \frac 12.
$$
For $\text{Re}(\omega_1), \text{Re}(\omega_2)>0$ and
$0<\text{Re}(u)<\text{Re}(\omega_1+\omega_2)$
one has the following integral representation
for the hyperbolic gamma function
$$
\gamma^{(2)}(u;\omega_1,\omega_2)=\exp\left(-\text{PV}\int_{\mathbb{R}}
\frac{e^{ux}}
{(1-e^{\omega_1 x})(1-e^{\omega_2 x})}\frac{dx}{x}\right),
$$
where `PV' means the principal value integral.

Different notations and names for slight modifications of this function are used
in the literature, most of them were explicitly described in Appendix A of \cite{S5}.
In \cite{Dimofte:2009yn}, the following ``quantum dilogarithm" is employed
\beq \Phi(z;\tau) = \frac{(-e(z+\tau/2);e(\tau))_\infty}{(-e((z-1/2)/\tau);e(-1/\tau))_\infty},\eeq
where $e(x) = e^{2 \pi \textup{i} x}$. One can easily find by comparison that
\beq \label{change}
\Phi(z;\tau) = \gamma\left(\frac{\omega_1+\omega_2}{2} + z\omega_2;\omega_1,\omega_2
\right)^{-1},\qquad \tau=\frac{\omega_1}{\omega_2}.
\eeq

Consider the so-called state integral for the figure eight knot
$\textbf{4}_{\textbf{1}}$ which was found first
by Hikami in \cite{Hikami1} and studied further in
\cite{Hikami2,Dimofte:2009yn,Dimofte:2011gm,Dimofte:2011jd}.
We stick to the notation of paper \cite{Dimofte:2009yn} where this
integral is given by formula (4.46) and has the form
\beq \label{initIntKnot}
I =\frac{e^{2 \pi \textup{i} u/\hbar  + u}}{\sqrt{2\pi \hbar }} \int_{-\infty}^{\infty}
\frac{\Phi((p-u)/2 \pi \textup{i};\hbar /\pi \textup{i})}
{\Phi(-(p+u)/2 \pi \textup{i};\hbar /\pi \textup{i})} e^{-2 p u/\hbar } dp.
\eeq
This integral describes also the partition function of non-supersymmetric
CS theory with the complexified gauge group $SL(2,\mathbb{C})$
living on the $3d$ manifold
$\mathcal{M} = S^3 \backslash \textbf{4}_{\textbf{1}}$
\cite{Dimofte:2009yn}.
Denoting $\omega_1=b,\omega_2=b^{-1},\tau=b^2$ and changing the variables
$
p \rightarrow 2 \pi \textup{i} p,\,
u \rightarrow 2 \pi \textup{i} u, \,
\hbar  \rightarrow \pi \textup{i} \tau
$
in (\ref{initIntKnot}), we obtain
\beq I =e^{2 \pi \textup{i} (2+b^2) u/b^2}
\int_{- \textup{i} \infty}^{\textup{i} \infty} \frac{\Phi(p-u;b^2)}
{\Phi(-p-u;b^2)} e^{-8 \pi \textup{i} p u/b^2} dp,
\eeq
where we drop the multiplier $\sqrt{2\pi}/\textup{i}\sqrt{\hbar }$ in front of the integral.
Using relation (\ref{change}), we can write
\beq I = e^{2 \pi \textup{i} (2+b^2) u/b^2}
\int_{- \textup{i} \infty}^{\textup{i} \infty}
\frac{\gamma\left(\frac{b+1/b}{2} - \frac{p+u}{b};b,b^{-1}\right)}
{\gamma\left(\frac{b+1/b}{2}
+ \frac{p-u}{b};b,b^{-1}\right)} e^{-8 \pi \textup{i} p u/b^2} dp.
\eeq
We apply the inversion formula
$\gamma(u,b+1/b-u;b,b^{-1}) \ = \ e^{\pi \textup{i} B_{2,2}(u;b,b^{-1})}$
to move the denominator $\gamma$-function to the numerator
and pass from the $\gamma$-function to the $\gamma^{(2)}$-function.
This yields another form of the integral:
\beq \label{fin} I = e^{2 \pi \textup{i} (2+b^2) u/b^2}
\int_{- \textup{i} \infty}^{\textup{i} \infty}
\gamma^{(2)}\left(\frac{b+1/b}{2} - \frac{p \pm u}{b};b,b^{-1}\right)
e^{-6 \pi \textup{i} p u/b^2} dp.\eeq

Consider now integral (6.77) from \cite{Dimofte:2011gm} (in the suggested there
normalization without the multiplier $2^{-1/2} e^{(4 \pi^2-\hbar ^2)/24\hbar ^2}$).
After changing the notation in it similar to the integral $I$,
we come to the following expression
\beq \label{fin1} \widetilde{I} = e^{-2 \pi \textup{i} u}
\int_{- \textup{i} \infty}^{\textup{i} \infty}
\gamma^{(2)}\left(\frac{b+1/b}{2} - \frac{p \pm u}{b};b,b^{-1}\right)
e^{6 \pi \textup{i} p u/b^2} dp.\eeq
One can see that the difference between expressions (\ref{fin}) and (\ref{fin1})
is in the coefficients in front of the integrals and in the sign of the exponent
of the integrand.

Let us take now the $n=1$ case of the integral $II^1_{n,(3,3)*a}(\mu;-;\lambda;\tau)$
defined on page 218 of \cite{BultPhD}.
Replacing the integration variable $x \rightarrow p/b$ in it and changing slightly
its normalizing multiplier, we come to the integral
\beq \label{finB} Z_{E}(\mu_1,\mu_2,\sigma)
= \int_{-\textup{i} \infty}^{\textup{i} \infty}
\prod_{i=1}^2 \gamma^{(2)}(\mu_i - p/b;b,b^{-1}) e^{\pi \textup{i} \sigma p/b} dp,
\eeq
where $\mu_1,\mu_2,$ and $\sigma$ are some free parameters.

Remarkably, our original integral of interest $I$ (\ref{fin})
is a special subcase of \eqref{finB}, which is obtained after
imposing the constraints
\beq \label{subst} \mu_1 = (b+1/b)/2 - u/b,
 \ \ \mu_2 = (b+1/b)/2 + u/b, \ \ \sigma = -6u/b.
\eeq

Using the results of \cite{DSV}, we see that
expression (\ref{fin1}) with arbitrary $\mu_1, \mu_2, \sigma$ describes
the partition function (that is why it is denoted as $Z_E$)
of $3d$ $\mathcal{N}=2$ theory
living on the squashed three-sphere with the $U(1)$ gauge group and
two quarks, which is referred to as the ``electric theory".
The global symmetry group is $SU(2) \times U(1)_A \times U(1)_R.$
We do not discuss the vector superfield having well known properties.
The matter content with the corresponding charges is presented in
the following table
\begin{center}
\begin{tabular}{|c|c|c|c|c|}
  \hline
   & $U(1)$ & $SU(2)$ & $U(1)_A$ & $U(1)_R$ \\
\hline
  $q$ & $-1$ & $f$ & $1$ & $\frac 12$ \\
 \hline
\end{tabular}
\end{center}

Integral (\ref{finB})
has the transformation formula described in Theorem 5.6.20 of \cite{BultPhD}:
\beqa \label{Th}
e^{\pi \textup{i} (4 \mu_1 \mu_2 - \mu_3^2 + (b+1/b)\mu_3
- (b+1/b)^2/4)/2 +\pi \textup{i} (b^2+1/b^2)/24}Z_E(\mu_1,\mu_2,2\mu_3-\mu_1-\mu_2)
= Z_M(\underline{\mu};\lambda),
\eeqa
where
\beq \label{ZM8} Z_M(\mu_1,\mu_2,\mu_3;\lambda) = \int_{-\textup{i} \infty}^{\textup{i} \infty}
\prod_{i=1}^3 \gamma^{(2)}(\mu_i - p/b;b,b^{-1}) e^{\pi \textup{i} \lambda p/b
 - 3 \pi \textup{i} p^2/2b^2} dp,\eeq
with $\mu_3$ being a new parameter introduced through the balancing condition
$
\mu_1+\mu_2+ \mu_3 = \lambda - (b+1/b)/2.
$
This condition relates fugacities associated
with the $SU(3)$ flavor group acting on quarks and the Fayet-Illiopoulos
term $\lambda$.

Expression (\ref{ZM8}) represents the partition function of a ``magnetic theory"
defined as the $3d$ $\mathcal{N}=2$ CS theory with $U(1)_{3/2}$ gauge group
and $3$ quarks.
The global symmetry group of the magnetic theory is $SU(3) \times U(1)_A \times U(1)_R.$
Note that the flavor groups of the electric and magnetic theories
differ although the number of independent variables
is the same for both statistical sums. The matter fields together with the
corresponding charges are presented in the table below
\begin{center}
\begin{tabular}{|c|c|c|c|c|}
  \hline
   & $U(1)_{3/2}$ & $SU(3)$ & $U(1)_A$ & $U(1)_R$ \\
\hline
  $q$ & $-1$ & $f$ & $1$ & $\frac12$ \\
 \hline
\end{tabular}
\end{center}
The duality between these two $3d$ theories is one of very many dualities
not considered in \cite{DSV} due to their abundance.

Now we can easily prove the equality of two forms of the figure-eight knot state integrals
\eqref{fin} and \eqref{fin1}, $I = \widetilde{I}$.
Evidently, expression (\ref{ZM8}) is symmetric in
parameters $\mu_1,\mu_2$, and $\mu_3$. If we substitute in the left-hand side of
\eqref{Th} restrictions \eqref{subst}, we obtain the integral $I$ up to
some factor. Now we permute the parameters in the left-hand side
$(\mu_1,\mu_2,\mu_3) \rightarrow (\mu_3,\mu_1,\mu_2)$
(which is permitted because of the identity) and substitute anew
the same restrictions \eqref{subst}. As a result we obtain the integral $\tilde I$
up to the same multiplier as before. Equating both expressions, we prove that $I=\tilde I$.

Moreover, we can use further this permutational symmetry
and replace in the left-hand side of \eqref{Th} $(\mu_1,\mu_2,\mu_3) \rightarrow
(\mu_2,\mu_3,\mu_1),$ and impose constraints \eqref{subst}.
As a result we come to one more form of the figure-eight knot state integral
\beq
I=\tilde I=\widehat{I} := e^{2\pi \textup{i}u(1-6u)/b^2}
\int_{- \textup{i} \infty}^{\textup{i} \infty}
\gamma^{(2)}\left(\frac{b+1/b}{2} - \frac{3u+p}{b}, \frac{b+1/b}{2}
+ \frac{u-p}{b};b,b^{-1}\right) dp,
\eeq
which was not considered in \cite{Dimofte:2011gm,Dimofte:2009yn,Hikami1,Hikami2}.

As observed in \cite{Dimofte:2011jd,Drukker:2010jp},
there is an extension of the AGT duality \cite{Alday:2009aq} to
the situation when the 6-dimensional space-time is descomposed as a $(3+3)d$ manifold
with the duality relation between the complexified CS theories
living on some $3d$ manifold $\mathcal{M}$ and $3d$
supersymmetric field theories. Our equality of partition functions
(\ref{Th}) gives an explicit example of such a duality.
In it the CS theory with $SL(2,\mathbb{C})$ gauge group
on $\mathcal{M}=S^3 \backslash \textbf{4}_\textbf{1}$ is dual
to the $3d$ theory with $U(1)$ gauge group and two flavors, which is
also dual to the $3d$ CS theory with $U(1)_{3/2}$ gauge group and three
flavors, as described  above.

Now we are coming to the main point of this section, namely,
to derivation of the identities presented above from the theory
of EHIs.
Identity (\ref{Th}) arises from the reduction of a
transformation formula of \cite{S2} for the elliptic extension of
 Euler-Gauss hypergeometric function \eqref{egauss}.
From the physical point of view EHIs  describe SCIs for $4d$
supersymmetric field theories and, analogously to \cite{Gaiotto:2011nm}, we
can claim that important ingredients of the knot theory are coming from the $4d$
space-time. In the considered example, the state integral model
for the figure-eight knot is obtained from
$4d$ $\mathcal{N}=1$ SYM theory with $SP(2)$ gauge group and 8 quarks,
which was studied in detail in \cite{SV1}.

The $V$-function obeys symmetry transformation \eqref{Sp}.
First, we reduce it to the level of hyperbolic $q$-hypergeometric integrals
by means of the reparametrization of variables
\begin{equation}
y = e^{2 \pi \textup{i} r z}, \qquad t_j = e^{2 \pi \textup{i} r \mu_j},
\quad j=1,\ldots,8, \qquad
p = e^{2 \pi \textup{i} b r}, \qquad q = e^{2 \pi \textup{i} r/b},
\label{limit_r}\end{equation}
(here the base parameter $p$ should not be mixed up with the integration
variable $p$ in \eqref{initIntKnot})
and the subsequent limit $r \rightarrow 0$.
In this limit the elliptic gamma function has the asymptotics \cite{Rui}
\beq \Gamma(e^{2 \pi \textup{i} r z};e^{2 \pi \textup{i} r b},
e^{2 \pi \textup{i} r/b}) \stackreb{=}{r \rightarrow 0}
e^{-\pi \textup{i}(2z-b-1/b)/12r} \gamma^{(2)}(z;b,b^{-1}).\eeq
Using it in the reduction, one obtains an integral lying on the top
of a list of integrals emerging as degenerations of the $V$-function
(we omit some simple diverging exponential multiplier appearing in this
limit),
\beq \label{top_int}
I_{h}(\mu_1,\ldots,\mu_8) = \int_{- \textup{i} \infty}^{\textup{i} \infty}
\frac{\prod_{i=1}^8 \gamma^{(2)}(\mu_i \pm z;b,b^{-1})}
{\gamma^{(2)}(\pm 2z;b,b^{-1})} dz,
\eeq
with the balancing condition $\sum_{i=1}^8 \mu_i = 2 (b+1/b)$.
It has the following symmetry transformation formula descending from the elliptic one
\beq \label{top}
I_{h}(\mu_1,\ldots,\mu_8) = \prod_{1 \leq i < j \leq 4}
\gamma^{(2)} (\mu_i + \mu_j,\mu_{i+4} + \mu_{j+4};b,b^{-1})
I_{h}(\nu_1,\ldots,\nu_8),
\eeq
where
$
\nu_i = \mu_i + \xi, \,\nu_{i+4} = \mu_{i+4} - \xi,  i=1,2,3,4,
$
and the parameter $\xi$ is
$$2\xi \ = \ \sum_{i=5}^8 \mu_i - b-1/b= b+1/b - \sum_{i=1}^4 \mu_i.$$

To get the desired transformation formula (\ref{Th}) one should use
the following asymptotic formulas when some of the parameters go to infinity
\beqa
\lim_{u \rightarrow \infty}e^{\frac{\pi \textup{i}}{2} B_{2,2}(u)} \gamma^{(2)}(u)
& = & 1, \text{ \ \ for } \text{arg }b < \text{arg } u <
\text{arg }1/b + \pi, \nonumber \\
\lim_{u \rightarrow \infty}e^{-\frac{\pi \textup{i}}{2} B_{2,2}(u)} \gamma^{(2)}(u)
& = & 1, \text{  \ \ for } \text{arg }b - \pi < \text{arg } u <
\text{arg }1/b.
\eeqa
The proof of formula (\ref{Th}) by van de Bult presented in \cite{BultPhD} is rather bulky.
Starting from the key transformation formula (\ref{top}) one has
to pass step by step from one level of complexity to another one in the list
of integrals obtained from $I_h$ by diminishing the number of
independent parameters. Therefore we are not presenting
it here explicitly although it is very straightforward.

\subsection{The trefoil knot}
Let us apply the same procedure to the state integral model of the
trefoil knot described by formula (6.59) in \cite{Dimofte:2011gm}
(where we omit a coefficient in front of the integral):
\beq \label{trefoil}
J = \int_{-\infty}^{\infty} \Phi\left(-\frac{p}{2\pi \textup{i}};
\frac{\hbar }{\pi\textup{i}}\right)
\Phi\left(\frac{p-c}{2\pi \textup{i}};\frac{\hbar }{\pi\textup{i}}\right)
e^{pu/2\hbar } dp.
\eeq
After rewriting this expression as in the figure-eight knot case
(replacing
$p \rightarrow 2 \pi \textup{i} p, c \rightarrow 2 \pi \textup{i} c,
 \hbar  \rightarrow \pi \textup{i} \tau$, etc),
we come to the integral
\begin{equation}
J = e^{\pi \textup{i} (1 + b^4 - 6 c^2)/12b^2}
\int_{-\textup{i}\infty}^{\textup{i}\infty}
\gamma^{(2)}\left(\frac{b+1/b}{2} + \frac{p}{b},
\frac{b+1/b}{2} - \frac{p-c}{b};b,b^{-1}\right) e^{ \pi \textup{i} p
(3c-p)/b^2} dp.
\end{equation}
Consider now the integral $II^1_{1,(3,3)a}(\mu,\nu;\lambda)$
on page 218 in \cite{BultPhD}. We choose the integration variable
in it $z=p/b$, impose the constraints
$\mu=(b+1/b)/2, \nu=(b+1/b)/2+c/b, \lambda=3c/b$, and denote the resulting
function as $\widetilde{Z}_E (\mu,\nu,\lambda)$:
\beq \label{TrE}
\widetilde{Z}_E (\mu,\nu,\lambda) = \int_{-\textup{i} \infty}^{\textup{i} \infty}
\gamma^{(2)}(\mu-z,\nu+z;b,b^{-1}) e^{\pi \textup{i} \lambda z -
\pi \textup{i} z^2} dz,
\eeq
which describes the partition function of a $3d$ $\mathcal{N}=2$ SYM theory with $U(1)$
gauge group and two quarks.  According to
Theorem 5.6.19 of \cite{BultPhD}, it obeys the following transformation formula:
\beq \label{TrefTh}
\widetilde{Z}_E (\mu,\nu,\lambda) = \widetilde{Z}_M
(\mu+\sigma',\nu-\sigma') e^{\pi \textup{i}
(\lambda^2 + (\mu+\nu)^2 - 2(b+1/b)(\mu+\nu))/4},
\eeq
where $4\sigma' = \nu - \mu - \lambda$ and
\beq
\widetilde{Z}_M (\alpha,\beta) =
\frac 12 \int_{-\textup{i} \infty}^{\textup{i} \infty}
\frac{\gamma^{(2)}(\alpha \pm y,\beta \pm y;b,b^{-1})}
{\gamma^{(2)}(\pm 2y;b,b^{-1})} e^{-4 \pi \textup{i} y^2} dy,
\eeq
which is the partition function of a $3d$ $\mathcal{N}=2$ CS theory
with $SU(2)_{1/2}$ gauge group and two quarks.
Comparing with \cite{Dimofte:2011gm}, we see that integral (\ref{TrE}) coincides
with the product wavefunction in
the transformed basis. To get the state integral model for the trefoil knot one has to
specify $\mu + \nu= b + 1/b.$
Then expression (\ref{trefoil}) simplifies (set $c=0$ in it and apply the inversion formula)
becoming a Gaussian integral which is easily evaluated.
Again, one can use equality (\ref{TrefTh}) for the connection
of $3d$ complexified CS theory living on
$\widetilde{\mathcal{M}} = S^3 \backslash \textbf{3}_\textbf{1}$
with $3d$ supersymmetric field theories.

\subsection{Some other integrals}
In the rest of this section we would like to consider some other hyperbolic
integrals which appear in this context \cite{BT,Bytsko:2006ut,Faddeev:2000if}
and describe their connection to EHIs.
There is nice Fourier transformation formula for the hyperbolic gamma
function \cite{BT,Faddeev:2000if} (in particular, in \cite{Dimofte:2011gm}
it is given by formula (6.54)). Let us define
\beq \label{FTr}
J_E = \int_{- \textup{i} \infty}^{\textup{i} \infty}
\gamma^{(2)}(\mu-z/b;b,b^{-1}) e^{\pi \textup{i}(2 \lambda z/b - z^2/b^2)/2} dz.
\eeq
To match the definition of \cite{Dimofte:2011gm} one should fix the parameters as
$\mu=(b+1/b)/2, \, \lambda=2x.$
Expression (\ref{FTr}) can be found in \cite{BultPhD}, where it is defined as
integral $II^0_{1,(3,2)a}(\mu;\lambda)$.
This integral is computable exactly, as described in Theorem 5.6.8 of
\cite{BultPhD},
\beqa
&& J_E=J_M:= \gamma^{(2)}((b+1/b)/4+\lambda/2-\mu/2;b,b^{-1}) \nonumber \\
&& \makebox[3em]{} \times
e^{\pi \textup{i} (-3\mu^2 + (\lambda-(b+1/b)/2)^2
+ 2 \mu (3\lambda+(b+1/b)/2))/4 - \pi \textup{i} (b^2+1/b^2)/24}.
\eeqa
To see the coincidence with formula (6.54) from \cite{Dimofte:2011gm}
one should take into account the inversion formula for the hyperbolic gamma functions.
Physically, the equality $J_E=J_M$ is obtained from the reduction of SCIs for
$4d$ $\mathcal{N}=1$ SYM theory with $SU(2)$ gauge group and 6 quarks
and its dual, and, mathematically, it emerges as a reduction of
 the elliptic beta integral \cite{S1}.

The equality $J_E=J_M$ defines one of the simplest examples of dualities between two $3d$
supersymmetric field theories. The electric theory is a $3d$ $\mathcal{N}=2$
CS theory with $U(1)_{1/2}$ gauge group and
one quark $Q$, while the magnetic theory is just a free $3d$ $\mathcal{N}=2$ theory of one
chiral field $X$. Again, such dualities were skipped in \cite{DSV}
because of their abundance, where for  brevity only the first steps of the reduction
procedure from $4d$ SCIs to $3d$ partition functions were considered explicitly.
The identities presented in this section lie further in the
reduction hierarchy of EHIs  to the hyperbolic level.

The equality of partition functions considered in \cite{Hosomichi:2010vh}
(later also discussed in \cite{Dimofte:2011jd,Terashima:2011qi})
is obtained as a reduction of the $V$-function identities as well
\cite{BultPhD}. The equality of statistical sums
of the initial theory and the mirror dual is taken from
\cite{Bytsko:2006ut}, where it was proven using the Fourier transformation
formula \cite{Faddeev:2000if}. The partition function  of the $3d$
mass-deformed $T[SU(2)]$ SYM theory coincides with the integral
$II^1_{1,(2,2)}(\mu_1,\mu_2,\nu_1,\nu_2;\lambda)$ from \cite{BultPhD}
(again we take $\omega_1=b, \omega_2=1/b$):
\beq \label{PT}
K(\mu_1,\mu_2,\nu_1,\nu_2,\lambda) = \int_{-\textup{i} \infty}^{\textup{i} \infty}
\prod_{i=1}^2 \gamma^{(2)}(\mu_i-z,\nu_i+z;b,b^{-1}) e^{\pi \textup{i} \lambda z} dz,
\eeq
where one should restrict the parameters to obtain the
expression from \cite{Hosomichi:2010vh} as follows
$$\mu_1=\nu_1=\frac{b+1/b}{4}-\frac{m}{2}+\mu, \ \ \
\mu_2=\nu_2=\frac{b+1/b}{4}-\frac{m}{2}-\mu, \ \ \
\lambda=-4 \xi.$$

Integral (\ref{PT}) has the transformation formula described in
Theorem 5.6.17  of \cite{BultPhD}:
\beqa  \label{Theorem1}
&& K(\mu_1,\mu_2,\nu_1,\nu_2,\lambda) = \widetilde{K}(\sigma_1, \ldots, \sigma_4)
e^{\pi \textup{i} (4 \widetilde{\sigma}^2 - 2 \mu_1 \mu_2
- 2 \nu_1 \nu_2)/2}
\nonumber \\ && \makebox[2em]{} \times
\gamma^{(2)}((\pm \lambda - \mu_1 - \mu_2 - \nu_1 - \nu_2)/2+b+1/b;b,b^{-1}),
\eeqa
where
\beqa \label{Ksigma} &&
\widetilde{K}(\sigma_1, \ldots, \sigma_4) = \frac 12
\int_{-\textup{i} \infty}^{\textup{i} \infty}
\frac{ \prod_{i=1}^4 \gamma^{(2)}(\sigma_i \pm y;b,b^{-1})}
{\gamma^{(2)}(\pm 2 y;b,b^{-1})} e^{-2 \pi \textup{i} y^2} dy,
\\ && \nonumber
\sigma_{1,2} = \mu_{1,2} + \widetilde{\sigma},\quad
\sigma_{3,4} = \nu_{1,2} - \widetilde{\sigma},\quad
4 \widetilde{\sigma} = \nu_1 + \nu_2 - \mu_1 - \mu_2 - \lambda.
\eeqa
There is a transformation formula for the integral $\widetilde{K}$
described in Theorem 5.6.14 (for $n=1$) in \cite{BultPhD}:
\beq \label{Theorem2}
\widetilde{K}(\sigma_1, \ldots, \sigma_4) =
\widetilde{K}(\rho_1, \ldots, \rho_4) \prod_{1 \leq i < j \leq 4}
\gamma^{(2)}(\sigma_i + \sigma_j;b,b^{-1}) e^{- \pi \textup{i} (b+1/b) \xi},
\eeq
where
$2\xi = b+1/b - \sum_{i=1}^4 \sigma_i, \ \rho_i = \sigma_i + \xi, i=1,2,3,4.$

Combining together formula (\ref{Theorem1}), symmetry transformation (\ref{Theorem2})
and, finally, again (\ref{Theorem1}) (taking into account that (\ref{Ksigma})
is symmetric in all the parameters $\sigma_i$), one gets the symmetry transformation
\beqa \label{SymForm}
&& K((b+1/b)/4-m/2 \pm \mu,(b+1/b)/4-m/2 \pm \mu,-4\xi) \gamma^{(2)}(-m;b,b^{-1})
\nonumber \\ && \makebox[4em]{}
= K(m/2 \pm \xi,m/2 \pm \xi,-4\mu) \gamma^{(2)}(m;b,b^{-1}).
\eeqa
Generalizing to arbitrary parameters $\mu_1, \mu_2,\nu_1, \nu_2$
one obtains formula (A.31) from \cite{Bytsko:2006ut}.
Described symmetry transformation formulas allow one to derive more identities
apart from (\ref{SymForm}), which should be explored separately.
Here our aim was to show that all known
examples of the equalities of partition functions from the literature are obtained
as reductions of the identities for EHIs
(actually, here we have discussed only the reduction of the elliptic beta
integral and the $V$-function).
There is also an interesting connection of the partition function of
mass-deformed $T[SU(2)]$ theory with the Liouville theory \cite{Hosomichi:2010vh},
where it coincides with the $S$-duality kernel
connecting conformal blocks \cite{Teschner:2003at}.
Note also that it can be derived from SCI of $4d$ $\mathcal{N}=2$ SYM theory
with $SU(2)$ gauge group and 4 hypermultiplets \cite{GY}.

We conclude this section by stating that the arguments given
above are quite general and applied to any state integral model.
Other examples for different knots presented in \cite{Hikami2} are obtained
from the reduction of SCIs of $4d$ $\mathcal{N}=1$ quiver supersymmetric
field theories and coincide with the partition functions of $3d$ $\mathcal{N}=2$
theories in which one restricts fugacities
associated with the matter content of the theory. The results of this
section may be useful for a better understanding of a generalization
of the AGT duality \cite{Alday:2009aq}, connecting $4d$ and $2d$ theories,
to the duality connecting $3d$ CS and $3d$ $\mathcal{N}=2$ supersymmetric field theories
\cite{DGG,Dimofte:2011jd,Drukker:2010jp,Hosomichi:2010vh}.

\section{Reduction to the $2d$ vortex partition function}\label{Vortex}

Dimensional reductions of field theories are usually considered
directly at the level of physical degrees of freedom. As discussed in
the previous section, often it is easier to make such reductions
at the level of collective objects such as
partition or correlation functions and topological indices.
In particular, partition functions of the field theories
on the squashed three-sphere $S_b^3$ can be derived from $4d$
SCIs \cite{DSV} (the case of ordinary  $S^3$ corresponds to the
limit $\omega_1=\omega_2^{-1}\to 1$).
An obvious question is whether one can proceed further
and reduce $3d$ partition functions
to $2d$ statistical sums? The squashed three-sphere
is isomorphic to $S^2 \times S^1$ and by shrinking
the radius of $S^1$ to zero one reduces this manifold
to $S^2$, which is a two-dimensional space-time.
One obtains in this way the vortex partition
function for a $2d$ supersymmetric sigma-model. This partition
function is the object of recent active studies
\cite{B1,Dimofte:2010tz,Gerasimov:2010td,Shadchin:2006yz,Yoshida:2011au}.
Its relation to the $3d$ Omega background is discussed in \cite{Dimofte:2011jd}.
From the mathematical point of view the $4d/3d$ correspondence of \cite{DSV}
is described by the reduction of EHIs  to
the hyperbolic $q$-hypergeometric integrals (see, e.g., \cite{ds:unit,rai:limits}).
Here we proceed with further reduction to the rational level \cite{rai:limits}
described by the integrals employing elementary functions and the
standard gamma function. In \cite{Nakayama:2011pa}, it was found
that introducing into $4d$ SCI of the surface operators
leads to the $2d$ $(4,4)$ SCFT
coupled to the $4d$ theory; here we obtain a more complete $2d$ picture.
A different type of $2d$ partition function associated with
SCIs of $\mathcal{N}=2$ theories was considered recently in \cite{GRRY2}.
A new $2d/3d/4d$ correspondence has been discovered in \cite{S5},
where it was shown that both $4d$ SCIs and $3d$ partition functions
of supersymmetric quiver theories describe statistical sums
of certain integrable models of $2d$ Ising-like spin systems with continuous
values of spins.

Let us discuss first the reduction of $4d$ SCIs to $3d$ partition functions  on the example
of Intriligator-Pouliot duality \cite{Intriligator1}.
As shown in Sect. 2 above and in \cite{Dolan:2008qi},
one can derive SCIs of $4d$ $\mathcal{N}=1$ SYM theories
with the orthogonal gauge groups from the corresponding $SP(2N)$-SCIs.
 But we can reduce the latter $4d$ SCIs
to $3d$ partition functions along the lines of \cite{DSV}.
This results in $3d$ dualities for both SYM \cite{A} and CS \cite{Giveon:2008zn}
theories and both $SP(2N)$ and $U(N)$ gauge groups.
We stress that $4d$ SCIs and dualities
are defined as a rule by unique relations for EHIs,
and at the $3d$-level one obtains the whole web of dualities/SCIs both
for SYM and CS theories based on different gauge groups.

More technically, we start from integral \eqref{eqR} describing SCI
of the electric theory of \cite{Intriligator1}
 \cite{Dolan:2008qi,SV2}. Reducing it to the hyperbolic level
\cite{ds:unit,rai:limits} one finds the following integral \eqref{eqR} \cite{BultPhD}:
\beqa \label{PF1} &&
Z = \frac{1}{N!} \int_{\mathcal{C}^N} \prod_{1 \leq i < j \leq N}
\frac{1}{\gamma^{(2)}(\pm(z_i-z_j);\omega_1,\omega_2)} \nonumber
\\ && \makebox[2em]{} \times
e^{2 \pi \textup{i} (\lambda+1/2) (\omega_1+\omega_2) \sum_{j=1}^N z_j/\omega_1\omega_2} \prod_{i,j=1}^{N}
\gamma^{(2)}(\mu_i - z_j,\nu_i + z_j;\omega_1,\omega_2)
\prod_{j=1}^N \frac{dz_j}{\textup{i} \sqrt{\omega_1\omega_2}},
\eeqa
where $\mathcal{C}$ is the Mellin-Barnes type integration contour.

In \cite{Willett:2011gp}, Willett and Yakov showed that this integral describes
the partition function \cite{Hama:2010av,Jafferis:2010un}
of the electric theory for Aharony duality \cite{A}, which is a $3d$
$\mathcal{N}=2$ SYM theory living on the squashed three-sphere with $U(N)$
gauge group, $N_f=N$ left quarks forming the fundamental representation
of $U(N)$, $N_f=N$ right quarks
forming the antifundamental representation of $U(N)$, and additional singlets $V_{\pm}$.
In (\ref{PF1}), parameters $z_j, j=1,\ldots,N,$ are the fugacities associated with the
gauge group $U(N)$, $\lambda$ is associated with the Fayet-Illiopoulos
term (the coefficient $4 (\lambda+1/2) (\omega_1+\omega_2)$ is introduced
for convenience). Parameters $\mu_i, \nu_i, i=1,\ldots,N,$
are the fugacities of $SU(N) \times SU(N)$ non-abelian global
symmetry group, which are normalized by taking
into account the abelian part of the global symmetry
$U(1)_A \times U(1)_J \times U(1)_R$.

Consider the limit $\omega_2 \rightarrow \infty$
using the hyperbolic gamma function asymptotics
$$ \gamma^{(2)}(z;\omega_1,\omega_2) \stackreb{=}{\omega_2 \rightarrow \infty}
\left(\frac{\omega_2}{2 \pi \omega_1}\right)^{\frac 12 - z}
\frac{\Gamma_{rat}(z/\omega_1)}{\sqrt{2 \pi}}.
$$
The $3d$ partition function (\ref{PF1}) then reduces to
\begin{equation}
Z^{lim} = \frac{\omega_2^{N/2}}{N! \omega_1^{3N/2}}
\left(\frac{\omega_2}{\omega_1}\right)^{-\sum_{i=1}^N (\mu_i+\nu_i)}
Z^{vortex},
\end{equation}
where $Z^{vortex} $ is
the function appearing after formula (2.6) in \cite{Gerasimov:2010td} for $N_f=N$:
\beqa \label{VPF}
Z^{vortex} = \int_{\mathcal{C}^N}
\frac{e^{2 \pi \textup{i} (\lambda+1/2) \sum_{j=1}^N \frac{z_j}{\omega_1}}}
{ \prod_{1 \leq i < j \leq N}
\Gamma_{rat}\left(\frac{z_i-z_j}{\omega_1},\frac{z_j-z_i}{\omega_1}\right)}
 \prod_{i,j=1}^N\Gamma_{rat}\left(\frac{\mu_i - z_j}{\omega_1},
\frac{\nu_i + z_j}{\omega_1}\right) \prod_{j=1}^N \frac{dz_j}{2 \pi \textup{i}}.
\eeqa
The multiplier $\prod_{i \neq j} \Gamma_{rat}((a_i-a_j)/\omega_1)$ standing
in front of the integral in \cite{Gerasimov:2010td} is not relevant for our
discussion and is omitted.

Expression (\ref{VPF}) defines the vortex partition function
for $2d$ $(2,2)$ supersymmetric field theory with $U(N)$ gauge group and $N_f=N$ flavors.
Its representation as a sum over Young diagrams can be obtained
from the partition function of $4d$ $\mathcal{N}=2$ SYM theory \cite{N,NO}
in the limit $\omega_2 \rightarrow \infty$ \cite{Shadchin:2006yz}.
More precisely, in this limit one should also normalize
the variable associated with the instanton parameter
to compensate additional divergences emerging for
$\omega_2 \rightarrow \infty$. In \cite{Gerasimov:2010td}, it was
realized that the latter sum over Young diagrams (instantons) can be
rewritten as a single contour integral \eqref{VPF}, which leads to a better understanding
of this function from the mathematical point of view.

This observation can be generalized to any number of flavors $N_f$ appearing
in \cite{Gerasimov:2010td}
by starting from the partition function  for $3d$ $\mathcal{N}=2$ SYM theory
with $U(N)$ gauge group, $N_f \neq N$ flavors, and
looking at the same limit $\omega_2 \rightarrow \infty$ accompanied by
pulling some of the parameters to infinity (i.e., by integrating out
some of the quarks). Technically, one should
use the asymptotic expansion of the gamma function
$\Gamma_{rat}(x) \rightarrow \sqrt{2 \pi} e^{-x} x^{x-1/2}$
for $x \rightarrow \infty.$ In principle one can get in the same
manner the vortex partition functions for $2d$ supersymmetric field
theories with symplectic and orthogonal gauge groups and different matter
fields (the contribution of adjoint matter field was considered in \cite{B1}).

We conclude by several remarks on the importance of the observation made
in this section. First, it may be very useful for checking a $2d$
analog of Seiberg's duality which was recently proposed and studied in
\cite{Hori:2011pd,Shifman:2010id}. Second, this reduction is close
to the one studied in the literature on connections of $3d$ Chern-Simons
theories with $2d$ supersymmetric field theories \cite{Dimofte:2010tz}
linking vortex partition function to the BPS invariants of dual geometries.
Finally, perhaps the most important, $4d$ SCIs for $\mathcal{N}=1$ SYM theories
are connected to $4d$ partition functions for $\mathcal{N}=2$
SYM theories in the discussed above limit.

\section{Conclusion}

In \cite{SV1,SV2}, we initiated the classification of EHIs
on different root systems and described all known
examples of such integrals for $A_N$, $BC_N$, and $G_2$ root systems
in association with $\mathcal{N}=1$ supersymmetric dualities.
In \cite{SV4}, for all irreducible root systems we described such integrals
associated with $\mathcal{N}=4$ SYM theories; there are
also two more particular examples associated with
$\mathcal{N}=1$ SYM $E_6$ and $F_4$ gauge group theories.

In the present paper we have described all known cases when $BC_n$-EHIs
and corresponding physical dualities with the symplectic gauge
groups are reduced to SCIs/dualities for orthogonal groups
by a restriction of parameters entering the integrals.
Remarkably, there are EHIs
for the $B_N$ and $D_N$ root systems which (currently) cannot be obtained
from integrals on the $BC_N$ root system --- they come from SCIs for $\mathcal{N}=1$ SYM
$SO(2N+1)$ or $SO(2N)$ gauge group theories with the matter fields in spinor
representation. Description of this type of integrals is one of the main
results of the present paper. Physical dualities of the corresponding gauge theories
lead to the conjectures on the equality of respective SCIs.
The latter conjectural identities for EHIs
use characters of the spinor representations,
and they were not predicted by the mathematical developments
prior to the supersymmetric duality ideas intervention. All of them
require now rigorous mathematical proofs.

In addition to SCIs for $\mathcal{N}=1$ dualities considered in this paper,
one can investigate SCIs for electric-magnetic dualities for extended
supersymmetric field theories: the quiver $\mathcal{N}=2$ SYM theories
with $SO/SP$ gauge groups \cite{SV4} or the $SP/SO$-groups
duality \cite{Goddard} in $\mathcal{N}=4$ SYM theory \cite{GPRR,SV4}.
Note that SCIs for extended supersymmetric theories can be obtained from
SCIs of $\mathcal{N}=1$ theories by adjusting the matter content appropriately
together with the hypercharges, as described in \cite{SV2,SV4}.
In the field theory lagrangians
one should fix also appropriately the superpotentials.

As described in \cite{SV2}, one of the physical applications of the
EHI   identities uses the reduction $p=q=0$,
which yields the Hilbert series counting gauge invariant operators
\cite{Hanany2,Pouliot:1998yv}.
Another interesting application of our identities is connected with
the Seiberg type dualities for $3d$ super-Yang-Mills
and Chern-Simons theories with orthogonal gauge groups.
Derivation of $3d$ partition functions out of $4d$ SCIs of \cite{DSV}
yields the most efficient way of obtaining $3d$-dualities.
Technically, the reduction to $3d$ theories is obtained after the
parametrization in $4d$ SCIs of the integration variables, global
symmetry fugacities, and bases $p$ and $q$ similar to \eqref{limit_r},
with the subsequent limit $r \rightarrow 0$. As a result, $4d$
SCIs defined on $S^3 \times S^1$
reduce to partition functions on
the squashed three-sphere $S_b^3$ \cite{Hama:2010av,Jafferis:2010un}.
In this limit the elliptic gamma function is reduced to the hyperbolic gamma function.
It is thus natural to expect that all the dualities considered in \cite{Kapustin:2011gh}
can be recovered by a reduction from the $4d$ SCIs considered in the present paper.
A more detailed description of the resulting hyperbolic
integrals was given in Sect. \ref{Knot}. The reduction procedure for $3d$
theories from $SP(2N)$ to $SO(n)$ gauge group is
similar to the one in $4d$ theories without spinor matter.
For that one needs the duplication formula for the hyperbolic gamma function
$$
\gamma^{(2)}(2z;\omega_1,\omega_2)
= \gamma^{(2)} (z,z + \omega_1/2, z + \omega_2/2, z
+ (\omega_1+\omega_2)/2;\omega_1,\omega_2).
$$
To get $SO(2N+1)$ partition functions it is necessary to restrict three chemical
potentials to $\omega_1/2, \omega_2/2, (\omega_1+\omega_2)/2$ (or two chemical
potentials to $\omega_1/2, \omega_2/2$) and for $SO(2N)$ case one should fix
four chemical potentials equal to $0,\omega_1/2, \omega_2/2,
(\omega_1+\omega_2)/2$ (or three chemical potentials equal to
$0,\omega_1/2, \omega_2/2$). This leads to a variety of $3d$
$\mathcal{N}=2$ supersymmetric dual theories (both SYM and
CS theories) without spinor matter. To construct $3d$ dualities for theories with
the spinor matter one should follow the algorithm suggested in \cite{DSV}.
As a final mathematical remark, we stress that all our computations are performed
analytically, i.e. we described exact (conjectural or proven) equalities
of the compared functions in all admissible domains of values of the parameters.

\

{\bf Acknowledgments.} We dedicate this work to D. I. Kazakov on the occasion
of his 60th birthday with the wishes of further scientific successes.
We are indebted to G. E. Arutyunov, A. A. Belavin, T. Dimofte,
F. A. H. Dolan, S. A. Frolov, S. Gukov, A. V. Litvinov, I. V. Melnikov,
and A. F. Oskin for valuable discussions. We thank the referee
for drawing our attention to papers \cite{MNS,NS}.
G. V. would like to thank  H. Nicolai for general support
and the Universities of Minnesota, Chicago, and Utrecht, CERN,
DESY, Nordita, and the Niels Bohr Institute in Copenhagen for invitations
and warm hospitality during visits to these Institutes.
V. S. was partially supported by the RFBR grants 09-01-93107-NCNIL-a
and 11-01-00980, NRU HSE Academic Fund Program (grant no. 12-09-0064) and
by the Max Planck Institute for mathematics.
This work is supported also by the Heisenberg-Landau program.

\appendix
\section{Characters of representations of orthogonal groups}\label{chSUSP}

In this Appendix we describe characters of representations of orthogonal
groups used in the paper. For needed $SU(N)$ and $SP(2N)$ group characters,
see Appendix A of \cite{SV2}, and invariant measures for all classical groups
are listed in Appendix B of that paper.

$SO(N)$-Groups with even and odd $N$ have
substantially different properties and should be considered separately.
The characters for their spinor representations are described most conveniently
by the expressions involving square roots of $z_j$-variables which
are not analytical. To overcome this obstacle we just double the root lengths
which results in the replacement in characters
variables $z_j$ by $z_j^2$ and assume in the integrals that $z_j$ lie
on the unit circle with positive orientation.
We remark that the adjoint representation
for orthogonal groups coincides with the $T_A$-representation.

{\bf $SO(2N)$ group.} The characters are expressed in terms
of $N$ independent variables $z_i, i=1,\ldots,N$. For the fundamental
representation one has
\beq
\chi_{f,SO(2N)}  =  \sum_{i=1}^N z_i^{\pm1}\equiv \sum_{i=1}^N (z_i+z_i^{-1}).
\eeq

The $T_S$-representation character is \beq \chi_{T_S,SO(2N)} \ =
\ \sum_{1 \leq i < j \leq N} z_i^{\pm1}z_j^{\pm1} + \sum_{i=1}^N
z_i^{\pm2} + N-1,\eeq
the $T_A$-representation character is
\beq \chi_{T_A,SO(2N)} \ = \ \sum_{1 \leq i < j \leq N}
z_i^{\pm1}z_j^{\pm1} + N.\eeq

The needed spinor representation characters are listed case by case.
For $SO(2N)$ groups there are two types of inequivalent spinors,
denoted as $s$ and $c$. For $SO(8)$, the spinor representation $s$ and $c$
are $8$-dimensional, self-conjugate, and their characters have the form
\beq \chi_{s,SO(8)} \ = \ z^{\pm1} + z^{-1} \sum_{1 \leq i < j \leq 4}
z_iz_j,\eeq where $z=\sqrt{z_1z_2z_3z_4}$.
For $SO(10)$, the $s$-representation is $16$-dimensional, it is complex conjugate to
$c$ (so that the character for $c$ can be obtained from the $s$-character by
the substitution $z \to 1/z$). Its character is
\beq
\chi_{s,SO(10)} \ = \ z + z^{-1} \sum_{j=1}^5 z_j + z \sum_{1 \leq i <
j \leq 5} z_i^{-1}z_j^{-1},\eeq
where $z=\sqrt{z_1z_2z_3z_4z_5}$.
For $SO(12)$, the $s$- and $c$-representations are $32$-dimensional,
self-conjugate, and have the character
\beq \chi_{s,SO(12)} \ =
\ z^{\pm1} + z^{-1} \sum_{j=1}^6 z_j + z \sum_{j=1}^6 z_j^{-1},
\eeq
where $z=\sqrt{z_1z_2z_3z_4z_5z_6}$.
For $SO(14)$, the $s$-representation is $64$-dimensional, it is complex-conjugate
to $c$, and its character is (with $z=\sqrt{z_1z_2z_3z_4z_5z_6z_7}$)
\beq \chi_{s,SO(14)} \ = \ z + z^{-1} \sum_{j=1}^7
z_j + z \sum_{1 \leq i < j \leq 7} z_i^{-1}z_j^{-1} + z^{-1} \sum_{1
\leq i < j < k \leq 7} z_iz_jz_k.
\eeq

{\bf $SO(2N+1)$ group.}
All the characters are expressed in terms of $N$ independent
variables $z_i, i=1,\ldots,N$. The fundamental representation
character is
\beq \chi_{f,SO(2N+1)} \ = \
\sum_{i=1}^N z_i^{\pm1} + 1.\eeq

The character for $T_S$-representation is
\beq \chi_{T_S,SO(2N+1)} \ =
\ \sum_{1 \leq i < j \leq N} z_i^{\pm1}z_j^{\pm1} + \sum_{i=1}^N
z_i^{\pm2} + \sum_{i=1}^N z_i^{\pm1} + N,\eeq
the character for the $T_A$-representation is
\beq \chi_{T_A,SO(2N+1)} \ = \ \sum_{1 \leq i
< j \leq N} z_i^{\pm1}z_j^{\pm1} + \sum_{i=1}^N z_i^{\pm1} + N.\eeq

The spinor representation characters are given for the lowest rank groups
only. For $SO(7)$, the spinor representation is
$8$-dimensional and its character is
\beq \chi_{s,SO(7)} \ = \
z^{\pm1} + z^{-1} \sum_{j=1}^3 z_j + z \sum_{j=1}^3 z_j^{-1},\eeq
where $z=\sqrt{z_1z_2z_3}$. For $SO(9)$, the spinor representation is
$16$-dimensional and its character is
\beq \chi_{s,SO(9)} \ = \
z^{\pm1} + z^{-1} \sum_{j=1}^4 z_j + z \sum_{j=1}^4 z_j^{-1} + z^{-1} \sum_{1 \leq i < j \leq
4} z_iz_j,\eeq
where $z=\sqrt{z_1z_2z_3z_4}$. For $SO(11)$,
the spinor representation is $32$-dimensional and its character
is
\beq \chi_{s,SO(11)} \ = \ z^{\pm1} + z^{-1} \sum_{j=1}^5
z_j + z \sum_{j=1}^5 z_j^{-1} + z^{-1} \sum_{1 \leq i < j \leq 5}
z_iz_j + z \sum_{1 \leq i < j \leq 5}
(z_iz_j)^{-1},\eeq
where $z=\sqrt{z_1z_2z_3z_4z_5}$. For $SO(13)$, the spinor representation
is $64$-dimensional and its character is (with $z=\sqrt{z_1z_2z_3z_4z_5z_6}$)
\beq \chi_{s,SO(13)} = z^{\pm1} + z^{-1} \sum_{j=1}^6
z_j + z \sum_{j=1}^6
z_j^{-1} + z^{-1} \sum_{1 \leq i < j \leq 6}
z_iz_j + z \sum_{1 \leq i < j \leq 6}
(z_iz_j)^{-1} + z^{-1} \sum_{1 \leq i < j < k \leq 6} z_iz_jz_k.
\eeq

\end{document}